\documentclass{aastex631}

\received{October 16th, 2025}
\revised{December 20th, 2025}

\submitjournal{AJ}

\shorttitle{Solar Neighborhood K Dwarfs}
\shortauthors{Hubbard-James et al.}

\usepackage{graphicx}
\usepackage{hyperref}
\usepackage{wrapfig}
\usepackage{tikz}
\usepackage{rotating}
\usepackage{subfigure}
\usepackage{xcolor}	
\usepackage{graphicx}	
\usepackage{threeparttable}
\usepackage{array}
\usepackage{booktabs}
\usepackage{rotating}
\usepackage{array}

\usepackage{amsmath}
\usepackage{amssymb,amsmath}
\let\tablenum\relax
\usepackage{siunitx}
\usepackage{gensymb}
\usepackage{xspace}

\newcommand\omicron{o}

\newcommand{\arcpt}{${{\lower3pt\hbox{$^{\prime\prime}$}}\atop{\raise4pt\hbox{.}}}$}
\newcommand{\msun}{$M_\odot$}
\newcommand{\teff}{$T_{eff}$\xspace}
\newcommand{\feh}{$[Fe/H]$\xspace}
\newcommand{\logg}{$log~g$\xspace}
\newcommand{\vsini}{$vsini$\xspace}

\begin{document}

\title{The Solar Neighborhood LV: Spectral Characterization of an Equatorial Sample of 580 K Dwarfs}

\author[0000-0003-4568-2079]{Hodari-Sadiki Hubbard-James} 
\altaffiliation{Visiting Astronomer, Cerro Tololo Inter-American Observatory. CTIO is operated by AURA, Inc., under contract to the National Science Foundation.}
\affil{Department of Physics and Astronomy, Agnes Scott College, Decatur, GA 30030, USA}
\affil{RECONS Institute, Chambersburg, PA 17201, USA}

\author[0009-0006-9244-3707]{Sebastian Carrazco-Gaxiola}
\altaffiliation{Visiting Astronomer, Cerro Tololo Inter-American Observatory. CTIO is operated by AURA, Inc., under contract to the National Science Foundation.}
\affil{RECONS Institute, Chambersburg, PA 17201, USA}
\affil{Department of Physics and Astronomy, Georgia State University, Atlanta, GA 30302, USA}

\author[0000-0002-9061-2865]{Todd J. Henry} 
\altaffiliation{Visiting Astronomer, Cerro Tololo Inter-American Observatory. CTIO is operated by AURA, Inc., under contract to the National Science Foundation.}
\affil{RECONS Institute, Chambersburg, PA 17201, USA}

\author[0000-0003-1324-0495]{Leonardo A. Paredes} 
\altaffiliation{Visiting Astronomer, Cerro Tololo Inter-American Observatory. CTIO is operated by AURA, Inc., under contract to the National Science Foundation.}
\affil{RECONS Institute, Chambersburg, PA 17201, USA}
\affil{Steward Observatory and Department of Astronomy, The University of Arizona, Tucson, AZ 85721, USA}

\author[0000-0002-1457-1467]{Azmain H. Nizak} 
\altaffiliation{Visiting Astronomer, Cerro Tololo Inter-American Observatory. CTIO is operated by AURA, Inc., under contract to the National Science Foundation.}
\affil{Department of Astronomy and Van Vleck Observatory, Wesleyan University, Middletown, CT 06459, USA}

\author[0009-0000-5136-6924]{Xavier Lesley-Saldaña}
\affil{Department of Astronomy, The Ohio State University, 140 West 18th Avenue, Columbus, OH 43210 USA}

\author[0000-0003-0193-2187]{Wei-Chun Jao}
\altaffiliation{Visiting Astronomer, Cerro Tololo Inter-American Observatory. CTIO is operated by AURA, Inc., under contract to the National Science Foundation.}
\affil{RECONS Institute, Chambersburg, PA 17201, USA}
\affil{Department of Physics and Astronomy, Georgia State University, Atlanta, GA 30302, USA}

\author[0009-0004-7539-8129]{Abigail Arbogast} 
\affil{Department of Physics and Astronomy, Agnes Scott College, Decatur, GA 30030, USA}

\correspondingauthor{Hodari-Sadiki Hubbard-James}
\email{hjames@agnesscott.edu}


\begin{abstract}

\noindent We present a spectroscopic characterization of 580 K dwarfs within 33 pc, observed with the CHIRON echelle spectrograph (R=80,000) on the SMARTS 1.5m telescope. This volume-limited sample is part of the RKSTAR survey of $\sim$4400 K dwarf primaries within 50 pc. Using Empirical SpecMatch and the diagnostic lines H$\alpha$ (6562.8 \AA) and Li I (6707.8 \AA), we derive stellar properties, activity status, and age indicators calibrated against 35 benchmark K dwarfs with ages from 20 Myr to 5 Gyr. We find that 7.4\% (43 stars) exhibit signatures of youth and/or chromospheric activity: 19 stars show lithium absorption indicating ages $<$1 Gyr, and 36 display H$\alpha$ emission. Kinematic analysis using BANYAN $\Sigma$ identifies 8 additional young stars through membership in the AB Doradus moving group and the Hyades cluster, bringing the total young/active population to 8.8\% (51 stars). Stellar parameters span 3600--5500 K in \teff, $-$0.60 to $+$0.55 dex in [Fe/H], and $<$10 to $>$25 km s$^{-1}$ in $v\sin i$. A metal-poor population ([Fe/H] $\leq -$0.50 dex) comprises 4\% of the sample. Galactic kinematics place 80\% in the thin disk and 18.4\% in the thick disk, with one halo member (HD 134439). Young and active stars are predominantly thin disk members, with two thick disk exceptions. Cross-matching with NASA's Exoplanet Archive reveals only 7.5\% (44 stars) host confirmed planets as of July 2025. Our results identify 529 mature, inactive K dwarfs as prime targets for terrestrial planet searches, providing a crucial resource for exoplanet habitability studies in the solar neighborhood.

\end{abstract}

\keywords{\href{http://astrothesaurus.org/uat/909}{Late-type stars (909)} ---\href{http://astrothesaurus.org/uat/1509}{Solar neighborhood (1509)} --- 
\href{http://astrothesaurus.org/uat/1558}{Spectroscopy (1558)} ---
\href{http://astrothesaurus.org/uat/1580}{Stellar activity (1580)} --- \href{http://astrothesaurus.org/uat/1580}{Stellar ages (1581)}} ---

\section{Introduction}\label{section:introduction}

K dwarfs, with surface temperatures between 3930--5270 K (our volume-limited sample extends this range to 3600--5500 K to ensure completeness near the K/M boundary) and masses of 0.59--0.88 M$_{\odot}$ \citep{henry93,gray09,pecaut13}, represent promising targets for exoplanet detection and characterization efforts. These stars constitute 11\% of the solar neighborhood population \citep{henry2024} and offer several advantages over their more commonly studied counterparts. K dwarfs have longer main-sequence lifetimes than more massive F and G dwarfs, providing extended periods for planetary formation and biological evolution. Compared to M dwarfs, K dwarfs produce less extreme ultraviolet radiation and exhibit reduced flare activity, potentially offering more stable environments for atmospheric retention on orbiting planets \citep{cuntz16,arney19}.

\begin{figure}[htb!]
\vspace{-0.4 cm}
    \includegraphics[width=9.5 cm, height=8.5 cm]{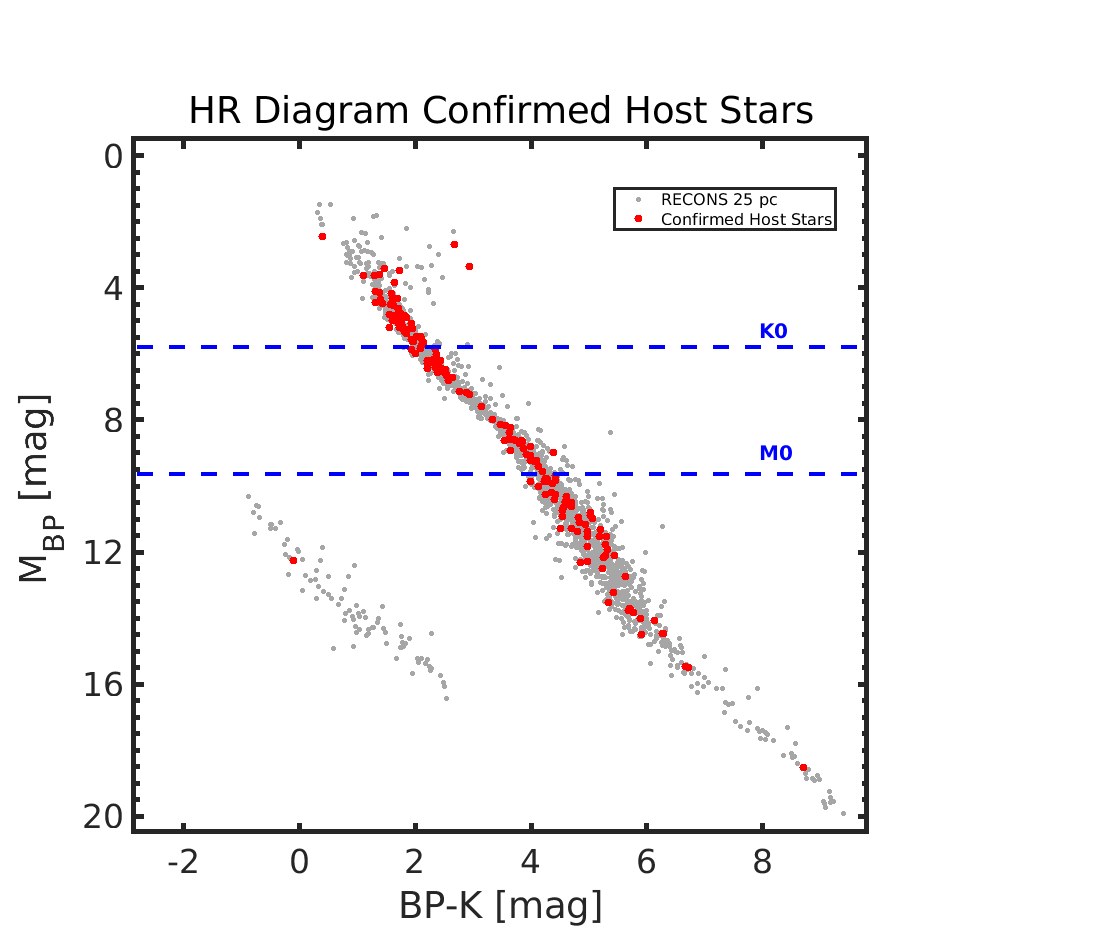}
    \includegraphics[width=9.5 cm, height=8.5 cm]{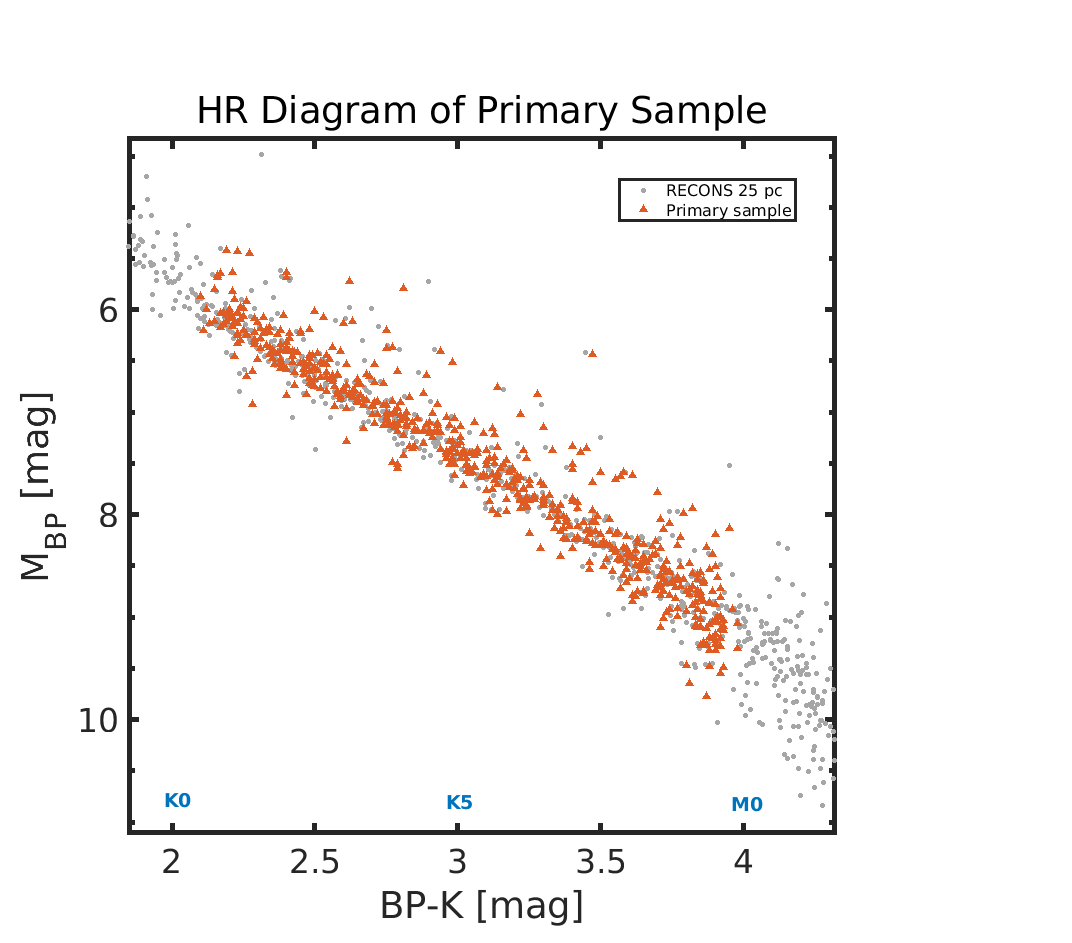}
    \caption{Left: HR diagram displaying the RECONS sample of stars within 25 pc (grey dots),with confirmed exoplanet hosts as of June 2025 from the NASA Exoplanet Archive \citep{nasaexoplanetarchive} highlighted in red. Dashed blue lines indicate the K0 and M0 dwarf boundaries used to select K dwarfs for this work. Right: HR diagram showing the same 25 pc sample in grey with our 580 K dwarfs (orange triangles) from the equatorial 33.3 pc survey. The sequence of objects above the main sequence indicates young and/or unresolved multiple stars, while subdwarfs appear along the lower envelope. Magnitude data are from $Gaia$ DR3 ($BP$) and 2MASS ($K=K_s$) with absolute magnitudes derived using $Gaia$ DR3 parallaxes.}
    \label{fig:hr_exo_primarysample}
\end{figure}

Despite these favorable characteristics, K dwarfs have been systematically underexplored in exoplanet surveys. Figure~\ref{fig:hr_exo_primarysample} illustrates this observational bias, showing that within 25 pc, mid-type K dwarfs (spectral types $\sim$K3V--K6V) host significantly fewer confirmed exoplanets than comparable samples of G and M dwarfs and K dwarfs similar to those more and less massive stars. This disparity reflects survey selection effects rather than an intrinsic lack of planetary systems --- relatively bright G dwarfs and early K dwarfs have been preferentially targeted due to their higher photon rates, enabling higher signal-to-noise observations, whereas late K dwarfs and M dwarfs offer larger planet-to-star size and mass ratios, facilitating transit detection for planets of a given size and resulting in larger radial velocity amplitudes for planets of a given mass \citep{arney19,richeyyowell19}. 

Recent theoretical and observational studies have highlighted K dwarfs as potentially optimal hosts for habitable planets. \citet{cuntz16} determined that early-type K dwarfs provide the most favorable conditions for detecting biosignatures, while \citet{arney19} identified a potential ``K dwarf advantage" for characterizing exoplanet atmospheres. Dedicated surveys targeting nearby K dwarfs, including the RECONS K Star (RKSTAR) survey \citep{paredes21,hubbardjames22} outlined below and the K Dwarfs Orbited By Habitable Exoplanets (KOBE) experiment \citep{lillobox22}, have yielded promising initial results. These efforts suggest that approximately 50\% of nearby K dwarfs host stellar or substellar companions, while approximately\textit{} half may contain habitable zone planets \citep{kunimoto20,paredes21,lillobox22}.

Central to evaluating exoplanet habitability is understanding the evolutionary state and activity level of the host star. Young and magnetically active stars produce enhanced ultraviolet emission, flares, and coronal mass ejections that can significantly impact planetary atmospheres through photoevaporation and atmospheric chemistry alteration \citep{segura10,luger15}. These high-energy processes are particularly problematic for terrestrial planets in habitable zones, where atmospheric stability over geologic timescales is crucial for maintaining liquid water \citep{ribas05,airapetian20}. Conversely, mature, quiescent stars provide stable radiation environments conducive to atmospheric retention and the development of detectable biosignatures \citep{meadows18}.

Stellar activity diagnostics, particularly the H$\alpha$ absorption line (6562.8 Å) and the Li I resonance line (6707.8 Å), have proven effective for identifying active and young stars across spectral types \citep{soderblom93}. H$\alpha$ emission serves as a tracer of chromospheric activity and magnetic heating, while lithium abundance provides a robust age indicator for stars younger than $\sim$1 Gyr through well-understood depletion mechanisms \citep{soderblom10}. These diagnostics are essential for constructing samples of optimal exoplanet host stars and understanding the relationship between stellar evolution and planetary system architecture.

In this paper, we present spectroscopic characterization of 580 K dwarfs within 33 pc, representing the largest uniform survey of high-resolution spectroscopy of nearby K dwarfs to date. Using CHIRON spectra from the SMARTS 1.5m and established spectroscopic diagnostics, we determine stellar properties, activity status, youth status, and kinematic populations to identify the most suitable targets for future exoplanet surveys. In $\S$\ref{section:sample}, we describe the construction of our volume-limited sample and the benchmark calibration set used to establish activity and age relationships. In $\S$\ref{section:spectroscopy}, we present the CHIRON observations and data reduction procedures, followed in $\S$\ref{section:analysis} by our spectroscopic analysis methods, including stellar parameter determinations, activity measurements, and age evaluations. In $\S$\ref{section:characterization}, we present the results of our comprehensive spectroscopic characterization, identifying active, young, and mature stellar populations, and in $\S$\ref{section:kinematics} we analyze the kinematic properties and Galactic population membership for stars in our sample. Our results provide crucial insights into the activity status and ages of the local K dwarf population, and establish a foundation for prioritizing targets in the ongoing search for potentially habitable worlds, as discussed in $\S$\ref{section:discussion} and $\S$\ref{section:conclusion}.

\section{The RECONS K Star (RKSTAR) Project}\label{section:sample}

\subsection{Four K Dwarf Surveys}
\label{subsec:surveys}

The RKSTAR Project is a RECONS\footnote{REsearch Consortium On Nearby Stars, www.recons.org} effort to survey the $\sim$4400 nearest K dwarfs within 50 parsecs of the Sun, a sample constructed primarily using $Gaia$ parallax and photometry measurements. The comprehensive RECONS effort includes four systematic surveys of these K dwarfs --- three are for stellar companions (and in the case of the Radial Velocity Survey, orbiting brown dwarfs and planets as well), while the fourth is a characterization survey that is the focus of the results here. The four surveys are:

The {\bf Wide Field Survey} investigates stellar companions with separations greater than $\sim$1\arcsec, utilizing $Gaia$ data \citep{gaiacollaboration18,gaiacollaboration22} and cataloged companions from, for example, the Washington Double Star (WDS) Catalog \citep{mason01}. To date, the Wide Field Survey includes over 1000 stellar companions with separations of $\sim$50--30000 AU in the entire RKSTAR 50 pc sample \citep{johns24}.

The {\bf Speckle Survey} reveals stellar companions with separations of $\sim$0.5--100 AU, thereby spanning distances similar to the scale of our Solar System. The primary instruments used are optical speckle cameras on 4m to 8m class telescopes, most notably the Differential Speckle Survey Instrument \citep{horch09,horch21}. This survey has detected over 160 stellar companions, with approximately 90 being new discoveries, with the majority of stellar companions found orbiting within 15 AU of the K dwarfs \citep{henry22}.

The {\bf Radial Velocity Survey} uses the CHIRON high-resolution spectrograph on the SMARTS 1.5m telescope at CTIO to reveal companions orbiting K dwarfs that are stars, brown dwarfs, and jovian exoplanets orbiting within $\sim$3 AU. This survey has resulted in the discovery of dozens of stellar companions \citep{johns24}. This survey is complemented by long-term work by others reporting companions with stars and brown dwarf companions at larger separations, as well as planets down to terrestrial masses.

The {\bf Characterization Survey}, which is the main focus of the work reported in this paper, utilizes CHIRON spectra to determine the stellar properties, activity status, ages, and kinematic motions of the nearby K dwarfs. Here we report on two distinct samples of K dwarfs --- a survey sample of 580 field stars within 33 pc in the equatorial region of the sky selected from the RKSTAR sample, and a benchmark comparison sample of 35 stars with reliable estimated ages \citep{hubbardjames22}. 

\subsection{Sample for this Portion of the Characterization Survey}
\label{subsec:define_survey}

In this paper, we report results from the initial portion of the Characterization Survey that began in 2017, focused on a sample of K dwarf primaries created using the results of {\it Hipparcos} and {\it Gaia} Data Release 2 (DR2). K dwarfs were defined to have $M_{BP}$ = 5.30 -- 9.90 mag and $BP-K_s$ = 2.00 -- 4.00 mag, where the $BP$ photometry comes from $Gaia$ and the $K_s$ photometry comes from 2MASS. Among the $\sim$4400 K dwarf systems in the full 50 pc sample, this work focuses on systems located within 33.3 pc, selected using a cutoff in parallax of 30 mas, and situated in the equatorial sky band, ranging from DEC $+$30$^\circ$ to $-$30$^\circ$. This approach results in a volume-limited sample that can be targeted at most major observatories in both hemispheres. 

The list was revised somewhat with $Gaia$ Data Release 3 (DR3), with particular attention paid to updated parallaxes and continued vetting for earlier-type primaries that knocked out K dwarfs that were companions to more massive stars or white dwarfs that were initially more massive than the K dwarfs. The focus on systems in which the K dwarf is the primary is central to our science goals: we aim to understand systems that formed with the primary star having a mass in the K dwarf range (roughly 0.6--0.9 \msun). These systems remain effectively unevolved over the age of the Galaxy, representing the outcomes of formation processes without evolutionary complications. This is particularly important for characterizing multiplicity statistics, including the frequency and properties of lower-mass stellar companions, brown dwarfs, and planets around K dwarfs. K dwarfs that are secondaries in systems with more massive components are maintained in a separate list but excluded from the statistical analyses presented here, as including them would mix distinct formation scenarios and complicate interpretation of companion demographics.

\begin{figure}[htb!]
    \hspace{-1 cm}
    \includegraphics[width=9 cm, height=9 cm]{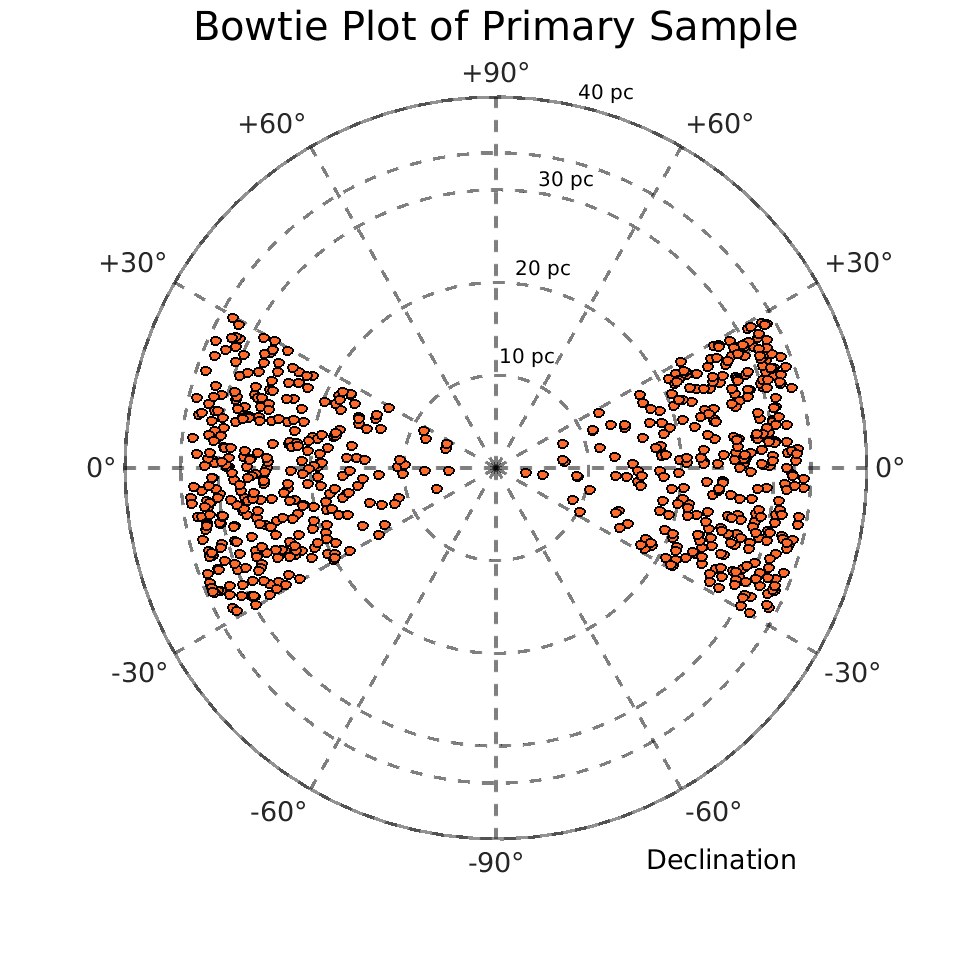}
    \hspace{0 cm}
    \includegraphics[width=9 cm, height=9 cm]{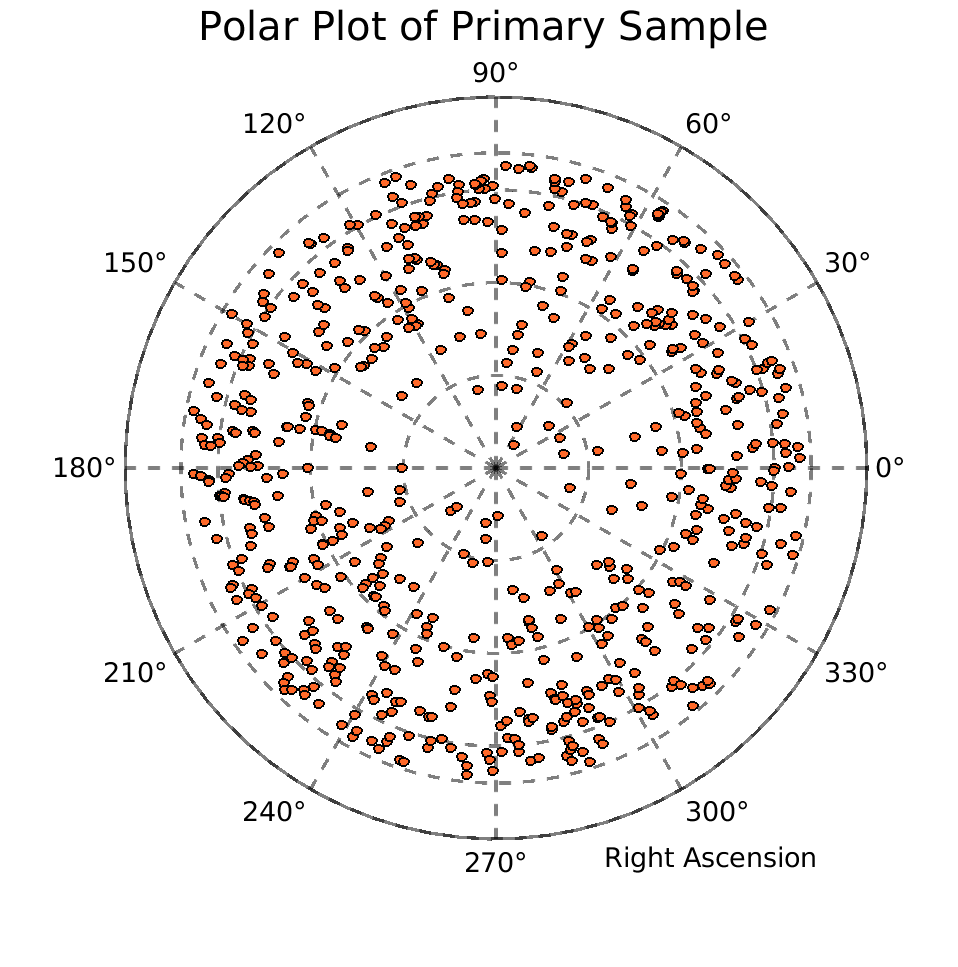}
    \caption{Left: Bowtie plot displaying declination (Dec) (circular direction) and distance (radial direction) for the survey sample of 580 K dwarfs. Right: Polar plot illustrating Right Ascension (R.A.) (circular direction) and distance (radial direction) for the survey sample of 580 K dwarfs. R.A. and Dec positions are based on J2000 coordinates, while distance values were derived from $Gaia$ DR3 parallax measurements. }
    \label{fig:primarysample}
\end{figure}

Table \ref{tab:samplesummary} provides a summary of the sample selection process. In May 2018, 687 K dwarfs were selected based on $Gaia$ DR2 parallax measurements and photometry \citep{gaiacollaboration16, gaiacollaboration18}, with a few additional stars from {\it Hipparcos} that were not in $Gaia$ DR2. Updated parallaxes from the $Gaia$ DR3 \citep{gaiacollaboration22} resulted in the exclusion of 22 K dwarf systems beyond the distance cutoff of 33.3 pc. Upon closer inspection, an additional 26 K dwarfs that are secondaries to earlier spectral type stars or white dwarfs were removed. Another 50 stars were removed due to $BP-K_s$ color cuts and additional quality control measures, and a final 9 for which we could not carry out complete data analyses. This leaves the survey sample of 580 K dwarf systems that are plotted in Figure~\ref{fig:primarysample}, where the left panel illustrates the sky distribution mapping the ``bowtie" configuration that depicts the declination and distance cutoffs, whereas the right panel is the more traditional polar plot. Note the increase in population density with increasing distance out to the horizon at 33.3 pc in both plots due to the projection of larger volumes onto the two dimensional maps representing cross-sectional cuts through space.
\vspace{-0.5 cm}
\renewcommand{\arraystretch}{1.0} 
\begin{table}[htbp]
\centering
\caption{Summary of the Sample Selection Process for the 580 K Dwarfs}
\label{tab:samplesummary}
\begin{tabular}{l|c}
\hline
\textbf{Selection Step} & \textbf{Number of K Dwarfs} \\
\hline
Initial list (May 2018, $Gaia$ DR2) & 687 \\
Excluded due to updated parallaxes ($Gaia$ DR3) & $-$22 \\
Removed due to secondary status & $-$26 \\
Removed due to color cuts and quality control & $-$50 \\
Removed due to incomplete data & $-$9 \\

\hline
\textbf{Final Survey Sample} & \textbf{580} \\
\hline
\end{tabular}
\end{table}
\renewcommand{\arraystretch}{1} 

A list of the 580 stars in the survey sample, along with their positions, proper motions, parallaxes, and photometry, can be found in Appendix \ref{append:append_generaltable}. Sample refinement remains ongoing, as parallaxes may be subject to minor changes in future $Gaia$ Data Releases, and new K dwarfs are likely to be added because $Gaia$ DR2 and DR3 may not have provided astrometric solutions in systems that are short period binaries exhibiting astrometric perturbations. Some additional stars will likely be removed to ensure only the inclusion of systems with K dwarf primaries. For example, if additional white dwarfs are discovered, those systems will be excluded from the sample because the white dwarf progenitor originally had a greater mass and was the primary star. It is important to note that systems with white dwarf primaries are valuable for age determination, so they will still be considered for age calibration work but will not be included in the statistics for K dwarf samples and their companions.

A supplementary benchmark sample of $\sim$100 K dwarfs with age estimates \citep{hubbardjames22} was created to provide measurements of various spectral features and space motions that can be used as guidelines for the larger survey samples. These stars were taken from moving groups, associations, or clusters, plus a handful of field stars within 25 pc that have ages determined via isochrone fitting. Table~\ref{tab:movinggroups} lists the various subsets used to construct this benchmark sample and the estimated ages of each group. The four associations utilized here are the $\beta$ Pictoris moving group ($\beta$ Pic, age $\sim$20 Myr), the Tucana-Horologium association (Tuc-Hor, $\sim$40 Myr), the AB Doradus moving group (AB Dor, $\sim$120 Myr), and the Hyades cluster ($\sim$750 Myr). The four field K dwarfs within 25 pc have age estimates made via model isochrone fits and have ages of 0.3--5.7 Gyr. References for the ages assigned to the groups and individual stars are noted at the end of Table~\ref{tab:movinggroups}. Additional information about this benchmark study can be found in \cite{hubbardjames22}.

\begin{deluxetable}{lcccccccc}
\tablecaption{Moving Groups (M.G.), Associations (Assoc.), Clusters, and Field K dwarfs in the Benchmark Sample \label{tab:movinggroups}}
\tablewidth{0pt}
\tablehead{
\colhead{Group Name} & \colhead{RA (J2000)} & \colhead{DEC (J2000)} & \colhead{Distance} & \colhead{Age} & \colhead{Members$^{b,c}$} & \colhead{K dwarfs} & \colhead{Observed} \\
& & & \colhead{(pc)$^a$} & & & & 
}
\startdata
$\beta$ Pic M.G. & 14 30 & $-$42 00 & $\sim$30 & $\sim$20 Myr$^d$ & 97 & 19 & 11 \\
Tuc-Hor Assoc. & 02 36 & $-$52 03 & $\sim$40 & $\sim$40 Myr$^d$ & 176 & 18 & 10 \\
AB Dor M.G. & 05 28 & $-$65 26 & $\sim$33 & $\sim$120 Myr$^d$ & 84 & 24 & 8 \\
Hyades Cluster & 04 26 & $+$15 52 & $\sim$42 & $\sim$750 Myr$^d$ & 177 & 47 & 10 \\
\\
Field K Dwarfs &&&&&&& \\
$\omicron^2$ Eri & 04 15 16.3 & $-$07 39 10 & 5 & 4.3 Gyr$^e$ & \ldots & \ldots & \ldots \\
20 Crt & 11 34 29.5 & $-$32 49 53 & 10 & 4.6 Gyr$^e$ & \ldots & \ldots & \ldots \\
PX Vir & 13 03 49.7 & $-$05 09 43 & 22 & 0.3 Gyr$^g$ & \ldots & \ldots & \ldots \\
$\epsilon$ Ind & 22 03 21.7 & $-$56 47 10 & 4 & 3.7--5.7 Gyr$^h$ & \ldots & \ldots & \ldots \\
\enddata
\tablenotetext{}{%
\begin{minipage}[t]{0.48\textwidth}
$^a$ Distance from $Gaia$ DR3 \\
$^b$ Membership list for $\beta$ Pic, Tuc-Hor, and AB Dor: \\
\cite{bell15} \\
$^c$ Membership list for Hyades: \cite{gagne18} \\
$^d$ \cite{gagne18}
\end{minipage}
\hfill
\begin{minipage}[t]{0.48\textwidth}
$^e$ \cite{mamajek08} \\
$^f$ \cite{luck17} \\
$^g$ \cite{stanfordmoore20} \\
$^h$ \cite{feng19}
\end{minipage}
}
\end{deluxetable}

\section{Spectroscopic Observations and Data Processing}\label{section:spectroscopy}

\subsection{High-Resolution Spectra from CHIRON}
\label{subsec:hires_spectroscopy}

The CHIRON high-resolution, cross-dispersed echelle spectrograph \citep{tokovinin13,paredes21} at the Small and Moderate Aperture Research Telescope System (SMARTS) 1.5-m telescope at Cerro Tololo Inter-American Observatory (CTIO) was used for the spectroscopic observations secured for this research. CHIRON covers an optical wavelength range of 4150--8800\AA, cross-dispersed into 59 to 62 spectral orders, depending on the selected mode, and can acquire targets as faint as $V$$\sim$18 through a fiber that is 2.7\arcsec~in diameter on the sky. CHIRON has four modes that provide resolutions ranging from 28,000 to 136,000 --- the choice of setup is determined by the scientific goals and the targets' brightnesses. For this work, we utilized CHIRON's slicer mode, which attains a resolution of R = 80,000 and offers minimal light loss with a fast 14-second CCD readout per image. CHIRON offers two wavelength calibration options: a ThAr comparison lamp and an iodine cell. In this study, we used the ThAr lamp for wavelength calibration, as it provides sufficient lines in the spectral orders needed for our science.

Observations at the 1.5-m with CHIRON are acquired by an onsite observer. Since 2017, operations have been led by RECONS team members at Georgia State, who create and manage nightly CHIRON observing queues, carry out the observing in tandem with CTIO staff, reduce the spectra using a modified pipeline based on that described in \cite{tokovinin13}, and deliver reduced data to the world \citep{paredes21}. All spectroscopic data used in this study were obtained between June 2017 and March 2022. Each visit to a star consisted of acquiring a single exposure, with an integration time of 900 seconds for stars with Johnson $V$ magnitudes brighter than 10.99 and up to 1800 seconds for fainter stars. These exposure times ensure a signal-to-noise ratio (SNR) in the spectra greater than 25 near a wavelength of 6740\AA~(see below), which is crucial for reliable analysis. To maintain overall data quality for the entire survey, stars with spectra having SNR $<$25, typically due to poor sky conditions or inconsistent tracking during an observation, were placed on a re-observation list to secure reliable equivalent width measurements for spectral features of interest. The final set of spectral observations has a mean S/N of 50 at the Li~I echelle order, with a standard deviation of 48. Only 44 spectra in the set have an S/N at Li~I lower than 25. 

Sets of bias and quartz lamp flat-field calibration frames were routinely taken before and after each observing night and used to set the background levels and correct for pixel-to-pixel variations on the CCD chip. To calibrate the wavelength scale for each spectrum, an observation was followed by a single ThAr lamp exposure with a duration of less than 1 second, which enables precise wavelength calibration of the acquired spectra. Table \ref{tab:table_obslog} in Appendix~\ref{append:tablechironobs} provides a comprehensive list of the CHIRON observations carried out for this characterization study. 

\subsection{Spectra Assembly and Normalization} 
\label{subsec:spectralassembly}

After the data were bias-subtracted and flat-fielded, the extracted spectra were blaze-corrected to flatten the spectra so that radial velocities, equivalent widths, and stellar properties could be derived. The first step for the blaze removal was to trim the first and last 100 pixels of the FITS flux-table file from each order. Major absorption features, such as the H${\alpha}$ line (at 6563~$\AA$) for order 38, were masked to remove their effects so that the continuum flux could be fit. Remaining points in each order were fit using a polynomial using Python's \textit{scikit-learn} package \citep{pedregosa2011} with a robust linear regression to reduce the effect of outlier points when fitting the continuum flux. A sixth-degree polynomial was fit to every order of interest for this survey, except for order 40 where the Li I doublet falls, for which a seventh-order polynomial was used. The blaze removal was performed on 17 echelle orders total: 14 orders for radial velocity calculations, plus three additional orders to analyze the H${\alpha}$, Li~I, and Ca~II~IRT$_{2}$ spectral lines. Finally, the original unmasked spectral order was divided by the blaze function fit to obtain a flattened spectrum, which is then normalized to 1.00 at the mean continuum value.

Astrometric data from Gaia DR3 and the time of the middle of the exposure were used to adjust for barycentric motion at the time of the observation. The barycenter velocities and time in Julian dates for the corrections were calculated using the python's open-source package \textit{barycorrpy}\footnote{\url{https://github.com/shbhuk/barycorrpy}} \citep{kanodia18, wright14}. A star's epoch systemic radial velocity, noted as the $\gamma$ velocity here, was calculated using steps 3 to 9 of the \textit{Radial Velocity Pipeline} described in \cite{paredes21}. Briefly, an RV was calculated using a cross-correlation function (CCF) of the stellar spectrum with appropriate K dwarf template spectra taken from the work of \citet{blancocuaresma14} and the peak location found by using a Cauchy–Lorentz function for the 14 CHIRON echelle orders numbered 10, 12, 13, 16, 17, 18, 20, 21, 22, 23, 24, 27, 30, and 35. Errors on the RVs are the standard deviations of the 14 measurements. The $\gamma$ velocities were then incorporated to offset all spectra to rest wavelengths. Once the spectra are deblazed and wavelength corrected, they are ready for scientific analysis. The detailed discussion of the $\gamma$ velocity results is given in $\S$\ref{subsec:chiron_gamma} and results can be seen in Figure~\ref{fig:gamma_mainsample}.

\subsection{K Dwarf Spectral Gallery}
\label{subsec:spectral_library}

Our spectroscopic observations enable the construction of a high-resolution activity and age spectral gallery of 580 nearby K dwarfs. We process this uniform, high-quality dataset to investigate stellar properties such as chromospheric activity, age indicators, metallicities, and surface gravities. The gallery will serve as a resource for broader astronomical studies, including exoplanet host characterization and comparative stellar astrophysics. Each spectrum was processed through the standardized pipeline described in $\S$\ref{subsec:spectralassembly}, and as mentioned, only spectra achieving a SNR greater than 25 at 6740 \AA{} are included. This standardized RECONS post-pipeline reduction produces a directory for each target spectrum. The output includes raw spectra, reduced files of wavelength/flux pairs in both CSV and FITS formats organized by individual stellar targets, and plot images of the selected echelle orders to aid visual confirmations. A first version of the gallery has been made publicly available\footnote{\url{https://hodarijames.github.io/spectral_library/page1.html}} and will continue to expand with additional spectra from the broader RKSTAR sample.

The gallery includes four key diagnostic features for each star: the Na I doublet at 5890/5896 \AA{} (surface gravity indicator), the H$\alpha$ line at 6563 \AA{} (chromospheric activity indicator), the Li I resonance line at 6708 \AA{} (youth indicator), and the Ca II infrared triplet line at 8542 \AA{} (chromospheric activity indicator). This paper focuses on the analysis of the H$\alpha$ and Li I features; future work will evaluate the Na I and Ca II triplet features when larger samples are available. Figure~\ref{fig:appendix_53_specialones} in Appendix~\ref{append:append_spectrallibrary} showcases all 53 young, active, or otherwise unique stars from our sample, illustrating the diversity captured across a broad range of effective temperatures and metallicities. Groups A--D highlight young and active K dwarfs identified spectroscopically through H$\alpha$ emission and lithium absorption features. Group E features stars identified kinematically as young (Hyades cluster members) as well as peculiar systems including a newly discovered spectroscopic binary (SB2s) and a halo star.

\section{Spectral Analysis}
\label{section:analysis}
\subsection{Stellar Parameters} 
\label{subsec:analysis_stellar parameters}

For this study, we determined fundamental stellar parameters using the Python algorithm Empirical SpecMatch (ESM) developed by \citet{yee17}. ESM derives stellar properties by comparing an input optical spectrum to a library of high-resolution (R$\sim$55,000), high signal-to-noise ratio (SNR $>$ 100) spectra of 404 well-characterized calibrator stars. These library stars were observed with the High Resolution Echelle Spectrometer (HIRES) on the 10-meter Keck telescope in Hawaii, as part of the California Planet Search. The stars' parameters span effective temperatures (\teff) from 3000 to 7000 K, metallicities (\feh) from $-$0.6 to $+$0.6 dex, stellar radii from 0.1 to 16 solar radii, and spectral types from F1 to M5. These parameters were derived through various independent methods, including interferometry, optical and near-infrared photometry, asteroseismology, and local thermal equilibrium (LTE) spectral synthesis \citep{yee17}. The ESM algorithm analyzes optical spectra through wavelengths 5100 to 5800 \AA{}, all of which are covered in our CHIRON observations at a R$\sim$80,000. This spectral range includes diagnostic features such as the magnesium b triplet (Mg I b, 5100 to 5340 \AA{}) while avoiding telluric line contamination found between 6270 and 6310 \AA{}. The spectral region around the Mg I b triplet is particularly diagnostic because line ratios in this region constrain effective temperature (\teff), the shapes of the lines provide measures of surface gravity (\logg), and specific iron lines yield metallicity (\feh).

The algorithm operates systematically, beginning with a correction of the target star's line-of-sight velocity to align its spectrum with reference spectra; this has already been accomplished for our spectra as described in $\S$\ref{subsec:spectralassembly}. The algorithm then performs a bootstrapping procedure to identify the target spectrum with the library spectrum exhibiting the highest median correlation peak. ESM then conducts pairwise matching with each library spectrum, fitting for rotational broadening (\vsini) and refining continuum normalization via cubic spline fits. The final stellar parameters are derived from a weighted linear combination of the five best-matching library spectra, using nonlinear least-squares minimization to reduce the unnormalized $\chi^2$ statistic. ESM provides parameter uncertainties based on the scatter in differences between the algorithm-derived and library values of the stellar parameters. Within its calibration range, typical uncertainties for K dwarfs are $\sim$100 K for \teff, 0.09 dex for \feh, and 0.60 dex for \logg. These values are added in quadrature to the errors from the four individual measurements that use different sections of the spectra for each star. While we expect some correlation between these ESM parameter uncertainties and 
the SNR of individual spectra, ESM was designed to work robustly across a range of spectral qualities. A detailed analysis of how parameter uncertainties scale with SNR 
and other spectral characteristics for our sample will be presented in a future methodological paper.

An important consideration in our analysis is the limited representation of mid-K dwarfs (4200--4800 K) with high metallicities ([Fe/H] $>$ 0.2 dex) in the ESM spectral library. This underrepresentation likely introduces systematic uncertainties and potentially underestimates metallicity values for stars in these ranges. Nonetheless, we include results for these stars in our analysis, but one must exercise caution when interpreting the derived parameters for stars in these temperature and metallicity ranges. Future expansions of the spectral library to include mid-K dwarfs could reduce these uncertainties. 

To validate our results, we compared ESM-derived parameters for five well-studied benchmark K dwarfs against independent determinations from the PASTEL catalog \citep{soubiran20}, which compiles stellar parameters from multiple high-resolution spectral analyses (Table~\ref{tab:esm_vs_pastel}). Minor discrepancies observed are primarily attributable to differences in modeling approaches, line lists, and adopted stellar physics. Our ESM-derived parameters generally match those from PASTEL within the uncertainties, with the single exception among the 15 quantities determined being the \feh value for $\omicron$$^2$ Eri for which we suspect the metallicity errors are underestimated because ESM has sparse coverage of library stars in this low metallicity region.

\begin{table}[htb!]
\centering
\caption{Comparison of Stellar Properties from Empirical SpecMatch (ESM) and the PASTEL Catalog}
\label{tab:esm_vs_pastel}
\begin{tabular}{lcccccc}
\hline\hline
Star & ESM $T_{\rm eff}$ & PASTEL $T_{\rm eff}$ & ESM [Fe/H] & PASTEL [Fe/H] & ESM $\log g$ & PASTEL $\log g$ \\
     & (K) & (K) & (dex) & (dex) & (dex) & (dex) \\
\hline
$\omicron$$^2$ Eri & $5109 \pm 105$ & $5133 \pm 43$ & $-0.43 \pm 0.09$ & $-0.29 \pm 0.01$ & $4.49 \pm 0.69$ & $4.52 \pm 0.02$ \\
HD 50281           & $4710 \pm 102$ & $4767 \pm 31$ & $+0.03 \pm 0.09$ & $+0.02 \pm 0.03$ & $4.54 \pm 0.69$ & $4.54 \pm 0.08$ \\
20 Crt             & $5220 \pm 110$ & $5196 \pm 23$ & $-0.47 \pm 0.10$ & $-0.40 \pm 0.02$ & $4.54 \pm 0.69$ & $4.60 \pm 0.04$ \\
PX Vir             & $5195 \pm 113$ & $5174 \pm 63$ & $-0.12 \pm 0.17$ & $-0.08 \pm 0.03$ & $4.56 \pm 0.70$ & $4.55 \pm 0.05$ \\
$\epsilon$ Ind     & $4617 \pm 104$ & $4641 \pm 21$ & $-0.09 \pm 0.09$ & $-0.13 \pm 0.03$ & $4.58 \pm 0.70$ & $4.54 \pm 0.22^a$ \\
\hline
\multicolumn{7}{l}{$^a$PASTEL $\log g$ recalculated to exclude outlier value of 2.87 dex.}
\end{tabular}
\end{table}

\subsection{Activity \& Youth Indicators}
\label{sec:act_you}

We selected four key spectral lines as indicators of stellar activity and youth for the K dwarf study: the Na I doublet at 5890 and 5896 \AA{}, H$\alpha$ at 6563 \AA{}, Li I at 6708 \AA{}, and the Ca II infrared triplet (IRT) line at 8542 \AA{}. This paper focuses on the H$\alpha$ and Li I lines, while future work with larger samples will incorporate the Na I and Ca II features. These lines were chosen based on their established sensitivities to surface gravity, chromospheric activity, and lithium depletion variations associated with stellar age and evolutionary status. For detailed discussions on the selection criteria, underlying physical processes, and prior usage of these lines as age and activity indicators, we refer the reader to \cite{hubbardjames22}, and references therein. Briefly, the Na I doublet lines (5890 and 5896 \AA{}) are sensitive to surface gravity changes, indicative of stellar evolutionary phases and corresponding age \citep{montes01,soderblom10}. The H$\alpha$ and Ca II IRT (8542 \AA{}) lines serve as effective tracers of chromospheric activity, which typically decreases with stellar age \citep{skumanich72,soderblom10}. The Li I resonance line at 6707.8 \AA{} is strongly correlated with stellar age due to lithium depletion processes occurring early in a star's lifetime \citep{soderblom93, white07, lopezsantiago10, binks14, riedel17}.

\subsection{Equivalent Width Measurement Methods}\label{sec:ew_measurement}

\begin{figure}[htb!]
    \hspace{0 cm}
    \centering
    \includegraphics[width=0.8\textwidth]{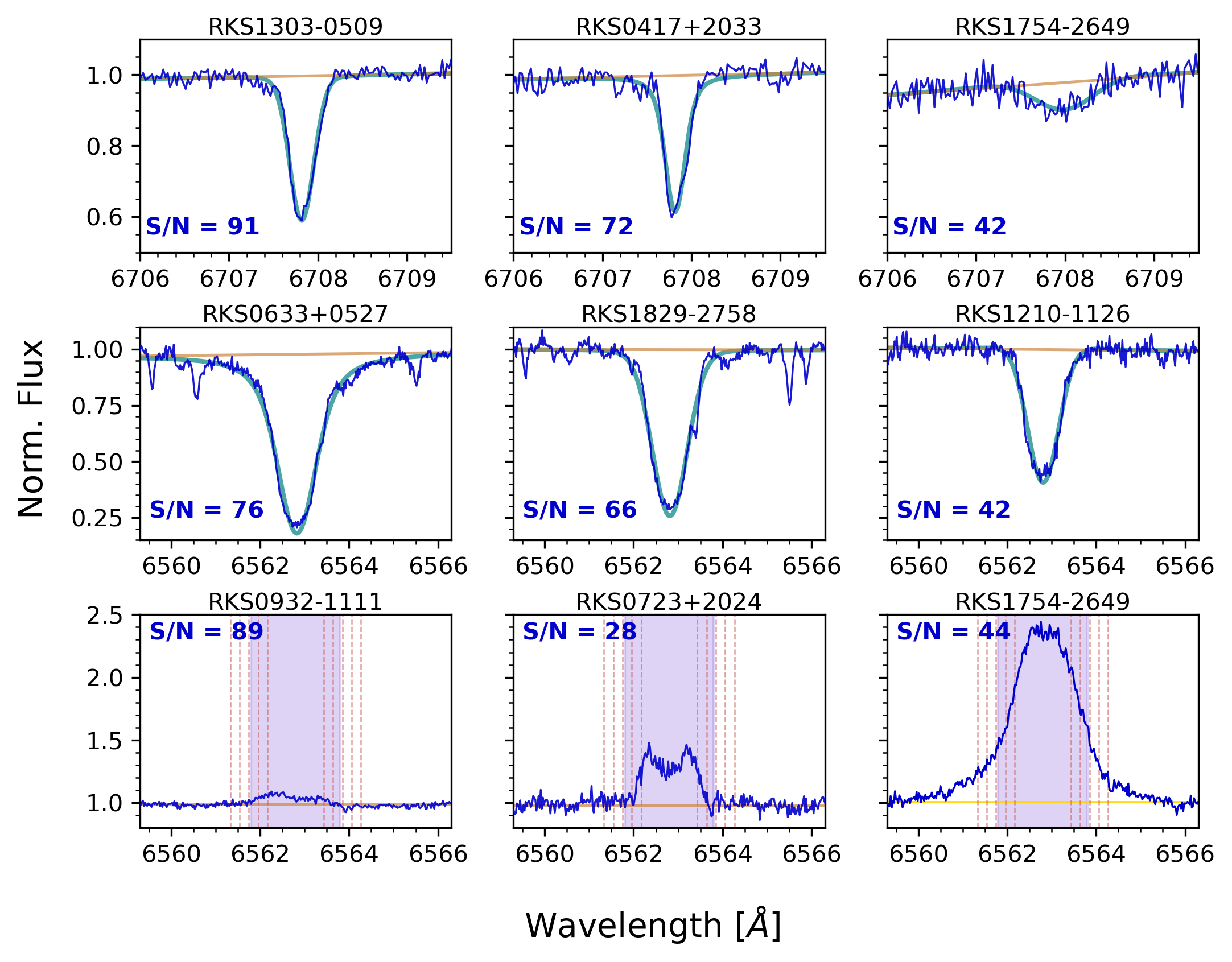}
    \hspace{0 cm}
    \caption{Examples of K dwarf spectra (blue) showing activity and age indicators with EW measurement methods. Voigt profiles (green), local continuum levels (yellow), and integration windows (purple) are shown. Top row: Li I absorption measured via Voigt fitting. Middle row: H$\alpha$ absorption measured via Voigt fitting. Bottom row: H$\alpha$ emission or filled-in profiles measured via the window method using \textit{specutils}. Final EWs were computed by integrating flux over a nominal 2.1\AA~window (purple) and four additional windows at ±10\% and ±20\% of the nominal width (dashed red lines).}
    \label{fig:sp_fits}
\end{figure}

To quantify stellar activity and youth, we measured equivalent widths (EWs) of key spectral lines using two main approaches, depending on the shape of the observed feature. Figure~\ref{fig:sp_fits} shows example spectra with the two methods used for various line shapes: Voigt profile fits and integration windows. Spectral line EW measurements for the H$\alpha$ and Li I lines --- the two lines analyzed in this paper --- are given in Appendix~\ref{append:append_stellarpropertiestable}.

{\it Smooth Absorption Features:} For absorption features such as the Li I line and most H$\alpha$ profiles, we used Voigt profile fitting, implemented using Python's \texttt{scipy.optimize.curve\_fit} module. This method models the spectral line with a Voigt function that accounts for both Gaussian and Lorentzian broadening. Each fit included a local linear continuum, and the EW was computed by numerically integrating the area between the fitted profile and the continuum using \texttt{numpy}. To estimate typical uncertainties in this method, we analyzed H$\alpha$ profiles in a subset of seven stars spanning the luminosities and colors of K dwarfs that had multiple high-SNR spectra. For each, we performed Voigt fits to five spectra and calculated the standard deviation of the resulting EWs. These tests showed typical errors ranging from 1 to 49 m\AA, with relative uncertainties of $\sim$1--7\%, so we adopt a 5\% error as typical for both H$\alpha$ and Li I absorption features given that they are similar in shape. These results are summarized in Table~\ref{tab:ew_errors}.

\renewcommand{\arraystretch}{1.0} 

\begin{table}[h]
\centering
\begin{threeparttable}
\caption{Description of Spectral Lines and Measurement Windows}
\label{tab:spectral_lines}
\begin{tabular}{cccccc}
\hline
\textbf{CHIRON} & \textbf{Line} & \textbf{Purpose} & \textbf{Lab $\lambda$} & \textbf{EW Window} & \textbf{Typical} \\
\textbf{Order}& &\textbf{Name} & \textbf{(\AA{})} & \textbf{Width} \textbf{(\AA{})} & \textbf{Error} \textbf{(\AA{})} \\
\hline
28 & Na I D1            & Gravity   & 5889.9 & 6.0 & 0.23 \\
   & Na I D2            & Gravity   & 5895.9 & 5.0 & 0.20 \\
38 & H$\alpha$          & Activity  & 6562.8 & 2.1 & 0.07 \\
40 & Li I               & Age       & 6707.8 & 1.4 & 0.03 \\
58 & Ca II IRT\tnote{a} & Activity  & 8542.0 & 7.0  & 0.25  \\
\hline
\end{tabular}
\begin{tablenotes}
  \small
  \item \hspace{1.25cm}[a] IRT: Infrared Triplet.
\end{tablenotes}
\end{threeparttable}
\end{table}

  
\renewcommand{\arraystretch}{1} 

{\it Complex Features:} For the subset of stars with complex H$\alpha$ 
profiles that cannot be reliably fit with Gaussian functions—including emission profiles, partially filled-in cores, and broader or more complex line shapes (e.g., rapidly rotating stars)—we used fixed-window integration methods. 
This approach provides consistent, reproducible measurements across diverse profile morphologies where defining variable window boundaries would be subjective. We carried out window-based integrations using the \texttt{specutils} package \citep{earl20} with wavelength coverage as outlined in Table~\ref{tab:spectral_lines}. EWs were calculated using the classical definition:

\begin{equation}
    EW = \int_{\lambda_{1}}^{\lambda_{2}} \left(1 - \frac{F(\lambda)}{F_C} \right) d\lambda,
\end{equation}

\noindent where $F(\lambda)$ is the observed flux, $F_C$ is the continuum level, and $\lambda_1$ and $\lambda_2$ define the spectral window. For H$\alpha$, we calculated the continuum as the average flux in the adjacent regions 6558.4 to 6560.4~\AA{} and 6565.2 to 6567.2~\AA{}, and for the Li I line we adopted a fixed continuum of $F_C = 1$. To quantify the uncertainty arising from window placement, each EW was calculated five times: once using the nominal window and four times using windows increased or decreased by 10 and 20 percent. The final EW was the average of these five measurements, and the standard deviation was adopted as the uncertainty, systematically capturing the sensitivity of the measurement to window definition. For EWs determined using windows, a summary of the spectral lines, their central wavelengths, window spans, and typical errors is provided in Table~\ref{tab:spectral_lines}.

\begin{table}
    \centering
    \begin{tabular}{l|cc|ccc|c}
    \hline
        RKSTAR ID    & M$_{BP}$ & $BP-K$  & $\langle$~EW~H$_{\alpha}$~$\rangle$ &      $\sigma$EW~H$_{\alpha}$ & Error & SNR at H$_{\alpha}$\\
                     & \textit{mag} & \textit{mag} & \textit{\AA} & \textit{\AA} & \textit{\%} &  \\
        \hline
        RKS2009$+$1648 & 5.92 & 2.26 & 1.165 & 0.001 & 0.8 & 92.8\\
        RKS2125$+$2712 & 6.49 & 2.30 & 1.109 & 0.014 & 1.3 & 73.6\\
        RKS0453$+$2214 & 6.93 & 2.75 & 0.844 & 0.049 & 5.8 & 57.4\\
        RKS2009$-$0307 & 7.36 & 3.07 & 0.777 & 0.016 & 2.0 & 50.4\\
        RKS0514$+$0039 & 7.93 & 3.24 & 0.713 & 0.025 & 3.5 & 43.5\\
        RKS1729$-$2350 & 8.53 & 3.46 & 0.555 & 0.012 & 2.4 & 51.1\\
        RKS1854$+$2844 & 9.00 & 3.83 & 0.420 & 0.030 & 7.1 & 28.7\\
        \hline
    \end{tabular}
    \caption{H$\alpha$ equivalent width (EW) data for seven K dwarfs of various luminosities and colors used to estimate the errors in EW measurements obtained with the Voigt fitting method.}
    \label{tab:ew_errors}
\end{table}

\subsection{Signal-to-Noise (SNR) Considerations}\label{sec:snr}

Reliable measurements of spectral line strengths depend on the SNR of each echelle order. For this study, we determined SNR individually for each of the orders containing the H$\alpha$ (order 38) and Li I (order 40) lines. We measured SNR values following the procedure established for CHIRON spectra by \citet{tokovinin13}: calculating the ratio of the mean flux to its standard deviation across three adjacent pixels located near the blaze peak of each order, where the signal is maximized and systematic variations in the continuum are minimized. This approach provides a practical estimate of the spectral quality achieved for each observation.

To establish appropriate SNR thresholds for reliable measurements in our 
survey, we conducted visual inspection of spectra spanning a range of SNR values. 
For H$\alpha$, we found that SNR greater than 15 was sufficient for reliable 
measurement, which preserved 97\% of the spectra taken for the survey K dwarfs. 
For the Li I line, distinguishing it from the nearby Fe I line at 6707.4~\AA{} 
required SNR greater than 30. This threshold was met in 83\% of the spectra. EWs 
from spectra with orders falling below the thresholds were marked as upper limits 
or excluded entirely, depending on the clarity of the line shape and the reliability 
of the EW measurement.

\subsection{Gamma Velocity Measurements and Kinematics} \label{subsec:gamma_kinematics}

We derived systemic radial velocities, here called $\gamma$ velocities, from the high-resolution CHIRON spectra for 572 of the 580 K dwarfs in the survey sample. 8 stars could not have their radial velocities measured due to rapid rotation or insufficient spectral quality. The methodology is described in $\S$\ref{subsec:spectralassembly}. $Gaia$ DR3 does not have values for 35 of the survey stars, and the DR3 values are generally $\sim$10 times less precise than our values from CHIRON that typically have errors of 0.1--0.3 km~s$^{-1}$, dependent on spectral quality and stellar rotation. Of course, long-term monitoring of $\gamma$ velocities can reveal companions such as low-mass stars, brown dwarfs, or exoplanets, even if these companions cannot be directly imaged or observed via eclipses or transits.

We calculated the Galactic $UVW$ space velocities for 572 stars in the survey using the \texttt{gal\_uvw} function from the Python AstroLib library to provide kinematic motions for the stars. Two additional stars were excluded from this analysis because their spectroscopic binary nature (double-lined spectroscopic binaries) prevented reliable radial velocity determinations required for the kinematic calculations. This process required using our computed $\gamma$ velocities combined with each star's celestial coordinates, proper motions, and parallax-derived distances from Gaia DR3 \citep{gaiacollaboration22}. The resulting space velocities represent Cartesian motions aligned with Galactic coordinates, where the $U$ axis is directed toward the Galactic center, the $V$ axis aligns with Galactic rotation, and the $W$ axis points toward the north Galactic pole. The calculated velocities presented in Appendix~\ref{append:append_kinematicstable} (Table~\ref{tab:table_kinematics}) are not corrected for solar motion relative to the Local Standard of Rest (LSR).

The $UVW$ motions allow us to evaluate potential membership of our sample stars in specific moving groups or stellar associations because members typically share similar $UVW$ motions. Furthermore, these velocities enable us to categorize our sample into the broader Galactic populations --- Thin Disk, Thick Disk, and Halo --- because each population exhibits characteristic kinematic properties. Thin Disk stars display relatively low random velocities and nearly circular Galactic orbits. Thick Disk stars exhibit larger velocities, particularly perpendicular to the Galactic plane, and have larger, more eccentric Galactic orbits. Halo stars, typically older and more metal-poor than disk stars, exhibit significantly higher random velocities with highly elliptical orbits and no distinct rotational preference. While kinematics alone cannot conclusively establish membership in specific associations or Galactic populations \citep[e.g.,][]{gagne18,riedel14}, these measurements combined with complementary analyses of age and metallicity provide essential context for understanding the structure, dynamics, and evolutionary history of stars in the solar neighborhood.

\section{Spectral Characterization of K Dwarfs}\label{section:characterization}

\subsection{Stellar Parameters Results: Temperatures, Metallicities, Surface Gravities, and Rotational Velocities}

The fundamental stellar parameters for our survey stars were derived using the Empirical SpecMatch (ESM) methodology described in $\S$~\ref{subsec:analysis_stellar parameters}. ESM's ability to match observed spectra to a library of well-characterized calibrator stars enables determinations of effective temperatures (\teff), metallicities (\feh), surface gravities (\logg), and projected rotational velocities (\vsini). These stellar parameters for the 580 survey stars are presented in Table~\ref{tab:stellar_properties} in Appendix~\ref{append:append_stellarpropertiestable}.

\begin{figure}[htb!]
\centering
    \includegraphics[width=0.85\textwidth]{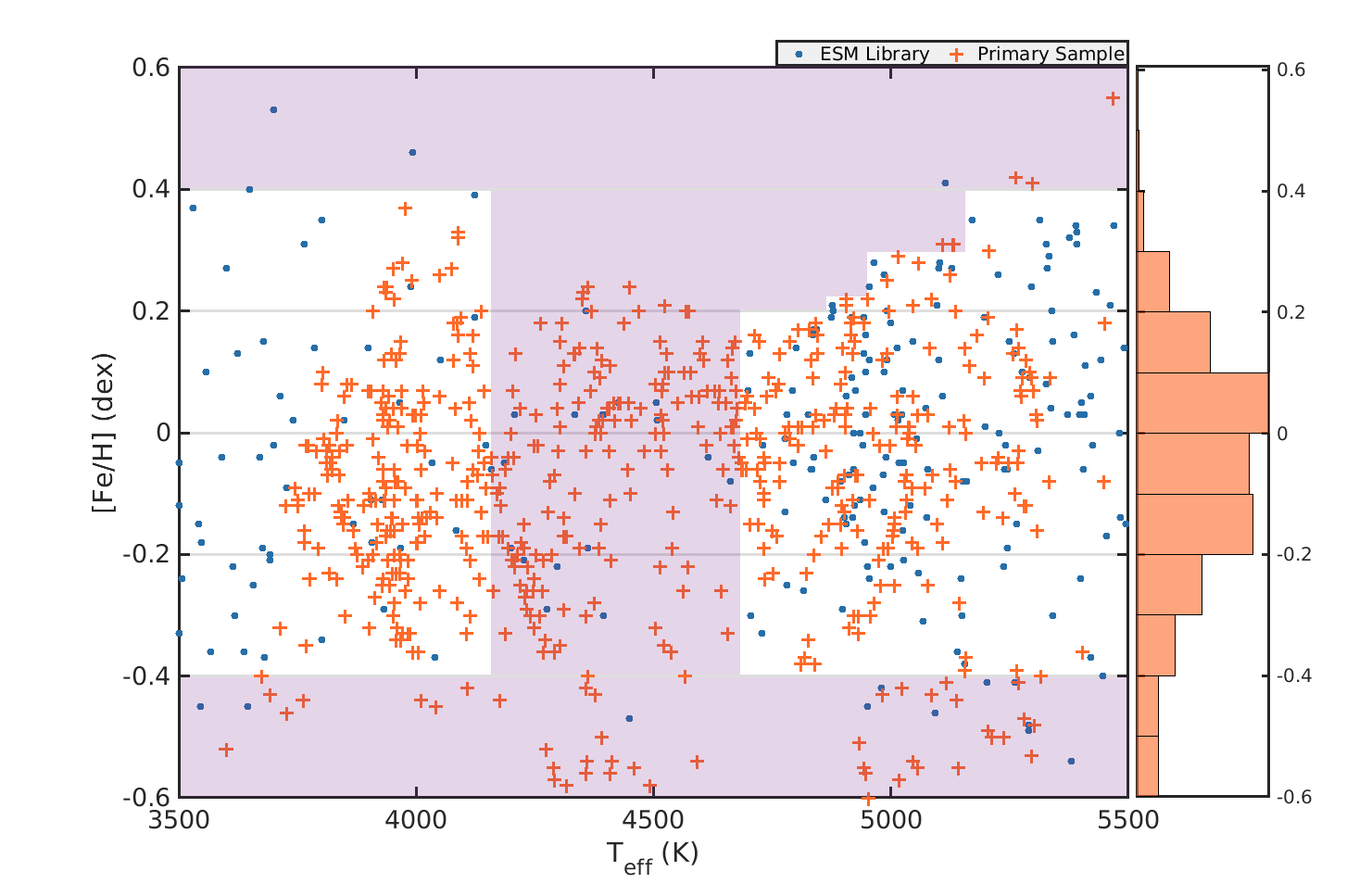}
    \caption{Distribution of \feh and \teff values determined via ESM for 571 of the 580 K dwarfs in our survey sample (orange plus symbols). The 215 stars within this \teff \& \feh range from the ESM library \citep{yee17} are shown as blue dots for comparison. The purple shaded regions highlight areas where the ESM library coverage is sparse, potentially affecting the reliability of derived parameters in these regions. The histogram on the y-axis shows the metallicity distribution of our sample in 0.1 dex bins, with a mean value of $-$0.02 dex.}
    \label{fig:feh_teff_primarysample}
\end{figure}

Figure~\ref{fig:feh_teff_primarysample} displays the distribution of our K dwarf sample in metallicity-temperature space. The survey stars are shown as orange plus symbols, while the 215 stars from the ESM library \citep{yee17} are marked by blue dots for comparison. Our K dwarfs generally span a temperature range of 3600 to 5500 K, with two higher temperature stars at 5582K and 6177K --- the latter star is a a fast rotator and the temperature from ESM is likely erroneous.  The stars exhibit metallicities from $-$0.6 to $+$0.4 dex, with one outlier near $+$0.6 dex. The histogram on the y-axis shows the metallicity distribution in 0.1 dex bins, revealing that a substantial fraction (413/580 = 71\%) of our sample possesses solar-like metallicities between $-$0.2 and $+$0.2 dex. The mean metallicity for the entire sample is $-$0.02 dex, confirming the predominance of solar-metallicity stars in the solar neighborhood. We adopt minimum errors of 100 K in \teff, 0.09 dex in \feh, and 0.60 dex for \logg based on \cite{yee17}.

Notable limitations evident in Figure~\ref{fig:feh_teff_primarysample} are three regions in the ESM library with sparse coverage, including metallicities greater than $+$0.4 dex, less than $-$0.4 dex and at all metallicities for temperatures between 4200 and 4700 K. These gaps, highlighted by purple shaded regions, may result in systematic biases against identifying stars in these parameter spaces. For example, the scarcity of library stars with \teff near 4500 K likely leads to K dwarfs being underrepresented near this temperature. Any future expansions of the ESM library should prioritize adding K dwarfs in these undersampled regions, potentially drawing from our own extensive spectroscopic dataset.

Despite these limitations, our analysis successfully characterizes a modest set of 20 (3.4\%) metal-poor stars in the sample with \feh $<$ $-$0.5 dex. These low-metallicity K dwarfs form an important subset for further work to understand Galactic stellar populations and chemical evolution. Significantly, none of these stars, nor those with metallicites of $-$0.4 to $-$0.5 dex are found to be young or active, as described in $\S$\ref{subsec:ha_activity} \& $\S$\ref{subsec:lithium_youth}, highlighting a clear dichotomy between the metal-poor and chromospherically active populations.

The distribution of rotational velocities in our sample provides additional insight into the stellar population. While ESM can measure \vsini values down to the instrumental resolution limit, we report only values exceeding \vsini = 7 km s$^{-1}$, following the approach of \citet{hubbardjames22}. A robust correlation between projected rotational velocity and stellar activity emerged from our analysis --- all six K dwarfs in Table~\ref{tab:stellarproperties_special_paper2} exhibiting \vsini $>$ 5 km s$^{-1}$ also demonstrate evident chromospheric activity, as determined by their H$\alpha$ measurements (see $\S$~\ref{subsec:ha_activity}). These rapid rotators are AK For, BD+20 1790, HD 29697, HD 118100, HD 175742, and LQ Hya.

\begin{figure}[htb!]
\centering
    \includegraphics[width=0.85\textwidth]{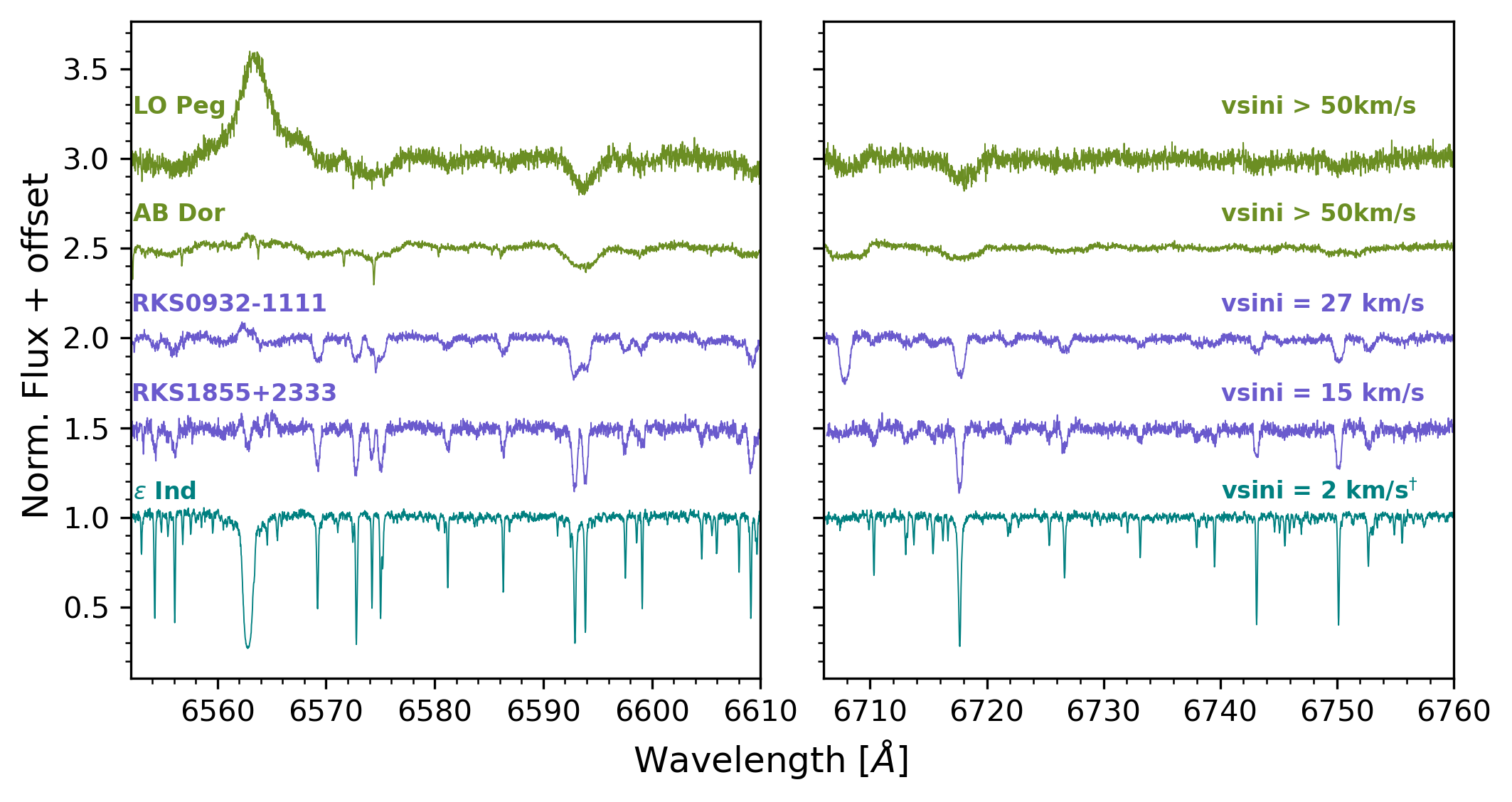}
\caption{A sequence of spectra showing the effects of rotation on line profiles. The left panel displays a $\sim$60 \AA{} window including H$\alpha$ at 6563 \AA{}, while the right panel shows $\sim$50 \AA{} window including Li I at 6707.8 \AA{} and a Ca I line at 6717 \AA{}. From bottom to top: $\epsilon$ Indi (\vsini $<$ 2 km s$^{-1}$, slow rotator), RKS1855+2333 (\vsini = 15 km s$^{-1}$) and RKS0932-1111 (\vsini = 27 km s$^{-1}$) representing fast rotators from our survey sample of 580 K dwarfs, and AB Dor (\vsini $>$ 50 km s$^{-1}$) and LO Peg (\vsini $>$ 50 km s$^{-1}$) representing extremely fast rotators from the benchmark sample used to calibrate age and activity relationships, as described in \citet{hubbardjames22}. The progressive line broadening and eventual smearing of spectral features with increasing rotation is clearly evident in both wavelength regions.}
    \label{fig:vsini_examples}
\end{figure}

Interestingly, while all rapid rotators show chromospheric activity, the converse is not true: among the 26 K dwarfs identified as chromospherically active in our sample, only six showed elevated \vsini values. This apparent discrepancy likely results at least in part from projection effects, where stars with rotational axes nearly aligned with our line of sight may be rotating rapidly but exhibit small projected velocities \citep{barnes03}. This underscores the complex relationship between stellar rotation and activity, and emphasizes the need for complementary diagnostics when characterizing stellar properties. Figure~\ref{fig:vsini_examples} illustrates the dramatic impact of rotation on spectral line profiles across two diagnostic wavelength regions, as seen in CHIRON spectra. The H$\alpha$ and Li I regions both show progressive line broadening with increasing \vsini, from the sharp, well-defined features of the slow rotator $\epsilon$ Indi to the severely broadened and smeared lines of the extremely fast rotators LO Peg and AB Dor from our benchmark sample. This sequence demonstrates how rapid rotation can complicate spectral analysis, particularly for activity indicators like H$\alpha$ and age diagnostics like Li I.

Finally, surface gravity measurements from ESM show the expected values for main-sequence K dwarfs, with \logg ranging from 4.40 to 4.85 dex for all but four of the 580 stars for which \logg could be derived. We adopt a minimum error of 0.60 dex for \logg measurements, based on the systematic error reported in \cite{yee17}. To ensure no cool subgiants contaminate the sample, we vetted stars using the HR diagram with a strict cutoff at $M_G$ = 5.5 mag at the bright end. Individual stars near this cutoff were examined to confirm they are main-sequence K dwarfs rather than slightly evolved objects.

The combination of the four derived stellar parameters provides a characterization of our K dwarf sample, establishing the foundation for interpretation of the measured chromospheric activity and age indicators.
In Table~\ref{tab:stellarproperties_special_paper2} we highlight detailed spectroscopic results and derived stellar properties for 53 noteworthy K dwarfs from our survey sample, including young stars, active stars, a double-lined spectroscopic binary, and the single halo star revealed.

\begin{deluxetable}{lcccccccc} 
\tabletypesize{\footnotesize}
\caption{Spectroscopic Results and Derived Stellar Properties of 53 K Dwarfs in the Survey Sample}
\label{tab:stellarproperties_special_paper2}

\tablehead{
\colhead{RKSTAR ID} & \colhead{$T_{\rm eff} \pm \sigma$} & \colhead{[Fe/H] $\pm\sigma$} & \colhead{log $g \pm \sigma$} & \colhead{$v\sin i$} & \colhead{EW[H$\alpha$]} & \colhead{EW[Li I]} & \colhead{Status} & \colhead{Fig~\ref{fig:sp_halpha_vs_lii} \#s} \\
\colhead{...} & \colhead{(K)} & \colhead{(dex)} & \colhead{(dex)} & \colhead{(km s$^{-1}$)} & \colhead{(\AA)} & \colhead{(\AA)} & \colhead{...} & \colhead{...}
}
\startdata
RKS0117$-$1530 & 5302 $\pm$ 108 & $-$0.48 $\pm$ 0.09 & 4.17 $\pm$ 0.63 & $<$5 & 0.63 & 0 & A & ... \\
RKS0121$+$2419 & 3942 $\pm$ 122 & $+$0.06 $\pm$ 0.13 & 4.68 $\pm$ 0.71 & ... & 0.26 & 0 & A / MG Hyades & ... \\
RKS0252$-$1246 & 5165 $\pm$ 103 & $+$0.11 $\pm$ 0.09 & 4.51 $\pm$ 0.68 & $<$5 & 0.92 & 0.21 & Y$+$A & 18 \\
RKS0417$+$2033 & 4446 $\pm$ 106 & $-$0.06 $\pm$ 0.12 & 4.64 $\pm$ 0.70 & $<$5 & 0.72 & 0.13 & Y & 13 \\
RKS0430$+$0058 & 4011 $\pm$ 102 & $+$0.03 $\pm$ 0.10 & 4.68 $\pm$ 0.70 & $<$5 & $-$0.14 & 0 & A & ... \\
RKS0436$+$2707 & 4671 $\pm$ 161 & $+$0.15 $\pm$ 0.10 & 4.55 $\pm$ 0.68 & $<$5 & $-$0.39 & 0 & A & ... \\
RKS0441$+$2054 & 4572 $\pm$ 104 & $-$0.22 $\pm$ 0.11 & 4.58 $\pm$ 0.69 & 8.0 & $-$0.18 & 0.06 & Y$+$A & 9 \\
RKS0536$+$1119 & 3936 $\pm$ 102 & $-$0.09 $\pm$ 0.10& 4.7 $\pm$ 0.69 & $<$5 & 0.06 & 0 & A & ... \\
RKS0626$+$1845 & 5269 $\pm$ 166 & $-$0.41 $\pm$ 0.14 & 4.41 $\pm$ 0.71 & $<$5 & 0.54 & 0 & A & ... \\
RKS0658$-$1259 & 4357 $\pm$ 129 & $-$0.30 $\pm$ 0.10& 4.66 $\pm$ 0.67 & $<$5 & 0.37 & 0.07 & Y$+$A & 10 \\
RKS0723$+$2024 & 4285 $\pm$ 363 & $-$0.20 $\pm$ 0.28 & 4.65 $\pm$ 0.74 & 8.0 & $-$0.55 & 0.16 & Y$+$A / MG AB Dor & 16 \\
RKS0734$-$0653 & 5056 $\pm$ 144 & $-$0.19 $\pm$ 0.22 & 4.57 $\pm$ 0.74 & $<$5 & 0.97 & 0.05 & Y & 7 \\
RKS0739$-$0335 & 4907 $\pm$ 113 & $+$0.04 $\pm$ 0.11 & 4.53 $\pm$ 0.72 & $<$5 & 0.52 & 0 & A & ... \\
RKS0819$+$0120 & 4965 $\pm$ 120 & $-$0.28 $\pm$ 0.12 & 4.56 $\pm$ 0.69 & $<$5 & 0.85 & 0.12 & Y$+$A & 12 \\
RKS0850$+$0751 & 3943 $\pm$ 120 & $-$0.25 $\pm$ 0.14 & 4.7 $\pm$ 0.70& $<$5 & 0.31 & 0 & A & ... \\
RKS0904$-$1554 & 4895 $\pm$ 111 & $+$0.08 $\pm$ 0.11 & 4.5 $\pm$ 0.70& $<$5 & 0.81 & 0.05 & Y & 8 \\
RKS0907$+$2252 & 5257 $\pm$ 113 & $+$0.13 $\pm$ 0.10& 4.49 $\pm$ 0.72 & $<$5 & 0.83 & 0.12 & Y$+$A & 11 \\
RKS0932$-$1111 & 6177 $\pm$ 254 & $+$0.01 $\pm$ 0.17 & 4.31 $\pm$ 0.77 & 27.4 & $-$0.06 & 0.19 & Y$+$A & 17 \\
RKS1000$+$2433 & ... & ... & ... & ... & 0.19 & 0 & A & ... \\
RKS1043$-$2903 & 5269 $\pm$ 101 & $+$0.14 $\pm$ 0.10& 4.51 $\pm$ 0.69 & $<$5 & 0.94 & 0.15 & Y$+$A & 15 \\
RKS1121$-$2027 & 4138 $\pm$ 111 & $-$0.13 $\pm$ 0.12 & 4.68 $\pm$ 0.70& $<$5 & 0.29 & 0 & A & ... \\
RKS1205$-$1852 & 3955 $\pm$ 108 & $-$0.08 $\pm$ 0.14 & 4.7 $\pm$ 0.70& $<$5 & 0.25 & 0 & A & ... \\
RKS1303$-$0509 & 5266 $\pm$ 113 & $-$0.05 $\pm$ 0.10& 4.54 $\pm$ 0.72 & $<$5 & 0.97 & 0.14 & Y & 14 \\
RKS1306$+$2043 & 4105 $\pm$ 113 & $-$0.19 $\pm$ 0.14 & 4.68 $\pm$ 0.70& $<$5 & 0.18 & 0 & A & ... \\
RKS1334$-$0820 & 4335 $\pm$ 118 & $-$0.10 $\pm$ 0.17 & 4.65 $\pm$ 0.68 & 8.1 & $-$0.32 & 0.02 & Y$+$A & 1 \\
RKS1414$-$1521 & ... & ... & ... & ... & 0.00 & 0 & A & ... \\
RKS1500$-$2905 & 3790 $\pm$ 108 & $-$0.03 $\pm$ 0.10& 4.71 $\pm$ 0.70& $<$5 & 0.17 & 0 & A & ... \\
RKS1633$-$0933 & 3909 $\pm$ 120 & $+$0.20 $\pm$ 0.16 & 4.67 $\pm$ 0.71 & $<$5 & $-$0.8 & 0 & A / MG AB Dor & ... \\
RKS1705$-$0147 & 4835 $\pm$ 130 & $+$0.01 $\pm$ 0.14 & 4.5 $\pm$ 0.73 & 6.2 & 0.30& 0.24 & Y$+$A & 19 \\
RKS1716$-$1210 & 4002 $\pm$ 110 & $-$0.02 $\pm$ 0.11 & 4.68 $\pm$ 0.70& $<$5 & 0.45 & 0.05 & Y & 5 \\
RKS1737$-$1314 & ... & ... & ... & ... & 0.00 & 0 & A & ... \\
RKS1754$-$2649 & 4249 $\pm$ 220 & $-$0.32 $\pm$ 0.27 & 4.68 $\pm$ 0.75 & ... & $-$2.51 & 0.05 & Y$+$A & 6 \\
RKS1818$-$0642 & 4673 $\pm$ 101 & $+$0.02 $\pm$ 0.10& 4.58 $\pm$ 0.70& $<$5 & 0.77 & 0.03 & Y & 2 \\
RKS1822$+$0142 & 4129 $\pm$ 110 & $-$0.07 $\pm$ 0.13 & 4.67 $\pm$ 0.70& $<$5 & $-$0.25 & 0 & A & ... \\
RKS1855$+$2333 & 5123 $\pm$ 161 & $+$0.01 $\pm$ 0.14 & 4.54 $\pm$ 0.73 & 14.8 & 0.02 & 0 & A & ... \\
RKS1910$+$2145 & 3931 $\pm$ 110 & $+$0.07 $\pm$ 0.11 & 4.67 $\pm$ 0.70& $<$5 & $-$0.41 & 0.03 & Y$+$A & 4 \\
RKS2041$-$2219 & 3953 $\pm$ 130 & $-$0.16 $\pm$ 0.20& 4.69 $\pm$ 0.71 & $<$5 & $-$1.51 & 0 & A & ... \\
RKS2105$-$1654 & 3820 $\pm$ 110 & $-$0.05 $\pm$ 0.11 & 4.7 $\pm$ 0.70& $<$5 & 0.18 & 0 & A & ... \\
RKS2108$-$0425 & 4566 $\pm$ 140 & $-$0.40 $\pm$ 0.15 & 4.6 $\pm$ 0.73 & $<$5 & 0.38 & 0 & A & ... \\
RKS2153$+$2055 & 5033 $\pm$ 107 & $-$0.07 $\pm$ 0.11 & 4.56 $\pm$ 0.70& $<$5 & 0.93 & 0.03 & Y$+$A & 3 \\
RKS2308$+$0633 & 3804 $\pm$ 108 & $+$0.10 $\pm$ 0.14 & 4.7 $\pm$ 0.69 & $<$5 & 0.08 & 0 & A & ... \\
RKS2335$+$0136 & 4112 $\pm$ 101 & $+$0.05 $\pm$ 0.13 & 4.66 $\pm$ 0.69 & $<$5 & 0.24 & 0 & A & ... \\
RKS2348$-$1259 & 4179 $\pm$ 113 & $-$0.12 $\pm$ 0.11 & 4.68 $\pm$ 0.69 & $<$5 & $-$0.19 & 0 & A & ... \\
RKS0706$+$2358 & 4281 $\pm$ 110 & $-$0.08 $\pm$ 0.13 & 4.66 $\pm$ 0.70& $<$5 & 0.67 & 0 & MG AB Dor & ... \\
RKS0820$+$1404 & 4146 $\pm$ 111 & $-$0.09 $\pm$ 0.11 & 4.67 $\pm$ 0.69 & $<$5 & 0.60& 0 & MG AB Dor & ... \\
RKS0104$+$2607 & 4161 $\pm$ 111 & $-$0.07 $\pm$ 0.13 & 4.67 $\pm$ 0.70& $<$5 & 0.57 & 0 & MG Hyades & ... \\
RKS0300$+$0744 & 5058 $\pm$ 106 & $+$0.28 $\pm$ 0.11 & 4.49 $\pm$ 0.70& $<$5 & 0.99 & 0 & MG Hyades & ... \\
RKS0320$+$0827 & 4520 $\pm$ 110 & $+$0.08 $\pm$ 0.12 & 4.6 $\pm$ 0.72 & $<$5 & 0.74 & 0 & MG Hyades & ... \\
RKS0322$+$2709 & 3947 $\pm$ 111 & $+$0.13 $\pm$ 0.20& 4.68 $\pm$ 0.69 & $<$5 & 0.42 & 0 & MG Hyades & ... \\
RKS0420$-$1445 & 4368 $\pm$ 101 & $+$0.07 $\pm$ 0.11 & 4.64 $\pm$ 0.69 & $<$5 & 0.64 & 0 & MG Hyades & ... \\
RKS2254$+$2331 & 3931 $\pm$ 113 & $+$0.02 $\pm$ 0.13 & 4.67 $\pm$ 0.72 & $<$5 & 0.48 & 0 & MG Hyades & ... \\
RKS1833$-$1626 & 4896 $\pm$ 557 & $-$0.19 $\pm$ 0.53 & 4.56 $\pm$ 0.77 & $<$5 & 0.83 & 0 & New SB2 & ... \\
RKS1510$-$1622 & ... & ... & ... & ... & 1.23 & 0 & Halo & ... \\
\enddata
\tablenotetext{}{Columns 6 and 7 --- EW[H$\alpha$]: Equivalent width of the H$\alpha$ line at 6563 \AA. 
EW[Li I]: Equivalent width of the Li I line at 6707.8 \AA}

\tablenotetext{}{\textbf{In the Status column:} Y = Young, A = Active, Y$+$A = both Youth and Active, MG = Young Moving Group, New SB2 = New spectroscopic binary, Halo = Halo star}
\end{deluxetable}

\subsection{Quiescent H${\alpha}$ Activity Level for K Dwarfs}\label{subsec:ha_quiescent}

Having established the fundamental stellar parameters for our sample, we now turn to characterizing chromospheric activity status across the K dwarf population. A critical first step in identifying active stars is establishing a baseline ``quiescent" level representing the typical H$\alpha$ absorption strength for mature, inactive K dwarfs. This baseline serves as a reference against which enhanced chromospheric emission can be measured.

\begin{figure}[htb!]
    \hspace{0 cm}
    \centering 
    \includegraphics[width=0.9\columnwidth,trim= 0cm 5.0cm 0cm 1cm,,clip]
    {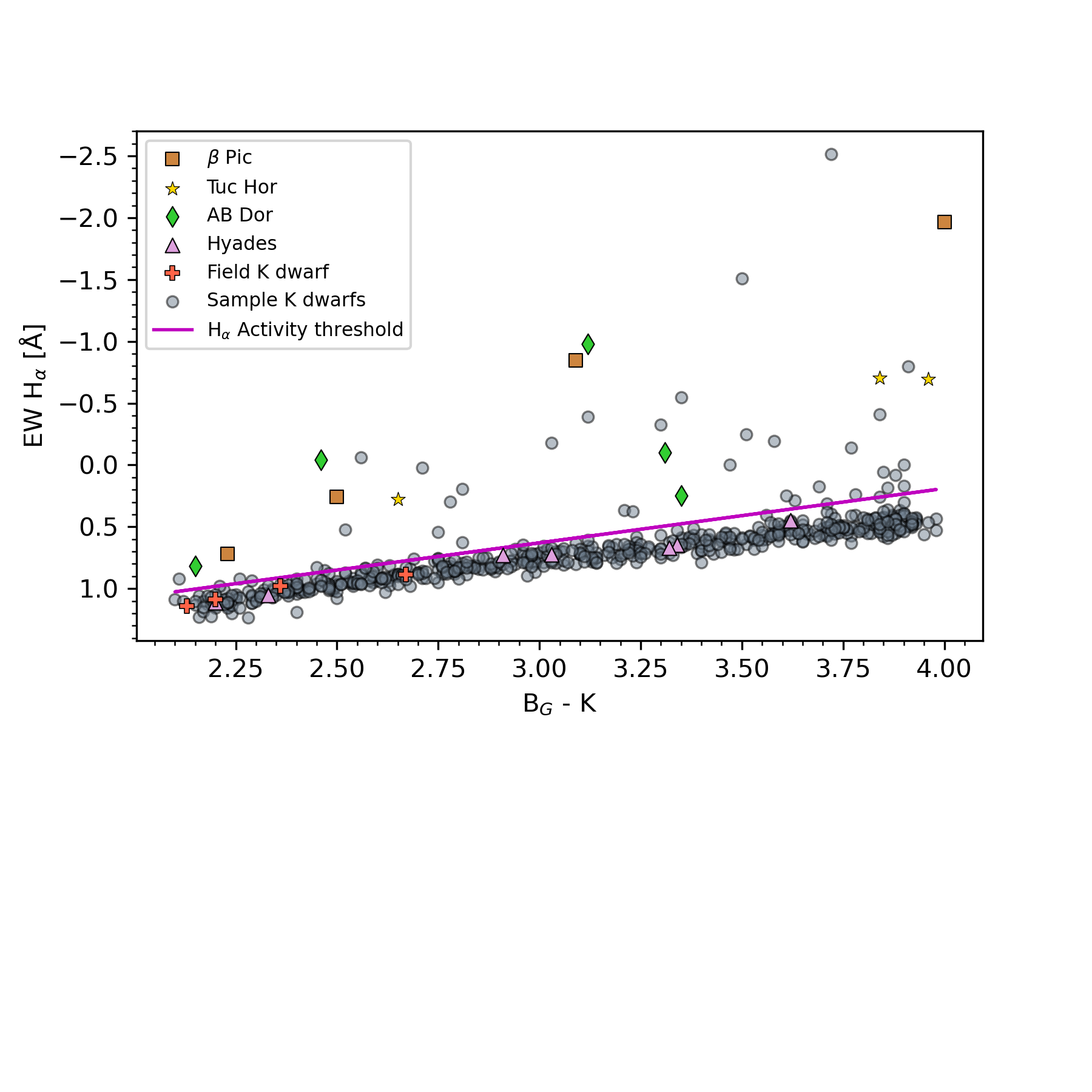}
    \hspace{0 cm}
    \caption{The equivalent width measurements for H$\alpha$ as a function of color. EW H$\alpha$ in emission is represented with negative values (toward the top of the plot). Gray circles represent 580 K dwarfs from the survey with H$\alpha$ values. The solid magenta line represents the typical quiescent level for K dwarfs, derived from the distribution of mature field stars, including Hyades members. Of the 580 K dwarfs, 36 (6.2\%) stars fall above the activity threshold (magenta line) and are deemed chromospherically active, while the remaining 544 (93.8\%) stars below this line are classified as calm. For comparison, K dwarfs with known ages from other methods are represented with markers other than circles: orange squares ($\beta$ Pic), yellow stars (Tuc-Hor), green diamonds (AB Dor), pink triangles (Hyades), and orange crosses (mature field K dwarfs with ages of 0.3--5.7 Gyr). These comparison stars are described in Section~\ref{section:sample} and Table~\ref{tab:movinggroups} and serve as age calibrators.}
        \label{fig:ewha_color}
\end{figure}

Figure~\ref{fig:ewha_color} presents the relationship between H$\alpha$ EW and stellar color ($BP-K$) for the 580 K dwarfs for which we could measure values. The H$\alpha$ strength exhibits a clear dependence on color, with redder K dwarfs showing progressively weaker absorption (more negative EW values). This trend arises from the decreasing continuum flux at H$\alpha$ wavelengths in cooler stars; for a given H$\alpha$ line flux, later-type stars yield smaller equivalent widths due to their fainter continua \citep{white07}.

Most stars fall along a quiescent population that forms a well-defined sequence in Figure~\ref{fig:ewha_color}. Through visual inspection of individual spectra, we excluded stars showing H$\alpha$ emission or core-filling from the distribution, then fit the remaining points to provide the baseline shown in solid magenta.  The remaining inactive stars, including all Hyades cluster members (750 Myr \citet{gagne18}) from our benchmark sample, were used to derive the quiescent activity level via a polynomial:

\begin{equation}\label{eq:quiescent}
    \text{EW H}{\alpha}^{\text{quiescent}} [\text{\AA}] = 3.2638 - 1.5798 \times (BP-K) + 0.3409 \times (BP-K)^2 - 0.0307 \times (BP-K)^3
\end{equation}

\noindent This quiescent level, shown as the magenta line in Figure~\ref{fig:ewha_color}, effectively traces the locus of mature, inactive K dwarfs across the full range of spectral types in our sample. The excellent agreement between our fit and the Hyades members provides confidence in our characterization, as these stars represent a population old enough to have settled into chromospheric quiescence, yet young enough as a cluster to provide a reasonable age estimate.

To distinguish chromospherically active stars from the quiescent population, we establish an activity threshold shown as the solid magenta line in Figure~\ref{fig:ewha_color}. This threshold is set 0.13 \AA{} above the typical quiescent level (i.e., more negative in EW H$\alpha$), derived from the distribution of mature field stars including Hyades members. The threshold accounts for measurement uncertainties and intrinsic scatter in the quiescent population while providing clear separation for genuinely active stars. Stars falling above this threshold (EW H$\alpha$ more negative than the magenta line) are classified as chromospherically active and form the basis for our activity analysis in the following section.

\subsection{H${\alpha}$ Results: Activity}\label{subsec:ha_activity}

We apply the quiescent H$\alpha$ description established in the previous section to identify chromospherically active K dwarfs within our sample. This classification is essential for distinguishing mature, stable stellar environments from those with enhanced magnetic activity that could impact planetary habitability through increased variability and likely boosted high-energy radiation levels. Active stars produce elevated levels of UV and X-ray emission that may erode planetary atmospheres, alter atmospheric chemistry, and affect the potential for surface liquid water, all of which are critical factors in assessing exoplanet habitability \citep{richeyyowell19, richeyyowell22}.

K dwarfs with H$\alpha$ EWs falling above the magenta line in Figure~\ref{fig:ewha_color} (more negative EW values) are designated as chromospherically active, while those below the line are considered calm. Among the 580 stars in our sample, we find 36 (6.2\%) that exhibit chromospheric activity, leaving 544 stars (93.8\%) classified as calm. This relatively small fraction of active stars is consistent with the expectation that most field K dwarfs in the solar neighborhood are mature, having ages greater than approximately 1 Gyr, by which time chromospheric activity has usually declined to quiescent levels.

\begin{figure}[htb!]
    \centering    \includegraphics[width=0.7\textwidth]{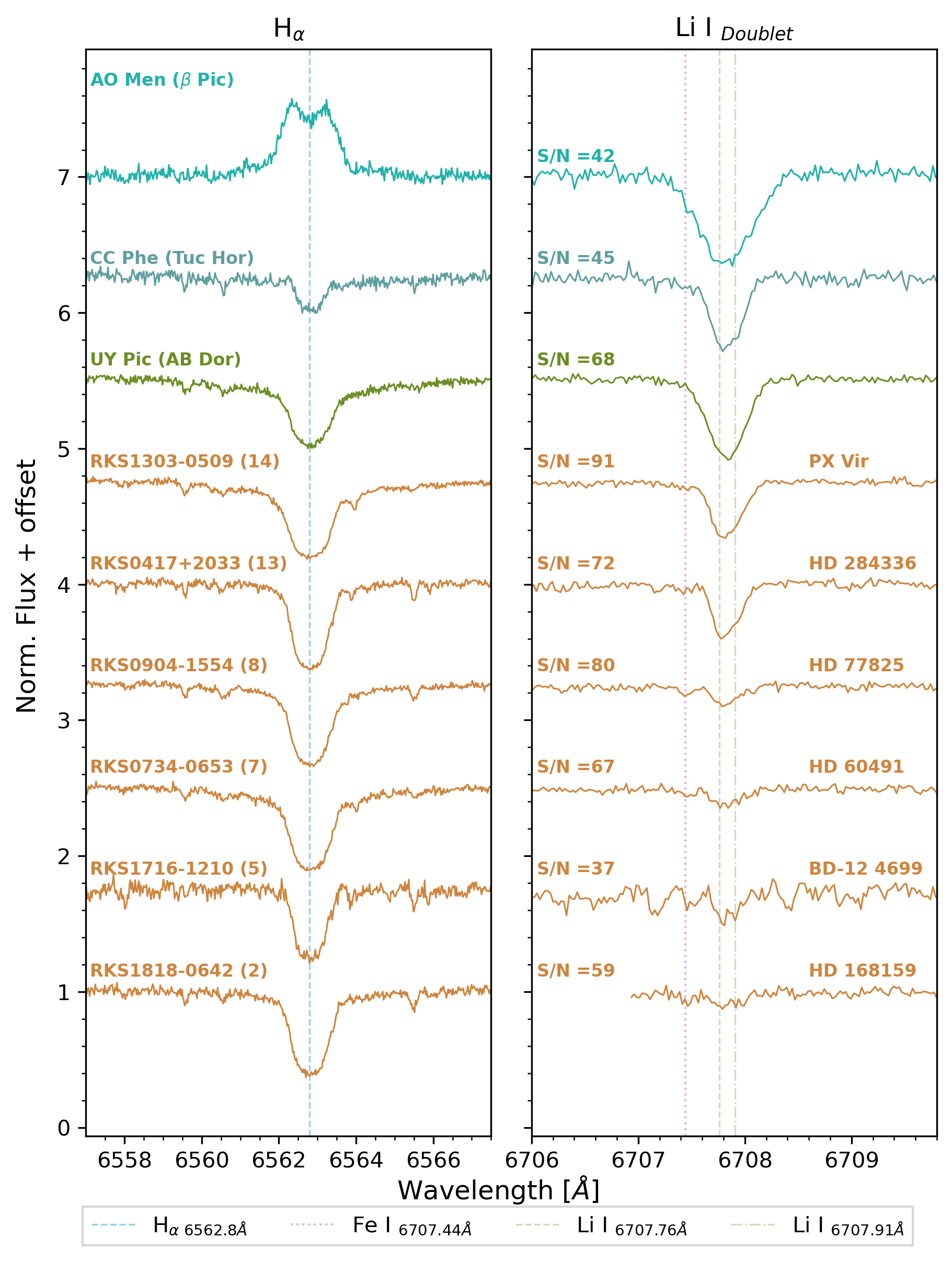}
    \caption{Panels showing the spectral regions of H${\alpha}$ and Li I for K dwarfs with known ages and young K dwarfs from our analysis. Spectra in colors other than orange represent stars with known ages studied in \citet{hubbardjames22}, where names from SIMBAD are given and the associated moving groups are in parentheses. Spectra in orange show six young K dwarfs from our analysis. For these spectra, the left panel includes the RKSTAR ID names and numbers used in Figure~\ref{fig:sp_halpha_vs_lii}, and the right panel includes names from SIMBAD. Vertical lines in different styles and colors show the centers of the lines of interest in the rest frame, given in the box below the main plot.}
    \label{fig:sp_young}
\end{figure}

\begin{figure}[htb!]
    \centering
    \includegraphics[width=0.7\textwidth]{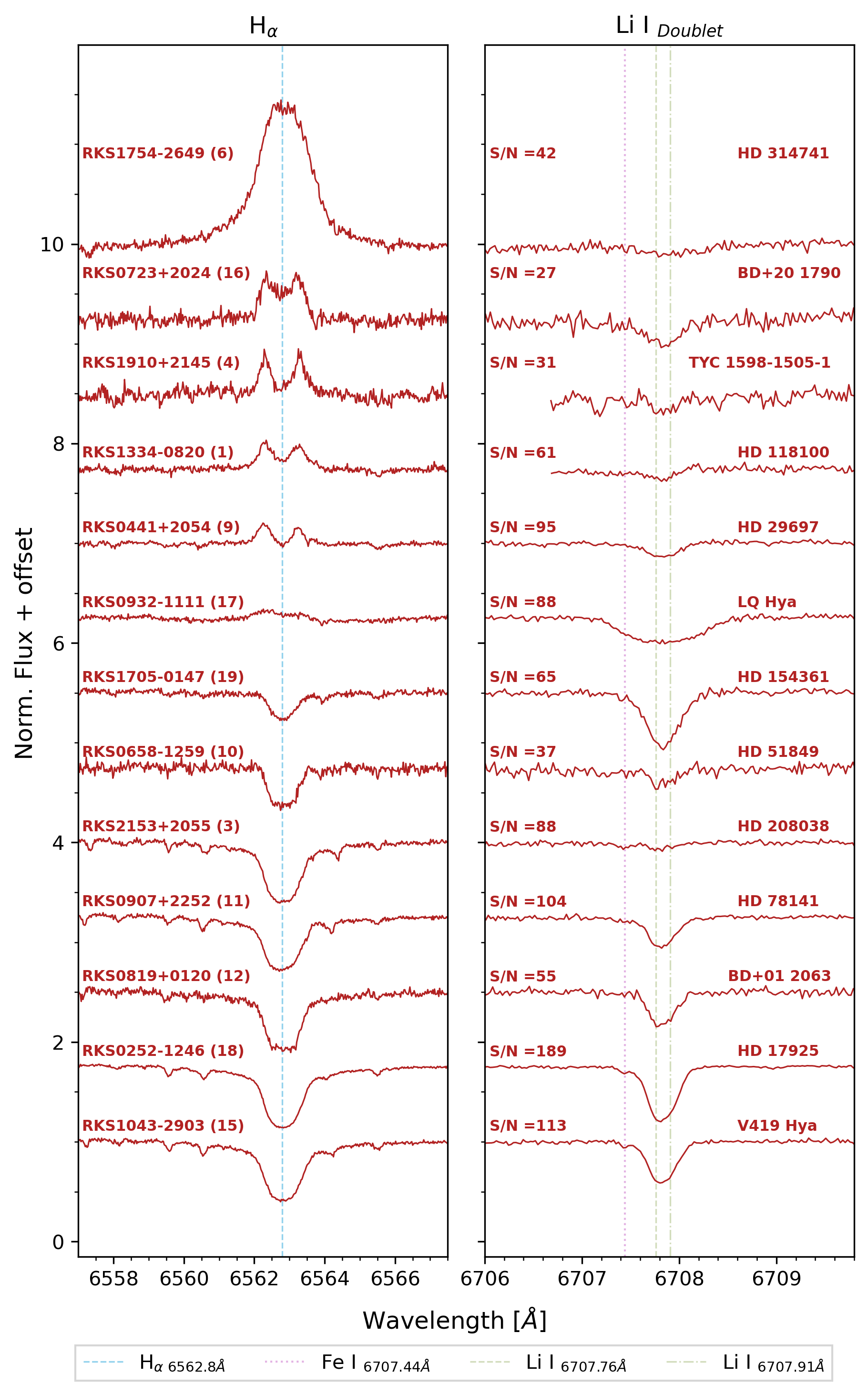}
    \caption{Panels showing the spectral regions of H${\alpha}$ and Li I for 13 young and active K dwarfs from our analysis. The left panel includes the RKSTAR ID names and numbers used Figure~\ref{fig:sp_halpha_vs_lii}, and the right panel includes names from SIMBAD. Vertical lines in different styles and colors show the centers of the lines of interest in the rest frame, given in the box below the main plot.}
    \label{fig:sp_youngactive}
\end{figure}

The spectral diversity within our active population reveals the complex nature of stellar chromospheric phenomena. Figures~\ref{fig:sp_young} and \ref{fig:sp_youngactive} display the H$\alpha$ and Li I spectral regions for representative active stars, illustrating a remarkable range of activity signatures. First, note that strong Li I absorption features do not necessarily imply strong H$\alpha$ emission, nor vice versa. The H$\alpha$ profiles in Figure~\ref{fig:sp_youngactive} are particularly instructive, showing a clear progression of activity levels. The uppermost spectra exhibit strong H$\alpha$ emission lines rising well above the continuum, characteristic of the most magnetically active stars with powerful chromospheric heating. These pure emission profiles indicate substantial non-thermal energy deposition in the chromosphere, likely from magnetic reconnection events \citep{cram85,hall08}. Moving down the sequence, we observe double-peaked emission profiles, which serve as a signature of chromospheric structure where the line core forms higher in the atmosphere than the wings \citep{montes97,lopezsantiago10}. Further down, the spectra transition to filled-in absorption profiles, where chromospheric emission partially compensates for photospheric absorption \citep{herbig85,strassmeier93}. The spectra near the bottom of the plot show only subtle signs of activity, with H$\alpha$ absorption lines that are slightly weakened by modest chromospheric emission. This continuum of activity signatures demonstrates that chromospheric activity is not a binary phenomenon but rather exists along a spectrum from highly active to completely quiescent \citep{soderblom93,mamajek08}.

In addition to H$\alpha$, the Ca II infrared triplet line at 8542 \AA{} provides a complementary diagnostic of chromospheric activity. Ca II core emission features for our sample stars can be viewed in the online spectral library\footnote{\url{https://hodarijames.github.io/spectral_library/page1.html}}, which includes all four diagnostic spectral features for each star. Visual inspection of the online library clearly shows Ca II core emission in several active stars, providing independent confirmation of their chromospheric activity. While our current analysis focuses on H$\alpha$ as the principal activity indicator, the Ca II infrared triplet represents a valuable complementary diagnostic that will be investigated in depth when the larger sample has been observed and analyzed, to be presented in a forthcoming paper (Carrazco-Gaxiola et al., in prep).

The spectral gallery presented in Appendix~\ref{append:append_spectrallibrary} (Figures~\ref{fig:appendix_53_specialones}--\ref{fig:groupE}) showcases H$\alpha$ and Li I features for all 53 young, active, or otherwise unique stars from our sample. Most stars display H$\alpha$ absorption profiles with varying degrees of chromospheric filling, while Figure~\ref{fig:groupD} (Group D) demonstrates prominent core emission features in four particularly active systems, with the central emission spike clearly visible in the H$\alpha$ panel.

The identification of chromospheric activity in 36 stars of our sample leaves 554 K dwarfs classified as chromospherically calm based on their H$\alpha$ measurements. This activity classification represents only one component of our comprehensive stellar characterization. While H$\alpha$ emission effectively identifies stars with active chromospheres, it does not distinguish between activity driven by youth versus activity maintained by other mechanisms such as close stellar companions or rapid rotation \citep{mamajek08}. To address this limitation and to provide a more complete picture of stellar ages within our sample, we next examine lithium abundance as an independent indicator of stellar youth. The relationship between H$\alpha$ activity and lithium detection aims to reveal which of our active stars are genuinely young versus those that maintain activity through alternative mechanisms.

\subsection{Lithium Results: Youth}\label{subsec:lithium_youth}

While chromospheric activity points to stellar magnetic phenomena often associated with young stars, lithium detection provides a more definitive diagnostic for stellar youth. Lithium is destroyed by nuclear burning processes in stellar interiors, with depletion timescales directly tied to stellar mass and age \citep{soderblom10}. G and K dwarfs younger than $\sim$1 Gyr retain lithium at a level where spectral features can be seen, making Li I a robust age diagnostic less susceptible to confounding factors that influence activity indicators. This distinction is critical for identifying stars that are intrinsically young versus those maintaining activity through other mechanisms.

In this study, we report equivalent widths of the Li I doublet feature detected using our Voigt profile fitting methodology described in $\S$~\ref{section:analysis}. We use lithium presence as a yes/no youth indicator, classifying stars with detectable Li I as young and confirming that the features are real via visual inspection for quality control. The Hyades cluster ($\sim$750 Myr) provides a useful age benchmark, as K dwarfs in the Hyades show no detectable lithium and have well-determined ages from main-sequence turnoff fitting \citep{perryman98}, white dwarf cooling timescales \citep{degennaro09}, and eclipsing binary analysis \citep{lebreton01}. Future papers in this series will explore quantitative age estimates using Li I depletion models \citep{soderblom10}. 

In the high resolution spectra provided by CHIRON, the Li I doublet at 6707.8 \AA{} is clearly resolved from the nearby Fe I line at 6707.4 \AA, eliminating line blending corrections required in previous optical lithium surveys \citep{white07}, except for the very strongest Li I absorption features in which the weak Fe I line is then negligible. We establish a detection threshold of EW Li I = 15 m\AA{} for reliable lithium identification, and classify stars with EW Li I $\geq$ 15 m\AA{} as young.

\begin{figure}[htb!]
    \hspace{0 cm}
    \centering 
    \includegraphics[width=0.9\columnwidth,trim= 0cm 8cm 0cm 0cm,,clip]{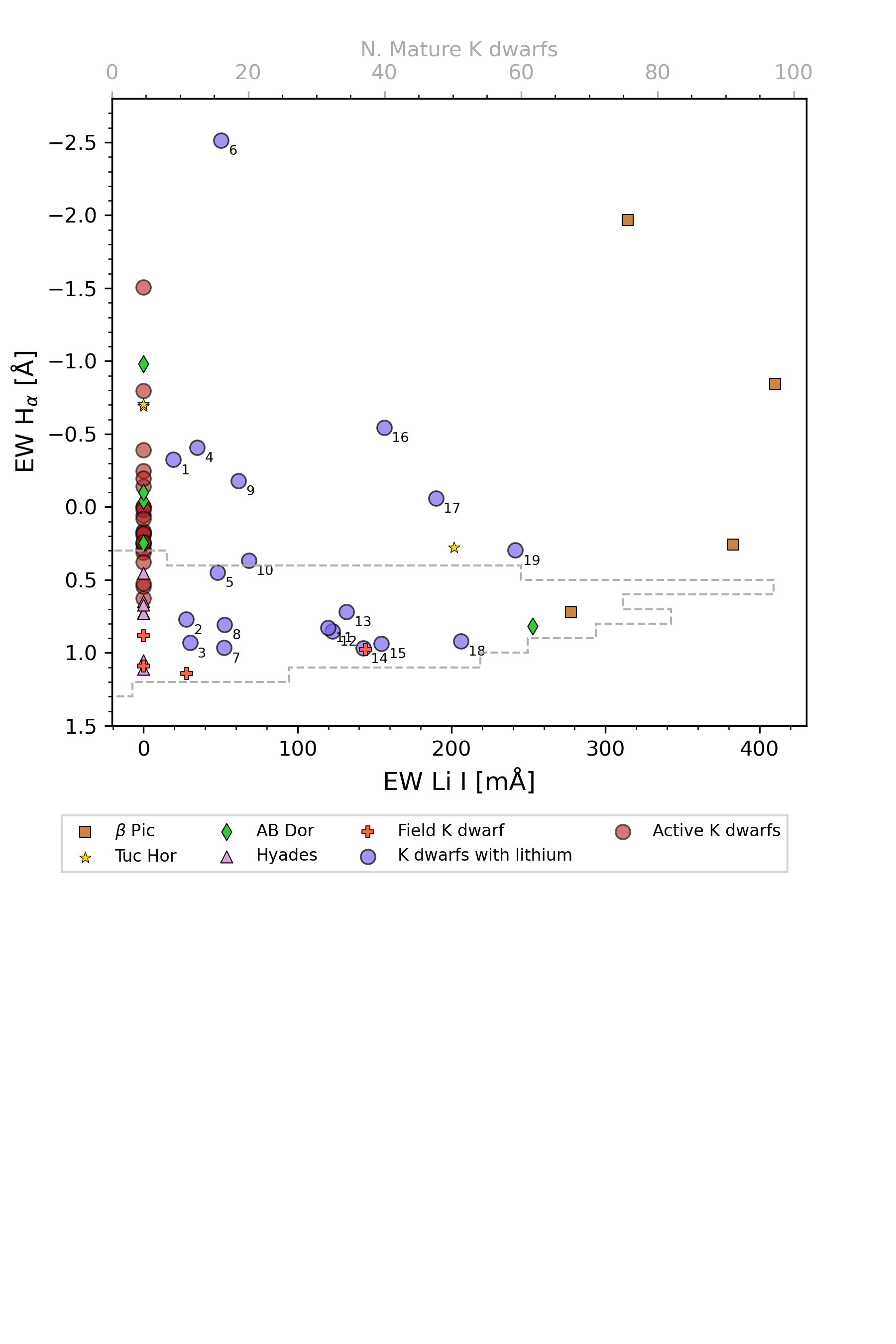}
    \hspace{0 cm}
  \caption{\small EW~H${\alpha}$ vs.~EW~Li~I for 43 chromospherically active and/or young K dwarfs from the survey, plus K~dwarfs with known ages for comparison. 
    Circles in red represent active K dwarfs with no discernable Li I features.
    Circles in purple represent young K dwarfs with Li I absorption features.
    Numbers at the lower right of each circle are sorted from weaker to stronger EW~Li~I (see Table 6 for star identifications).
     Stars with known ages from other methods are represented with markers other than circles: orange squares ($\beta$ Pic), yellow stars (Tuc-Hor), green diamonds (AB Dor), pink triangles (Hyades), and orange crosses (mature field K dwarfs with ages of 0.3--5.7 Gyr). These comparison stars are described in Section~\ref{section:sample} and Table~\ref{tab:movinggroups} and serve as age calibrators. PX Vir (0.3 Gyr) is the youngest field star.
    The overplotted histogram in gray represents the number of mature K dwarfs in our sample which have no Li I detection and weak EW H$\alpha$ relative to the activity threshold. The overplotted histogram in gray represents the numbers of mature K dwarfs in our sample with no Li I detections.}
    \label{fig:sp_halpha_vs_lii}
\end{figure}

Figure~\ref{fig:sp_halpha_vs_lii} illustrates the relationship between lithium detection and chromospheric activity, showing EW H${\alpha}$ vs.~EW Li I for active and/or young K dwarfs from the survey along with comparison stars of known ages from \cite{hubbardjames22}. Applying our detection threshold to the complete sample of 580 K dwarfs, we identify 19 stars (3.3\%) with measurable abundances of lithium. These young stars appear as purple circles in Figure~\ref{fig:sp_halpha_vs_lii}, numbered 1--19 as Li I strength increases. Red circles represent active K dwarfs from $\S$~\ref{subsec:ha_activity} that do not show Li I features, while the gray histogram shows the distribution of mature quiescent stars that comprise the bulk of the population.

The spectral evidence for lithium detection for the 19 survey K dwarfs is displayed in Figures~\ref{fig:sp_young} and \ref{fig:sp_youngactive}. Clear Li I absorption features are visible at 6707.8 \AA, with RKSTAR IDs and figure numbers shown in the left panels and SNR values along with names from SIMBAD in the right panels. Figure~\ref{fig:sp_young} shows H${\alpha}$ and Li I spectral regions for 3 comparison stars with known ages from young moving groups ($\beta$ Pic, Tuc-Hor, and AB Dor) (see Section~\ref{section:sample} and Table~\ref{tab:movinggroups}) and 6 young stars in orange from our analysis that are not demonstrably active. Figure~\ref{fig:sp_youngactive} presents H${\alpha}$ and Li I spectral regions for 13 additional young K dwarfs that do exhibit H${\alpha}$ features indicative of activity.

As mentioned previously, there is not a one-to-one correlation between H$\alpha$ and Li I EWs. Of the 19 lithium detections, 13 also exhibit chromospheric activity while 6 show lithium without significant H${\alpha}$ emission. An additional 24 stars display activity without lithium, indicating older yet still magnetically active systems. In total, 43 stars (7.4\%) are classified as young and/or active, leaving 537 stars (92.6\%) that are mature and quiescent.

Comparisons to stars with known ages validate our methodology. Moving group K dwarfs ($\beta$ Pic, Tuc-Hor, AB Dor) and Hyades members show expected lithium depletion trends. Of the 580 stars observed in the survey, PX Vir emerges as the single youngest field star (0.3 Gyr), demonstrating that isolated young K dwarfs are rare, but present, in the solar neighborhood. The 537 mature, calm stars represent prime targets for terrestrial planet searches, offering now-stable radiation environments given that their ages are likely beyond 1 Gyr.

\section{Kinematics}\label{section:kinematics}

\subsection{Gaia Astrometry}\label{subsec:gaia_astrometry}

Following our comprehensive spectroscopic characterization of stellar properties, activity status, and ages, we now examine the kinematic properties of our K dwarf sample to understand their dynamical context within the Galaxy. Stellar kinematics provide crucial insights into Galactic structure and evolution, as different stellar populations exhibit characteristic velocity distributions that reflect their formation histories and subsequent dynamical evolution \citep{freeman02,bland-hawthorn16}. The well-established relationships between stellar kinematics, age, and metallicity \citep{edvardsson93,nordstrom04,casagrande11} make this analysis essential for placing our spectroscopic results in a broader Galactic context. By determining three-dimensional space motions for our sample, we can associate individual stars with moving groups, stellar associations, and broader Galactic populations, while also identifying stars with unusual motions that may indicate non-standard evolutionary histories or membership in disrupted stellar systems.

Our kinematic analysis utilizes high-precision astrometric measurements from $Gaia$ Data Release 3, including celestial coordinates, proper motions ($\mu_{R.A.}$, $\mu_{Decl}$), and trigonometric parallaxes ($\varpi$) \citep{gaiacollaboration22}. These astrometric parameters, combined with systemic radial velocities ($\gamma_{REC}$) derived from our CHIRON spectroscopic observations, enable accurate determination of $UVW$ space velocities in the Galactic coordinate system, where $U$ points toward the Galactic center, $V$ aligns with Galactic rotation, and $W$ points toward the north Galactic pole. Essential kinematic data for 572 of the 580 stars in our survey, including both $Gaia$ and CHIRON radial velocity measurements ($\gamma_{Gaia_{DR3}}$ and $\gamma_{REC}$), measurement uncertainties ($\sigma_{\gamma_{REC}}$), derived $UVW$ velocities, and population assignments are compiled in Table~\ref{tab:table_kinematics} in Appendix~\ref{append:append_kinematicstable}. These kinematic measurements form the foundation for identifying Galactic population membership and understanding the relationships between the stellar properties characterized in $\S$~\ref{section:characterization} and the broader context of Galactic chemical and dynamical evolution.

\subsection{CHIRON $\gamma$ Velocities}\label{subsec:chiron_gamma}

Using the methodology described in $\S$~\ref{subsec:gamma_kinematics}, we derived systemic radial velocities ($\gamma$ velocities) for 574 of the 580 K dwarfs in the survey sample. The results are given in Table~\ref{tab:table_kinematics} in Appendix~\ref{append:append_kinematicstable}, spanning values of $+$116 to $-$133 km s$^{-1}$, with one additional star at $+$310 km s$^{-1}$ that turns out to be a halo star. These measurements, combined with $Gaia$ DR3 astrometry, provide the foundation for our kinematic analysis and enable direct comparison with space-based radial velocity determinations.

\begin{figure}[htb!]
\centering
    \includegraphics[width=\textwidth]{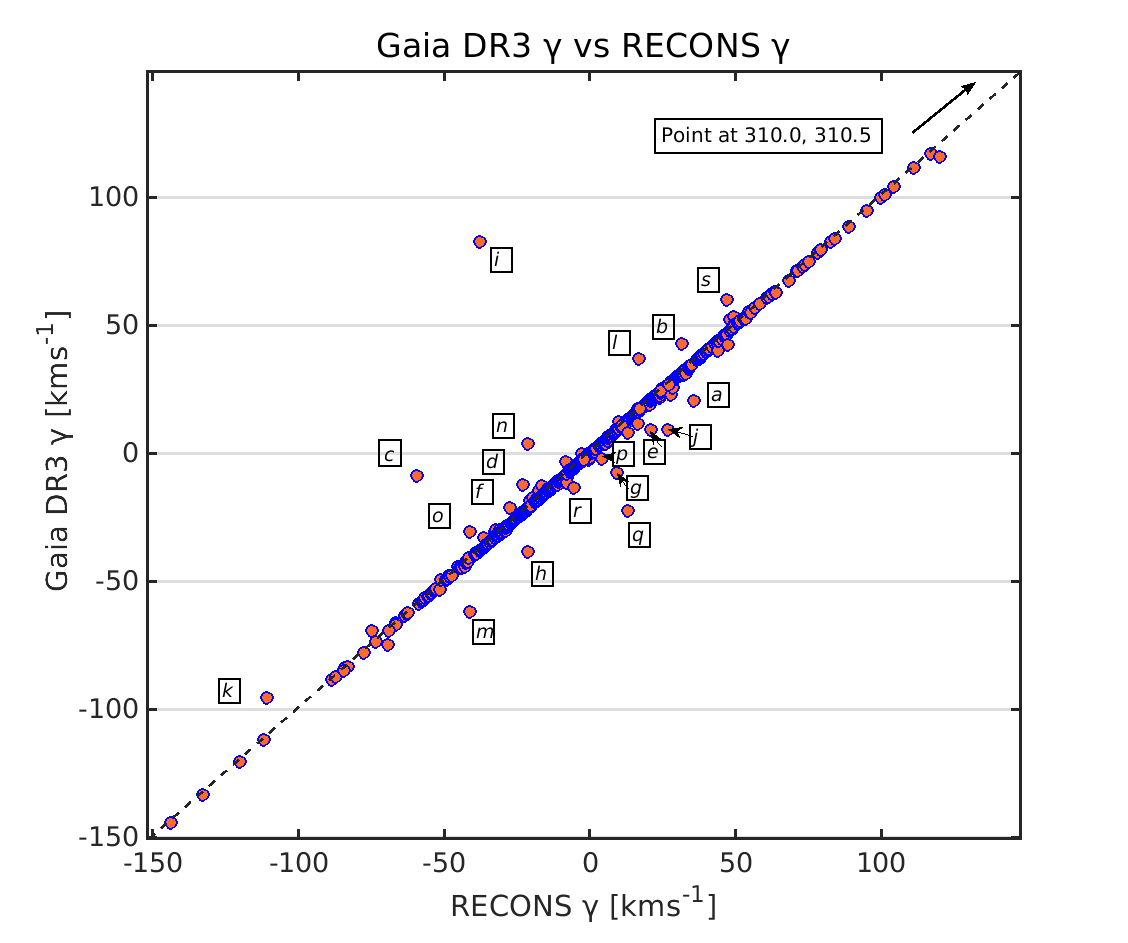}
\caption{Comparison between our RECONS $\gamma$ velocities from CHIRON observations and $Gaia$ DR3 radial velocities showing an overall 1-to-1 correlation, with only 18 stars in our survey sample showing significant deviations, labeled $\boxed{a}$ through $\boxed{s}$, excluding $\boxed{i}$. Star identifications and classifications are discussed in the text.}
    \label{fig:gamma_mainsample}
\end{figure}

Figure~\ref{fig:gamma_mainsample} compares our CHIRON-derived $\gamma$ velocities with corresponding measurements from $Gaia$ DR3. The majority of stars (554, or 95\%) follow the expected 1-to-1 correlation, validating our measurement methodology and the consistency between ground-based and space-based radial velocity determinations. The remaining 18 stars exhibit significant deviations from this relationship, marked with boxed letters `a' through `s' (excluding `i', see below) and are identified in the Figure~\ref{fig:gamma_mainsample} caption. The 18 outliers represent a diverse population of stellar systems with velocity discrepancies between CHIRON and Gaia measurements. These are labeled as follows: $\boxed{a}$ RKS0236-0309 (variable star), $\boxed{b}$ RKS0258+2646 (unknown), $\boxed{c}$ RKS0626+1845 (binary and active), $\boxed{d}$ RKS0907+2252 (young and active), $\boxed{e}$ RKS1108-2816 (unknown), $\boxed{f}$ RKS1253+0645 (binary, known SB1), $\boxed{g}$ RKS1303-0509 (young), $\boxed{h}$ RKS1504-1835 (binary, known SB1), $\boxed{j}$ RKS1528-0920 (binary, known SB2), $\boxed{k}$ RKS1555+1602 (variable star), $\boxed{l}$ RKS1605-2027 (binary, known SB2), $\boxed{m}$ RKS1833-1626 (new SB2), $\boxed{n}$ RKS1855+2333 (active), $\boxed{o}$ RKS2041-2219 (active), $\boxed{p}$ RKS2108-0425 (active), $\boxed{q}$ RKS2119-2621 (binary, known SB1), $\boxed{r}$ RKS2308+0633 (active), $\boxed{s}$ RKS2345+2933 (binary, known SB1). These outliers include known binary systems observed at different orbital phases, chromospherically active stars with activity-induced line profile variations, young stars with similar spectroscopic complications, and newly discovered spectroscopic binaries.

\begin{figure}[ht!]
\centering
\includegraphics[width=0.7\textwidth]{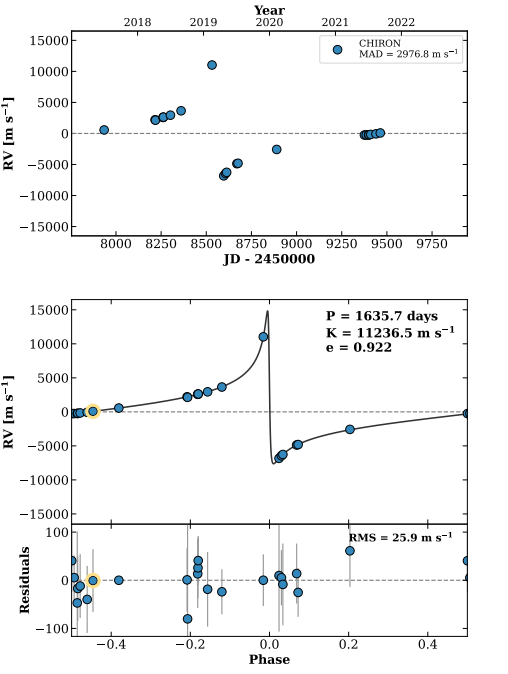}
\caption{Radial velocity measurements of RKS1518-1837 (BD$-$18 4031) showing obvious variations over the 2018-2021 observing period (top panel) and phase-folded orbital solution revealing a companion in a highly elliptical orbit with eccentricity $e = 0.922$ and period of 1636 days (bottom panel). This companion is likely a white dwarf, so the K dwarf is not included in our survey sample. This system provides an instructive example of binary detection through radial velocity monitoring. The data are from Paredes (2022).}
\label{fig:wdbinary_rvplots}
\end{figure}

We include one additional notable outlier (RKS1518-1837, BD$-$18 4031, labeled $\boxed{i}$) from our broader RECONS observations that, while the K dwarf was excluded from the survey sample during quality control, provides an instructive example of radial velocity discrepancies. Among the most intriguing examples from our broader observations, this star shows a substantial radial velocity difference between our measurements and $Gaia$ values. Follow-up observations revealed that this star hosts a companion in a highly eccentric orbit ($e = 0.922$) with a period of 1636 days (4.5 years), as illustrated in Figure~\ref{fig:wdbinary_rvplots}. The large semi-amplitude of more than 11 km s$^{-1}$ suggests that the companion may be a white dwarf, although the orbital inclination remains unknown, preventing a definitive determination of the mass. Thus, because the white dwarf progenitor would originally have been the more massive star in the system, this K dwarf was removed from the survey sample. This case shows how radial velocity discrepancies can reveal hidden stellar multiplicity and contribute to our understanding of the binary fraction of K dwarfs.

\subsection{Using $UVW$ Space Motions to Identify Young K Dwarfs}\label{subsec:uvw_space_motions}

The three-dimensional space motions ($UVW$) for stars in our sample provide an independent method for identifying young stars through kinematic association with nearby moving groups and stellar clusters. This kinematic approach serves as a crucial complement to the spectroscopic age indicators discussed in $\S$~\ref{section:characterization}, as stellar kinematics preserve the signature of stellar formation environments even after spectroscopic youth markers have faded or become undetectable. Young stellar associations retain coherent space motions for hundreds of millions of years, allowing kinematic identification of group members that may have depleted their lithium or ceased exhibiting strong chromospheric activity \citep{soderblom10,mamajek14, gagne18}.

\begin{figure}[htb!]
    \centering
    \includegraphics[width=0.32\textwidth]{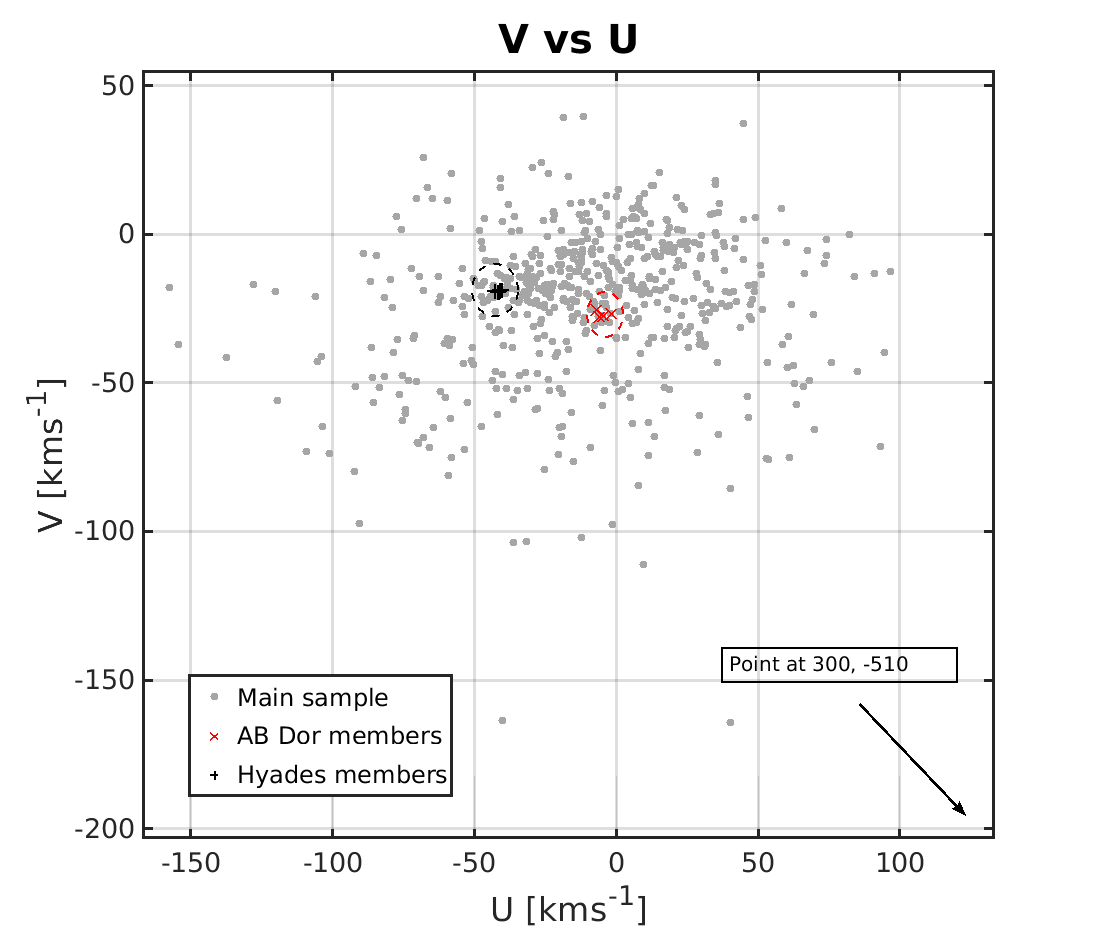}
    \includegraphics[width=0.32\textwidth]{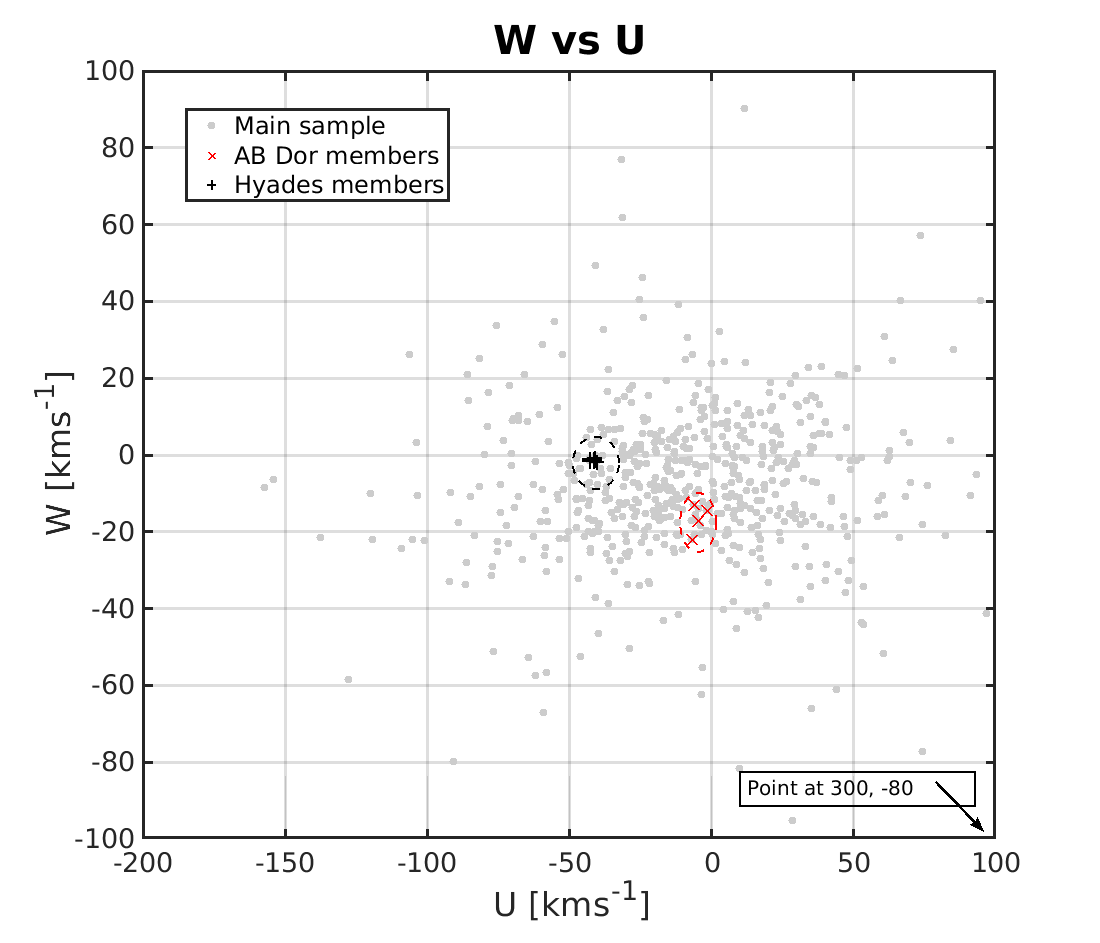}
    \includegraphics[width=0.32\textwidth]{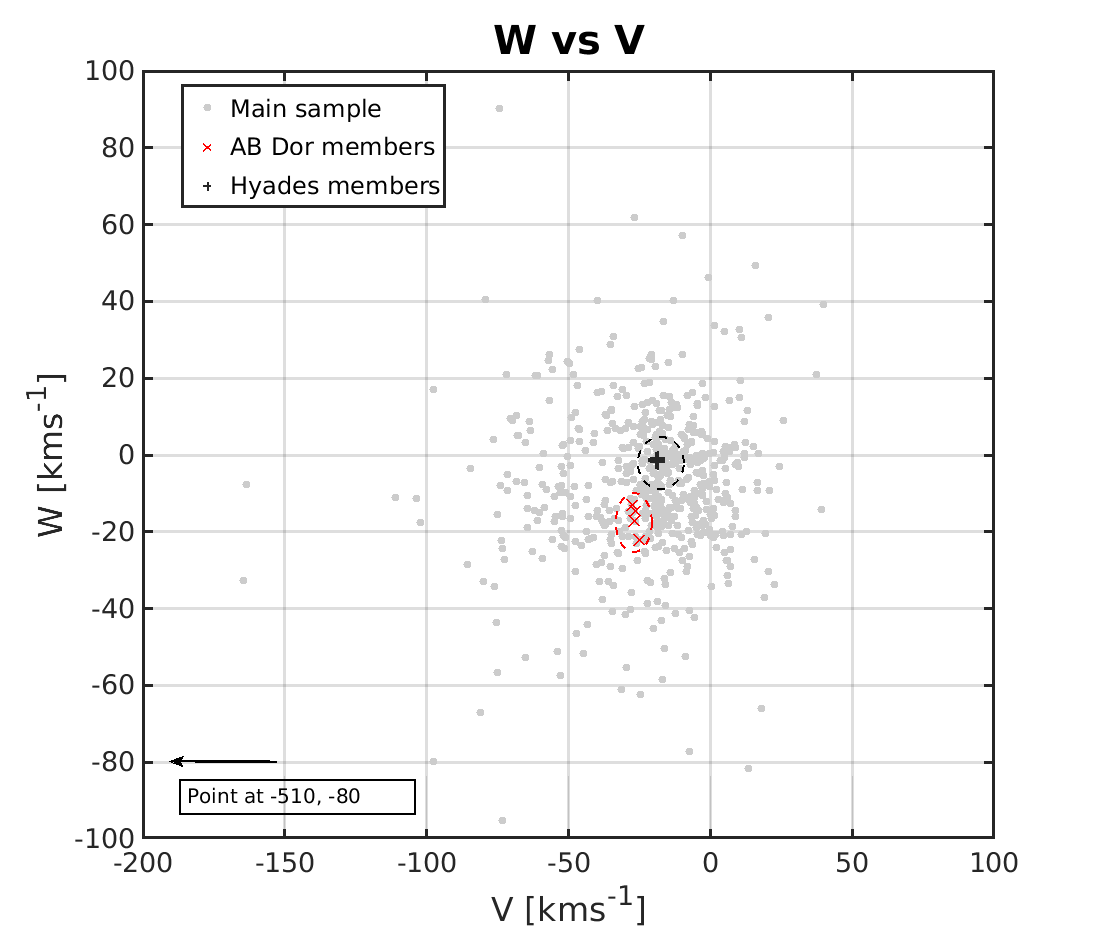}
    
\caption{Galactic space motion diagrams for 572 K dwarfs showing kinematic identification of moving group members: (a) $V$ vs $U$ velocities, (b) $W$ vs $U$ velocities, and (c) $W$ vs $V$ velocities. Field stars are shown as  grey circles, Hyades cluster members as black crosses, and AB Doradus moving group members as red Xs. The black ellipse (Hyades) and red ellipse (AB Dor) represent 2$\sigma$ velocity dispersion boundaries based on the mean UVW velocities and velocity dispersions from \citet{riedel14}. These ellipses encompass approximately 95\% of expected group members. Moving group member identifications are provided in Table 6.}
\label{fig:rvuvwplot_mainsample}
\end{figure}

Figure~\ref{fig:rvuvwplot_mainsample} illustrates the distribution of our sample in Galactic velocity space using three diagnostic projections: $V$ vs $U$, $W$ vs $U$, and $W$ vs $V$ for 572 K dwarfs. The plots can be used to reveal stars that share common space motions with well-characterized young stellar associations, providing an age-dating method independent of stellar atmospheric diagnostics. Using established kinematic criteria and the BANYAN $\Sigma$ methodology \citep{gagne18}, we identified 11 stars with space motions consistent with known moving groups. Our membership assignment required two criteria: (1) BANYAN $\Sigma$ membership probability exceeding 95\%, and (2) $UVW$ velocities falling within 2$\sigma$ of the mean group velocities, where $\sigma$ represents the intrinsic velocity dispersion of each association \citep{riedel14}. The 2$\sigma$ ellipses shown in Figure~\ref{fig:rvuvwplot_mainsample} visualize these kinematic boundaries and encompass approximately 95\% of expected group members based on their velocity dispersions. 

This approach identified seven stars exhibiting kinematics matching the Hyades cluster ($UVW = [-41, -19, -1]$ km s$^{-1}$; $\sigma = [2.0, 2.0, 2.0]$ km s$^{-1}$) and four stars showing motions consistent with the AB Doradus moving group ($UVW = [-7, -27, -13]$ km s$^{-1}$; $\sigma = [1.3, 1.2, 1.6]$ km s$^{-1}$) \citep{riedel14}. The $UVW$ space motions for the stars in our sample, including moving group assignments, are provided in Table~\ref{tab:table_kinematics} in Appendix~\ref{append:append_kinematicstable}. Remarkably, 8 of the 11 kinematically-identified young stars were {\it not} detected through our spectroscopic analysis described in $\S$~\ref{section:characterization}, demonstrating the complementary nature of kinematic and spectroscopic age indicators. These six Hyades members and two AB Doradus members are listed in Table~\ref{tab:stellarproperties_special_paper2} with the designation ``Y-MG" to distinguish them from stars showing spectroscopic youth signatures. The remaining three moving group members (RKS0121+2419, RKS0723+2024, and RKS1633-0933) were independently identified as young or active through our spectroscopic analysis.

This kinematic approach effectively increases our count of young or active stars from 43 (identified spectroscopically in $\S$~\ref{section:characterization}) to 51, representing 8.8\% of our survey sample and highlighting the importance of multi-faceted approaches to stellar age determination. The identification of moving group members through kinematics reveals additional young stars that have depleted their lithium or lack strong chromospheric activity signatures but retain the kinematic memory of their birth environments --- such stars would be missed in lithium and activity surveys alone. These results underscore the complex relationship between stellar age, activity, and observable diagnostics, emphasizing that comprehensive stellar characterization requires both spectroscopic and kinematic analysis. The result leaves us with 529 mature and inactive K dwarfs in our sample, shown in Table~\ref{tab:population_summary} as 91.2\% of our stars.

\subsection{Identifying Galactic Populations via $UVW$ Space Motions}
\label{subsec:galactic_populations}

We now examine the broader Galactic population structure within our K dwarf sample. This classification utilizes the $UVW$ space motions compiled in Table~\ref{tab:table_kinematics} in Appendix~\ref{append:append_kinematicstable} to distinguish between the primary stellar populations of the Milky Way: the thin disk, thick disk, and halo components. Each population exhibits characteristic velocity distributions that allow us to place our K dwarf sample in the context of Galactic structure and evolution.

\begin{figure}[htb!]
\centering
\includegraphics[width=0.8\textwidth]{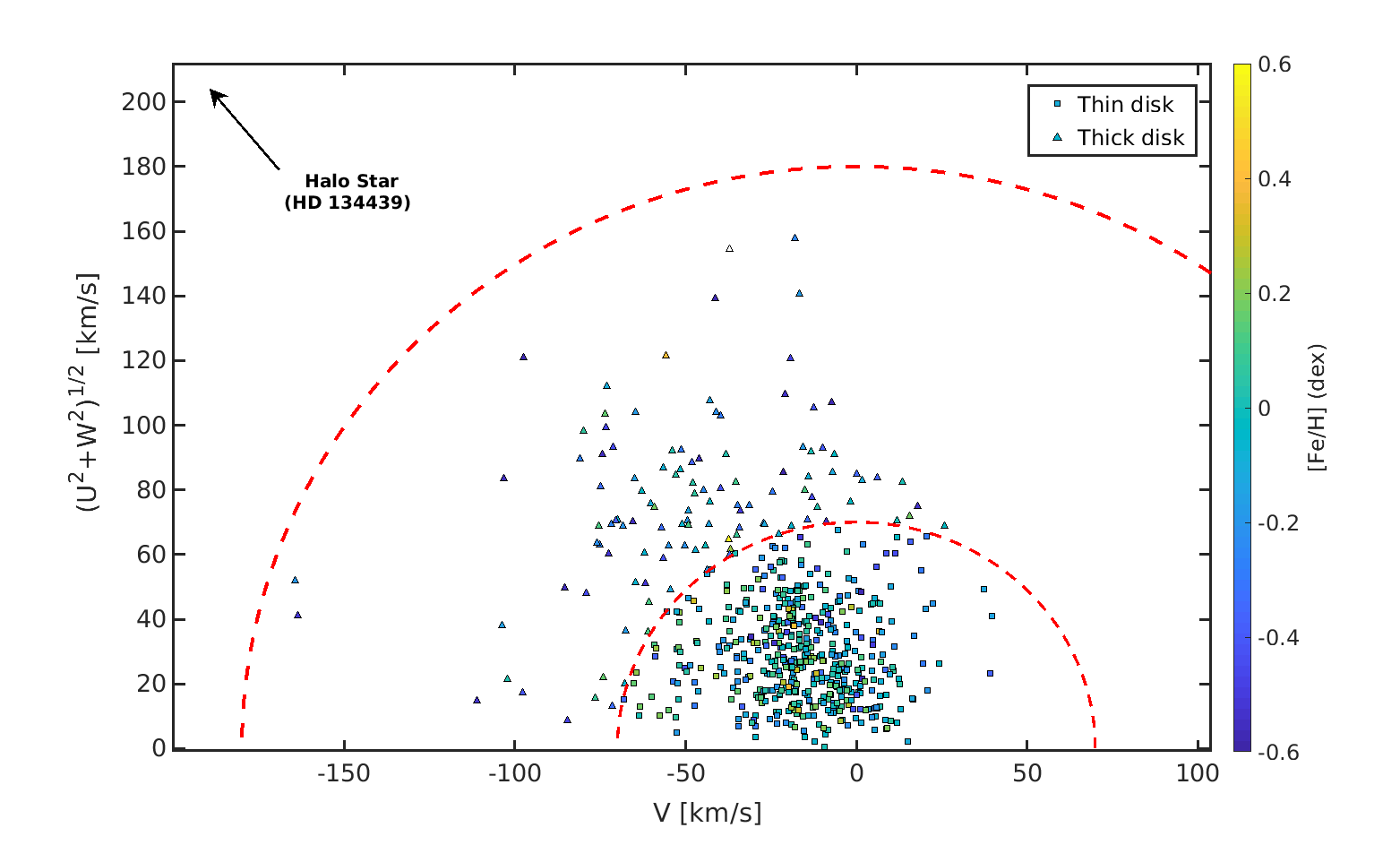} 
\caption{A Toomre diagram of 572 K dwarfs with $UVW$ values helps distinguish Galactic populations. The dotted curves at 75 km s$^{-1}$ and 180 km s$^{-1}$ roughly delineate the thin disk, thick disk, and halo populations. The thin disk (squares), thick disk (triangles), and halo (off the plot) contain 464, 107, and 1 stars, respectively. The adjacent color bar indicates the metallicity (\feh) for each star.}  
\label{fig:toomre_diagram}
\end{figure}

Figure~\ref{fig:toomre_diagram} presents a Toomre diagram for our characterized sample, plotting the total random velocity $\sqrt{U^2 + W^2}$ against the $V$ component of Galactic motion. This diagnostic effectively separates stellar populations based on their kinematic properties. Dotted curves at velocities of 75 km s$^{-1}$ and 180 km s$^{-1}$ from \citet{bensby03} and \citet{hinkel14} provide provisional demarcations between populations. In the sample of 572 stars with $UVW$ values, we find 464 thin disk stars (80\%), 107 thick disk stars (18.4\%), and 1 halo star (0.2\%); these are represented graphically as squares and triangles, with the single point for the halo star well off the plot, in Figure~\ref{fig:toomre_diagram}. An additional 8 stars could not be reliably classified due to measurement limitations. The adjacent color bar indicates the metallicity (\feh) for each star, revealing the well-known metallicity-kinematics correlation.

\begin{figure}[htb!]
\centering
\includegraphics[width=0.48\textwidth]{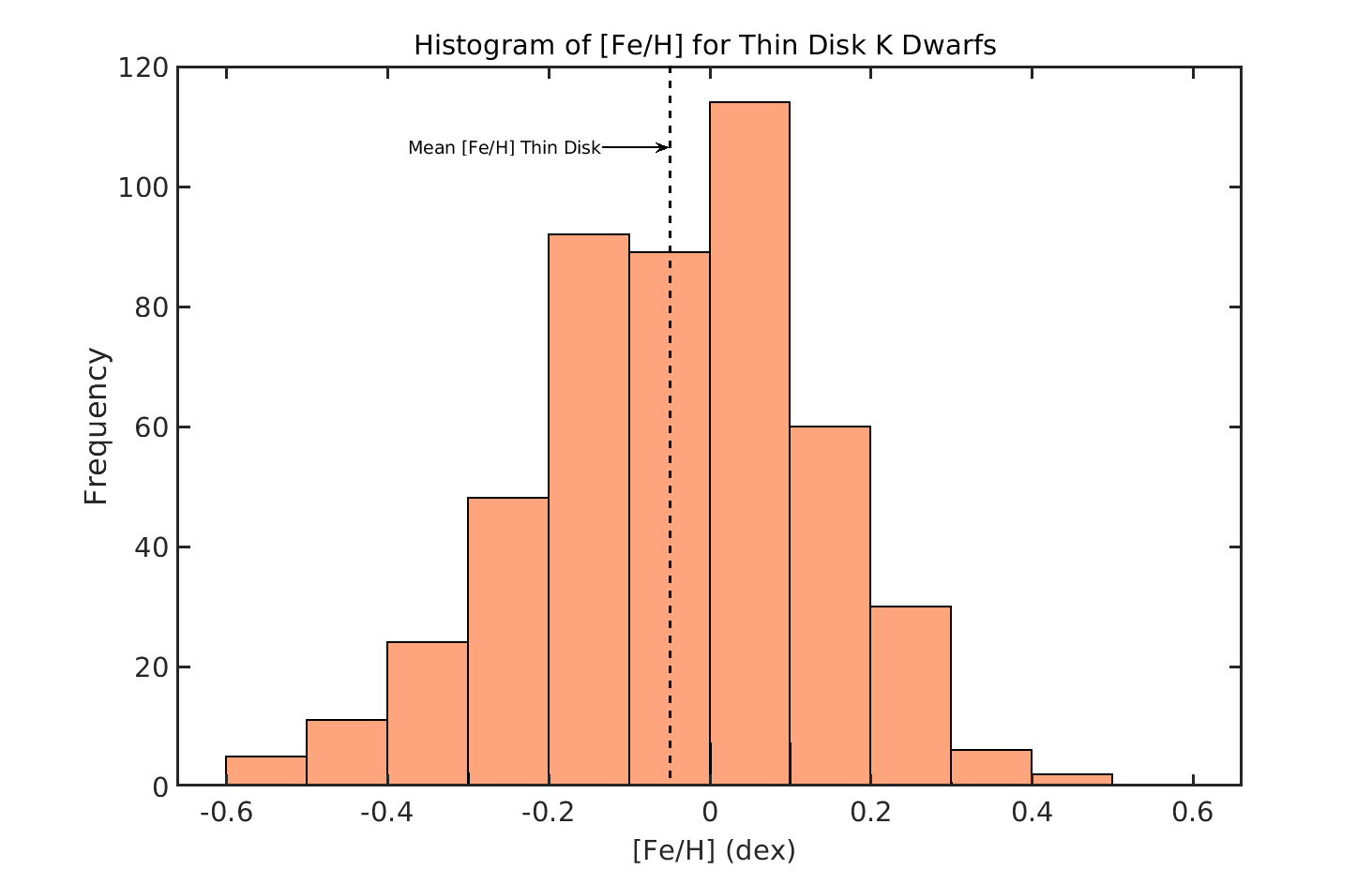}
\includegraphics[width=0.48\textwidth]{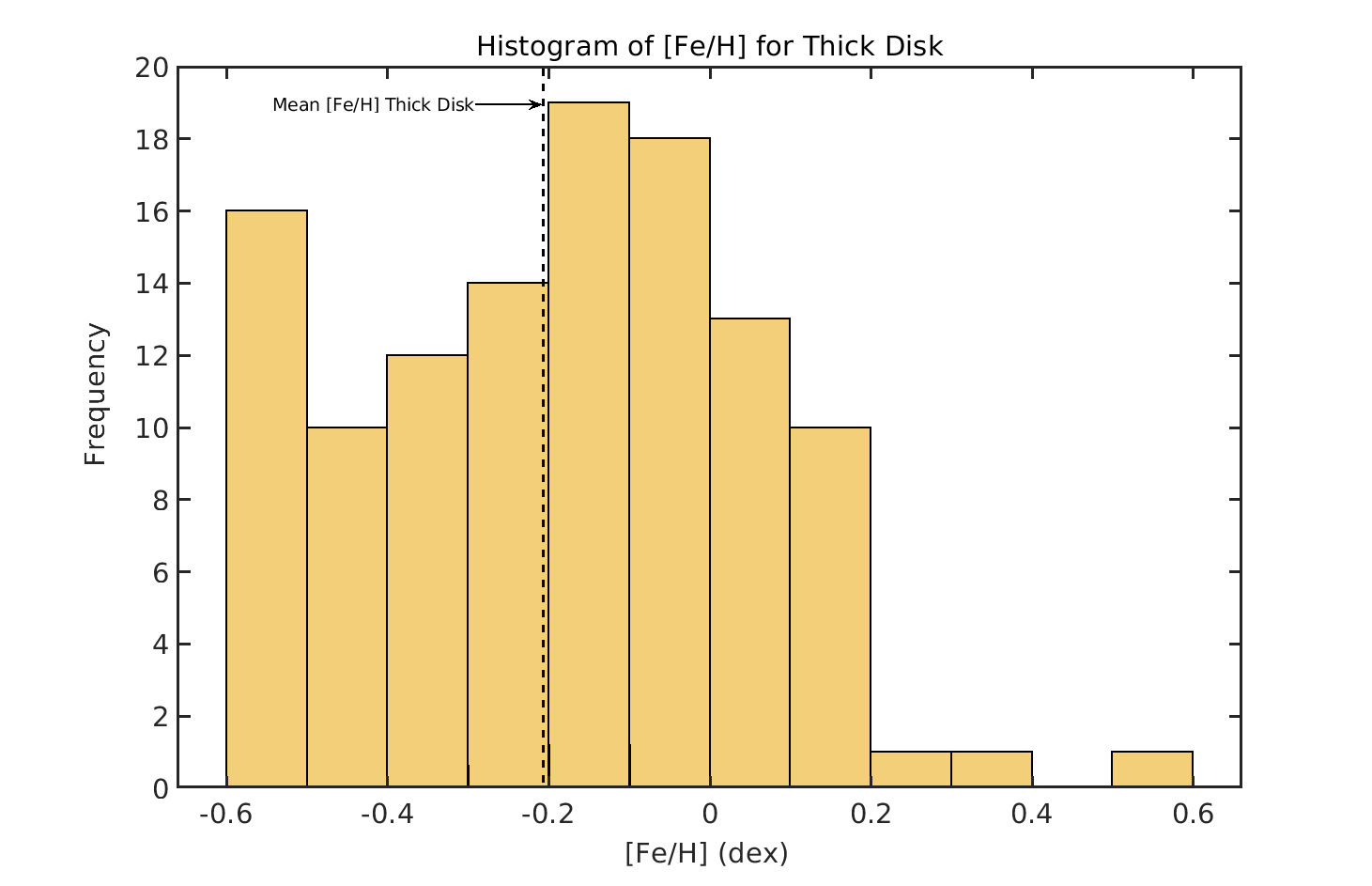}
\caption{Histograms illustrating the \feh distribution of K dwarfs in our sample in two distinct populations within the Milky Way Galaxy. The left histogram represents the thin disk population, with a mean \feh value indicated by a vertical dashed line at $-0.05$ dex. The right histogram represents the thick disk population, with a mean \feh value indicated by a vertical dashed line at $-0.21$ dex.}
\label{fig:feh_histograms}
\end{figure}

The metallicity distributions of these populations, shown in Figure~\ref{fig:feh_histograms}, reveal significant differences between the thin and thick disk components. Our thin disk sample has a mean \feh of $-0.05$ dex, while the thick disk has a mean of $-0.21$ dex, indicating the thick disk population is approximately 30\% more metal-poor. These proportions align remarkably well with previous studies: our 18.3\% thick disk fraction compares favorably with the 18\% reported by \citet{hinkel14} for a sample of over 2000 nearby stars, and agrees with the comparable value of $\sim$20\% from \citet{adibekyan13}. However, our metallicity difference of $\sim$0.16 dex between populations is somewhat smaller than the 0.2--0.3 dex differences reported for nearby stars \citep{feltzing03,reddy06}, and is predictably smaller than the $\sim$0.5 dex differences found in studies covering larger Galactic volumes that include stars with more extreme metallicity values \citep{ivezic08}.

A striking result from our kinematic analysis is the strong correlation between stellar age and Galactic population membership. Combining our spectroscopic results from $\S$~\ref{section:characterization} with the kinematic identifications from $\S$~\ref{subsec:uvw_space_motions}, we identify 51 young or active K dwarfs in our sample. Of these, only two belong to the thick disk population: HD 196998 (RKS2041-2219, DG Cap) and HD 7808 (RKS0117-1530). HD 196998 is an SB1 system that exhibits radial velocity variations and was identified as active in our benchmark study \citep{hubbardjames22}, while HD 7808 was identified as active through our spectroscopic analysis. The remaining 49 young or active stars are thin disk members, reinforcing the expected relationship between stellar age and kinematic properties.

The most remarkable object found in our kinematic analysis is HD 134439 (RKS1510-1622, also known as LHS 53 or GJ 579.2), which appears as the lone outlier in Figure~\ref{fig:toomre_diagram}. This high proper motion ($\mu = 3.68\arcsec$ yr$^{-1}$), high velocity ($\gamma$ = $+$310 km s$^{-1}$) K dwarf at a distance of 29.4 pc is a known metal-poor subdwarf with \feh $< -1.0$ dex, likely accreted from a disrupted dwarf galaxy \citep{lim21}. Its extreme kinematics place it firmly in the halo region of velocity space, making it the only halo star in our sample. As noted in $\S$~\ref{subsec:analysis_stellar parameters}, this star exhibits a unique spectrum that falls outside the boundaries of the abridged ESM library shown in Figure~\ref{fig:appendix_53_specialones}, preventing standard stellar parameter determination and highlighting its unusual nature among nearby K dwarfs.

\section{Discussion}
\label{section:discussion}

\subsection{Population Summary}

The combined results of our spectroscopic and dynamic analyses reveal the activity and age population characteristics of the nearby K dwarfs in unprecedented detail. Table~\ref{tab:population_summary} summarizes our stellar classifications. The overwhelming majority of K dwarfs in the solar neighborhood --- 91.2\% (529 stars) --- are chromospherically quiescent (showing no H$\alpha$ excess), mature (showing no Li I feature), and are not part of a young moving group or cluster. This sample of ``calm, mature" stars is an exceptional resource for future exoplanet surveys targeting potentially habitable worlds.

The remaining 8.8\% (51 stars) of K dwarfs in the survey exhibit activity, have Li I absorption features, or are members of young moving groups or clusters. Of these, 24 (4.1\%) are active based on H$\alpha$ measurements but do not appear to be young, and 14 (2.4\%) are presumably young because they have Li I features or are members of young groups or clusters but are quiescent in H$\alpha$. A modest number of 13 stars (2.2\%) show both H$\alpha$ activity and Li I absorption. This latter group demonstrates that stellar activity and age are related but not matched phenomena, with some young stars maintaining relatively quiet chromospheres while some older stars sustain magnetic activity, perhaps through mechanisms such as tidal interactions caused by close companions.

Kinematically, our sample reflects the expected Galactic population structure of the solar neighborhood as shown in Table \ref{tab:population_summary}, with 464 stars (80.0\%) belonging to the thin disk, 107 stars (18.4\%) to the thick disk, and one halo member (0.2\%). The stellar properties span effective temperatures from 3600 to 5500 K and metallicities from $-$0.6 to $+$0.4 dex, with a mean metallicity of $-$0.02 dex indicative of the solar neighborhood's composition. Notably, 413 stars (71.2\%) exhibit solar-like metallicities ($-$0.2 to $+$0.2 dex), while 20 stars (3.4\%) are metal-poor with [Fe/H] $< -$0.5 dex. The metal-poor population shows no overlap with the young or active stars, highlighting a clear separation between chemically distinct stellar populations and their evolutionary states.

\begin{table}[htb!]
\centering
\caption{Population Summary of 580 K Dwarf Survey Sample}
\label{tab:population_summary}
\begin{tabular}{lcc}
\hline\hline
Population Category & Number & Percentage \\
\hline
\multicolumn{3}{l}{\textbf{Activity/Age Classifications:}} \\
Mature, quiescent (calm) & 529 & 91.2\% \\
Active only (H$\alpha$, no Li) & 24 & 4.1\% \\
Young only (Li or kinematic) & 14 & 2.4\% \\
Young and active (both) & 13 & 2.2\% \\
Unclassified & 0 & \nodata \\
\hline
\multicolumn{3}{l}{\textbf{Galactic Populations:}} \\
Thin disk & 464 & 80.0\% \\
Thick disk & 107 & 18.4\% \\
Halo & 1 & 0.2\% \\
Unclassified & 8 & 1.4\% \\
\hline
\multicolumn{3}{l}{\textbf{Metallicity:}} \\
Metal-rich ([Fe/H] $>$ $+$0.2 dex) & 32 & 5.5\% \\
Solar metallicity ($+$0.2 to -0.2 dex) & 408 & 70.3\% \\
Lower metallicity ($<$ $-$0.2 to $-$0.5 dex) & 111 & 19.1\% \\
Metal-poor ([Fe/H] $<$ $-$0.5 dex) & 20 & 3.4\% \\
Unclassified & 9 & 1.6\% \\

\hline
\end{tabular}
\end{table}

\subsection{Exoplanet Host Stars: Current Status and Metallicity Trends}
\label{subsec:hoststars}

K dwarf stars have garnered significant attention as potentially optimal hosts for habitable worlds, often termed ``super-habitable" for planets, due to their favorable balance of properties: longer main-sequence lifetimes (17--70 Gyr) than solar-type stars, reduced activity compared to M dwarfs \citep{richeyyowell19}, and habitable zones positioned at distances that avoid tidal locking \citep{kasting14,cuntz16}. As illustrated in Figure \ref{fig:hr_exo_primarysample}, K dwarfs have been systematically underexplored in exoplanet surveys compared to G and M dwarfs, making direct occurrence rate comparisons difficult with current data. Our comprehensive characterization of 580 nearby K dwarfs provides crucial empirical data to evaluate these claims and identify the most promising stellar hosts for future exoplanet surveys and habitability assessments.

\begin{figure}[htb!]
\centering
\includegraphics[width=0.70\textwidth]{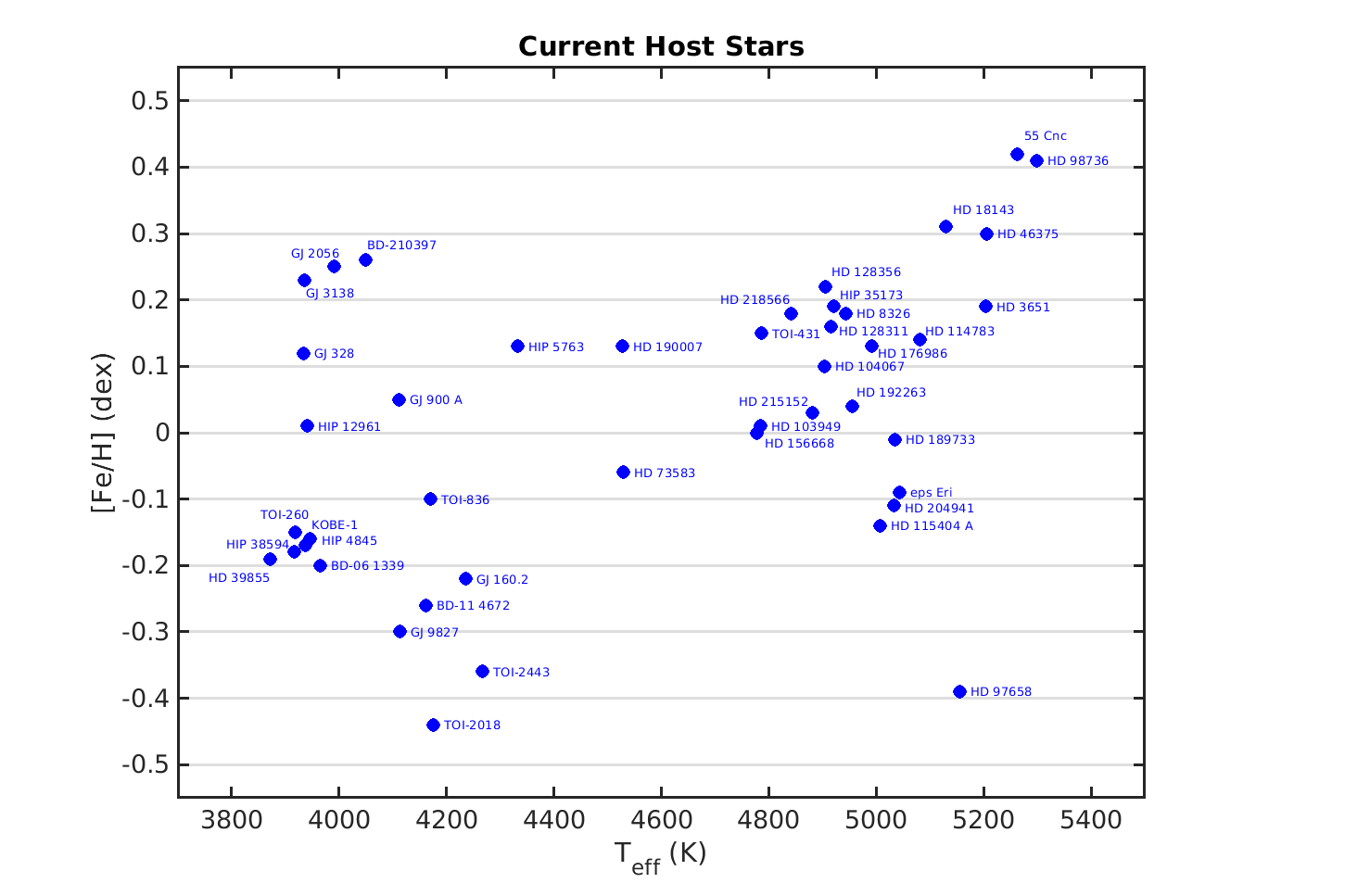}
\includegraphics[width=0.70\textwidth]{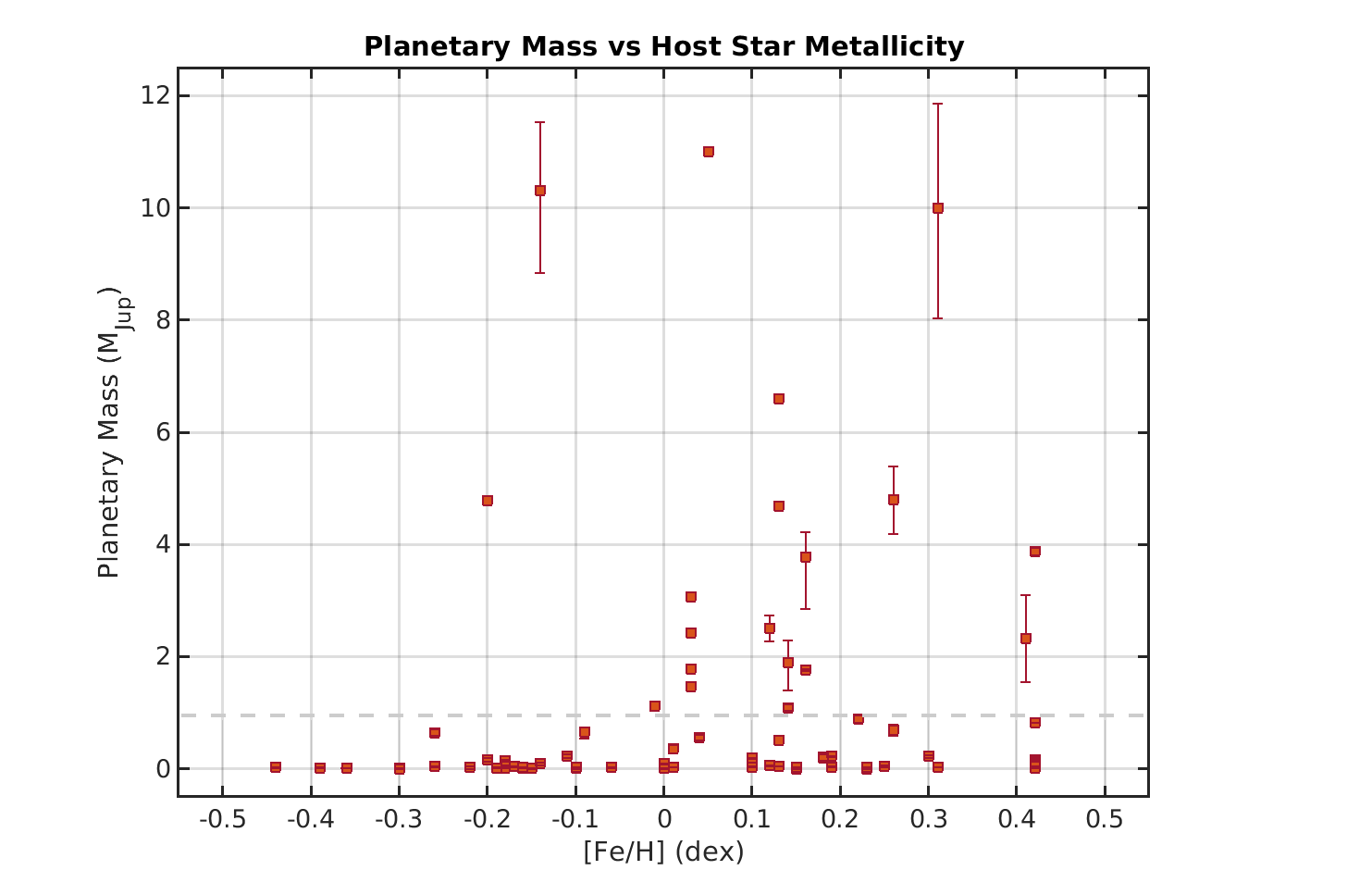}
\caption{Exoplanet host properties for 44 nearby K dwarfs in our sample reported to have planets in the NASA Exoplanet Archive \citep{nasaexoplanetarchive} as of July 2025. Top: Distribution of the 44 host stars in \feh{} vs.~\teff{} space, with stellar parameters determined via ESM. Star names are labeled using NASA Exoplanet Archive identifiers. Bottom: Planetary mass vs. host star metallicity for the 72 planets orbiting these 44 K dwarfs. The horizontal dashed line at $M_p = 1$ $M_{Jup}$ marks Jupiter's mass for reference. Error bars on planetary masses are shown where available from the NASA Exoplanet Archive.}
\label{fig:feh_teff_hoststars}
\end{figure}

According to the NASA Exoplanet Archive \citep{nasaexoplanetarchive}\footnote{https://exoplanetarchive.ipac.caltech.edu/}, as of July 2025, 44 of our 580 K dwarfs (7.5\%) host a total of 72 confirmed planetary companions. Given that many of these stars have not yet been systematically searched for planets, this is certainly an underestimate of the true planetary occurrence rates around these stars. In fact, K dwarfs have been systematically overlooked in major planet-hunting surveys because they are fainter than G dwarfs at equivalent distances and less amenable to ultra-precise radial velocity measurements than M dwarfs that are observed to reveal terrestrial planets \citep{lillobox22}.

Figure~\ref{fig:feh_teff_hoststars} (top panel) reveals that exoplanet host stars are distributed across the full range of stellar metallicities in our sample, from metal-poor systems like HD 26965 and HD 96758 with \feh = $-$0.4 to metal-rich hosts such as 55 Cancri and HD 98736 with \feh = $+$0.4.\footnote{55 Cancri hosts five confirmed planets, including 55 Cancri f in the liquid-water zone, although with $M_p \sin i =$ 45.8$\pm$12.7 M$_{\oplus}$, this world likely lacks a solid surface conducive to surface-based life \citep{fischer08}.} While K dwarf exoplanet hosts span the complete metallicity distribution, the bottom panel of Figure~\ref{fig:feh_teff_hoststars} demonstrates that planetary mass shows a clear correlation with host star metallicity. Jupiter-mass planets ($M_p \gtrsim 1$ $M_{Jup}$) preferentially orbit metal-rich hosts with [Fe/H] $> 0$, consistent with the well-established giant planet-metallicity correlation \citep{fischer05,buchhave14}. In contrast, lower-mass planets (Neptune-mass and below) show no strong metallicity preference and are found around both metal-rich and metal-poor K dwarfs across our sample.

This differential metallicity dependence has important implications for habitability prospects. Previous studies have demonstrated that planets of various sizes form around stars with diverse metallicities, though with different formation mechanisms: terrestrial and small planets show weak or no metallicity correlations \citep{buchhave12,mann13,gaidos16}, while giant planet formation shows a strong preference for metal-rich environments through core accretion \citep{fischer05,buchhave14}. Our K dwarf sample exhibits this same pattern, with Jupiter-mass planets concentrated above [Fe/H] $> 0$ while lower-mass rocky worlds appear across the full metallicity range observed in Figure~\ref{fig:feh_teff_hoststars}. This suggests that even the metal-poor K dwarfs in our sample (21 stars with \feh $\leq -$0.5 dex, comprising 3.6\% of the population) could harbor terrestrial planetary systems, potentially expanding the Galactic real estate available for life beyond the metal-rich thin disk population.

\subsection{Stellar Activity and the K Dwarf Habitability Paradigm} \label{subsec:activity_paradigm}

Recent work has raised questions about whether K dwarfs truly offer the most favorable conditions for habitable worlds.\citet{richeyyowell19} conducted the first comprehensive investigation of X-ray and UV evolution in K stars, finding that these stars maintain elevated high-energy flux levels longer than previously assumed. Their follow-up spectroscopic analysis \citep{richeyyowell22} revealed that K dwarf UV emission remains nearly constant for the first 650 Myr before declining, in stark contrast to early M dwarfs, which begin UV decline after only a few hundred megayears. Most critically, they demonstrated that K dwarfs experience ``rotational stalling" during their first gigayear, maintaining rotation periods near 10 days and sustaining chromospheric and coronal activity levels comparable to young M dwarfs. This prolonged active phase appears to extend far beyond the typical timescales needed for at least initial atmospheric creation on terrestrial planets, potentially subjecting habitable zone worlds to damaging UV radiation for hundreds of millions of years longer than anticipated.

These findings suggest that the K dwarf advantage may be more nuanced than originally proposed. While K dwarfs ultimately settle into long-lived, quiescent phases ideal for hosting stable atmospheres, the extended juvenile active period may delay rather than prevent the emergence of habitable conditions. It could be that planets around K dwarfs require additional time to develop and retain thick atmospheres capable of supporting liquid water and potentially life \citep{airapetian20}. Thus, this delayed habitability timeline does not necessarily eliminate K dwarfs as premier planetary hosts. The extraordinary main-sequence lifetimes of K stars (17--70 Gyr) provide ample time for atmospheric recovery and the development of complex ecosystems, even if there are extended active phases. Furthermore, planets with substantial initial volatile inventories or efficient atmospheric replenishment mechanisms might maintain habitability throughout the active period \citep{meadows18}. 

\section{Conclusions}\label{section:conclusion}

Here we present a large spectroscopic survey of nearby K dwarfs, characterizing 580 stars within 33.3 pc using high-resolution CHIRON spectra. Our comprehensive analysis of stellar properties, activity status, youth indicators, and kinematic populations yields crucial insights for prioritizing targets in the search for potentially habitable worlds around some of the most promising stellar hosts in the solar neighborhood.

One of the major outcomes of this project is the establishment of a comprehensive, high-resolution activity and age spectral gallery of 580 nearby K dwarfs\footnote{\href{https://hodarijames.github.io/spectral_library/page1.html}{RKSTAR Survey Activity and Age Spectral Gallery}}. This gallery represents one of the most uniform datasets of nearby K dwarfs available to the astronomical community and expands on previous work of \cite{montes98} and the ``Library of High-Resolution Spectra for Late-Type Stars'',\footnote{\href{https://www.sea-astronomia.es/sites/default/files/archivos/proceedings10/via_lactea/POSTERS/poster_montesd1.pdf}{Library of Spectra for Cool Stars (Poster 2013)}} as detailed in \cite{montes13}. The publicly available gallery provides critical resources for future studies of stellar magnetic activity cycles, Galactic chemical evolution, and the identification of optimal exoplanet host stars in the solar neighborhood.

We find that 51 (8.8\%) of the K dwarfs are active and/or young based on H$\alpha$ emission, lithium absorption, or kinematic youth indicators. Spectra for all 51 are shown in Appendix~\ref{append:append_spectrallibrary}, with additional information about some of this sample's most interesting members provided in the Systems Worthy of Note section (Appendix~\ref{append:append_systems}). The remaining 529 (91.2\%) stars show no spectroscopic signatures of youth or activity and do not appear to be members of young moving groups or clusters. These calm, mature stars have passed through their active juvenile phases and settled into stable, long-lived configurations ideal for maintaining planetary atmospheres over geological timescales.

Our kinematic analysis reveals that the local K dwarf population is dominated by thin disk members (464 stars, 80\%), with a substantial thick disk component (107 stars, 18.4\%), and one confirmed halo star. The metallicity distribution spans \feh  = $-$0.6 to $+$0.5 dex, with 21 stars (3.6\%) classified as metal-poor \feh $\leq -$0.5 dex. Only two of the 51 young or active stars belong to the thick disk population (HD 7808 and HD 196998), reinforcing the expected correlation between stellar age and Galactic population membership. The remaining 49 active stars are thin disk members, consistent with the younger ages of this population.

The NASA Exoplanet Archive lists 44 stars (7.5\%) in the sample with reported planets, and undoubtedly more await discovery. Our K dwarf hosts span the full metallicity range sampled, from metal-poor to metal-rich systems. Importantly, while Jupiter-mass planets in our sample preferentially orbit metal-rich hosts ([Fe/H] $> 0$), lower-mass planets show no such metallicity preference, consistent with formation mechanisms that favor giant planet growth in metal-rich environments while permitting terrestrial planet formation across diverse metallicities. If this pattern holds for larger K dwarf samples, it has important implications for Galactic habitability prospects, suggesting that terrestrial planetary systems may exist around K dwarfs across diverse stellar populations, including ancient, metal-poor thick disk and halo stars. The presence of confirmed planets around both thin and thick disk K dwarfs in our sample provides initial support for this possibility.

Overall, the 529 mature, inactive K dwarfs comprise a vetted, rich list of targets for terrestrial planet detection and habitability assessment.  These stars span the full range of stellar parameters in our sample: effective temperatures from 3600--5500 K, \feh from $-$0.6 to $+$0.5 dex, and represent both thin disk and thick disk populations. This diversity ensures that future planet searches of these stars will probe planetary systems across a wide range of formation environments and stellar ages. The proximity of these targets, all within 33.3 pc, makes their planets prime candidates for detailed atmospheric characterization. 

\vspace{-0.25cm}

\section*{Acknowledgments}

We thank the dedicated staff at Cerro Tololo Inter-American Observatory, particularly Roberto Aviles and Rodrigo Hinojosa, for their expertise in executing the CHIRON observations that form the foundation of this work. We are grateful to our RECONS colleagues Andrew Couperus, Tim Johns, Aman Kar, Madison LeBlanc, and Eliot Vrijmoet for offering valuable insight along the way as this project came to fruition.

This work was supported by the National Science Foundation under grant AST-1910130. H.-S.H.-J. acknowledges support from the Georgia State University Provost's Dissertation Fellowship and the Southern Regional Education Board (SREB) Dissertation Year Award.

We have used data from the CHIRON spectrograph on the SMARTS 1.5 m telescope, which is operated as part of the SMARTS Observatory by RECONS (www.recons.org) members. This work has made use of data from the European Space Agency (ESA) mission Gaia (https://www.cosmos.esa.int/gaia), processed by the Gaia Data Processing and Analysis Consortium (DPAC, https://www.cosmos.esa.int/web/gaia/dpac/consortium). Funding for the DPAC has been provided by national institutions, in particular the institutions participating in the Gaia Multilateral Agreement. This research has made use of the NASA Exoplanet Archive(https://exoplanetarchive.ipac.caltech.edu/), which is operated by the California Institute of Technology, under contract with the National Aeronautics and Space Administration under the Exoplanet Exploration Program. This research has made use of the Two Micron All Sky Survey (2MASS), which is a joint project of the University of Massachusetts and the Infrared Processing and Analysis Center/California Institute of Technology, funded by the National Aeronautics and Space Administration and the National Science Foundation. This research has made use of the SIMBAD database, operated at CDS, Strasbourg, France, and NASA's Astrophysics Data System.

\vspace{0.2cm}
\noindent\textbf{Facility:} CTIO:1.5m (CHIRON).
\noindent\textbf{Software:} astropy \citep{astropy13,astropy18}, barycorrpy \citep{kanodia18}, Empirical SpecMatch \citep{yee17}, matplotlib \citep{hunter07}, SciPy \citep{virtanen20}, and TOPCAT \citep{taylor05}.

\section*{ORCID iDs}

Hodari-Sadiki Hubbard-James \href{https://orcid.org/0000-0003-4568-2079}{https://orcid.org/0000-0003-4568-2079}

Sebastian Carrazco-Gaxiola \href{https://orcid.org/0009-0006-9244-3707}{https://orcid.org/0009-0006-9244-3707}

Todd J. Henry \href{https://orcid.org/0000-0002-9061-2865}{https://orcid.org/0000-0002-9061-2865}

Leonardo A. Paredes \href{https://orcid.org/0000-0003-1324-0495}{https://orcid.org/0000-0003-1324-0495}

Azmain H. Nizak \href{https://orcid.org/0000-0002-1457-1467}{https://orcid.org/0000-0002-1457-1467}

Xavier Lesley-Saldaña \href{https://orcid.org/0009-0000-5136-6924}{https://orcid.org/0009-0000-5136-6924}

Wei-Chun Jao \href{https://orcid.org/0000-0003-0193-2187}{https://orcid.org/0000-0003-0193-2187}

Abigail Arbogast \href{https://orcid.org/0009-0004-7539-8129}{https://orcid.org/0009-0004-7539-8129}

\clearpage

\appendix
\section{Systems Worthy of Note}\label{append:append_systems}

\noindent
This appendix highlights a selection of the most notable and scientifically interesting systems from our sample of 51 young and/or active K dwarfs, as well as several calm stars with unique characteristics.
\newline

\noindent
\textit{RKS0252-1246}: \boxed{18} {\bf HD 17925} was initially identified as a young star due to its strong Li I feature \citep{cayrel89}. Moreover, using the BANYAN $\Sigma$ tool, this star is not linked to any specific moving group or cluster, suggesting that it is a field star.
\newline

\noindent
\textit{RKS0417+2033}: \boxed{13} {\bf HD 284336} is a previously unidentified young star that we have now classified as such. Our spectral examination, evident in Figure~\ref{fig:sp_young}, reveals a noticeable presence of the Li I absorption line, typically indicative of stellar youth. Although not characterized by high rotational speed, as per our calculated \vsini values, HD 284336 is found within the thin disk population of the galaxy, a feature worth noting. This new discovery warrants further investigation.
\newline

\noindent
\textit{RKS0436+2707}: {\bf HD 283750, also known as V833 Tau or BD +26 730}, is a renowned flare star in our solar neighborhood. Its activity and status as a single-lined spectroscopic binary (SB1) with a two-day period were first noted by \cite{hartman81}. HD 283750 is among the most active stars in our study, exhibiting an EW[H$\alpha$] of -0.34 \AA{} and an EW[Ca II] ratio of 0.21. Interestingly, despite its elevated activity and binary status, its $\gamma$ velocity—measured by RECONS—aligns with the Gaia DR3 value, as shown by its position on the 1-to-1 line in Figure \ref{fig:gamma_mainsample}.
\newline

\noindent
\textit{RKS0441+2054}: \boxed{9} {\bf HD 29697 or V834 Tau} is a known active star and is also one of the seven RVV K dwarfs that were analyzed with our benchmark sample. It is described in more detail in \cite{hubbardjames22}.
\newline

\noindent
\textit{RKS0658-1259}: \boxed{10} {\bf HD 51849}, identified as an active star with a stellar companion, was classified as an SB1 with a separation of 0.5\arcsec by \cite{tokovinin19}. Our research exhibits evidence of both activity and youthfulness in this proximate binary system. Figure~\ref{fig:sp_youngactive} reveals a minor Li I absorption line, while Figure \ref{fig:sp_halpha_vs_lii} indicates that HD 51849 doesn't align with the old locus of stars.
\newline

\noindent
\textit{RKS0723+2024}: \boxed{16} {\bf BD+20 1790} serves as a notable member of the AB Dor Moving Group. Recognized as a young star, it boasts an estimated age of 120 Myr. BD+20 1790 was flagged in our study as both youthful and active, as supported by spectroscopic and kinematic assessments. Notably, it displays a heightened rotational velocity with a \vsini measurement of 7.99 km s$^{-1}$, reported in Table~\ref{tab:stellarproperties_special_paper2}. Its distinctive space motions further classify BD+20 1790 as a thin disk star.
\newline

\noindent
\textit{RKS0734-0653}: \boxed{7} {\bf HD 60491}, categorized as a young star and potential member of the Ursa Majoris moving group by \cite{montes01}, only meets one of the two kinematic criteria for association membership. Our analysis, as well as the BANYAN $\Sigma$ tool, does not identify HD 60491 as part of any moving group. However, we do note the existence of a minor Li I absorption line in Figure~\ref{fig:sp_young}, despite a lack of chromospheric activity at H $\alpha$ or Ca II.
\newline

\noindent
\textit{RKS0819+0120}: \boxed{12} {\bf BD+01 2063}, acknowledged as a young star and a member of the Carina-near (estimated age $\sim$200 Myr) association \citep{ujjwal20}, is also placed in the Carina moving group by the BANYAN $\Sigma$ tool. Our analysis reveals a robust Li I absorption line, as shown in Figure~\ref{fig:sp_youngactive}. Additionally, we also observe chromospheric activity for this K dwarf, highlighted in Table~\ref{tab:stellarproperties_special_paper2}.
\newline

\noindent
\textit{RKS0904-1554}: \boxed{8} {\bf HD 77825}, previously identified as a young star and a member of the Castor Moving Group (estimated age 200 Myr) by \cite{montes01}, does not correlate with any moving group or association according to the BANYAN $\Sigma$ tool. Our observations depict a faint but discernible Li I absorption line as seen in Figure~\ref{fig:sp_young}. 
\newline

\noindent
\textit{RKS0907+2252}: \boxed{11} \& \boxed{d} {\bf HD 78141}, identified as a young star by \cite{stanfordmoore20}, is estimated to be between 50-400 Myr old, as evidenced by a strong Li I signature. Our study corroborates this, depicting a profound Li I absorption feature in Figure~\ref{fig:sp_youngactive}. Notably, HD 78141 deviates from the 1-to-1 line in Figure \ref{fig:gamma_mainsample}, likely due to its classification as an SB1 star with a period of 160 days \citep{griffin16}.
\newline

\noindent
\textit{RKS0932-1111}: \boxed{17} {\bf LQ Hya} is a well-known young and chromospherically active K dwarf, initially reported by \cite{fekel86}, and classified as a BY Draconis variable. The spectrum of LQ Hya in Figure~\ref{fig:sp_youngactive} reveals not only a Li I absorption line and notably broader spectral lines compared to its K dwarf counterparts, attributable to rotational broadening. Our measurements give LQ Hya a \vsini value of 27.44 km s$^{-1}$. 
\newline

\noindent
\textit{RKS1043-2903}: \boxed{15} {\bf V419 Hya} is a recognized BY Draconis variable, with a debris disk reported by \cite{plavchan09} using data from the Spitzer Space Telescope. While our analysis categorizes V419 Hya as youthful due to a strong Li I absorption line, and minor signs of chromospheric activity as seen in Figure~\ref{fig:sp_youngactive}. 
\newline

\noindent
\textit{RKS1121-2027}: {\bf HD 98712A}, part of a known binary system, displays chromospheric activity potentially linked to its close companion as reported by \cite{paredes22}. Our study identified this K dwarf as active based on both EW[H $\alpha$] and EW[Ca II] ratio tests. The absence of the Li I feature adds credence that this activity is due to the presence of a close companion and not youth.
\newline

\noindent
\textit{RKS1303-0509}: \boxed{14} {\bf PX Vir} is also a member of our benchmark sample and was described in detail in \cite{hubbardjames22}.
\newline

\noindent
\textit{RKS1306+2043}: {\bf BD+21 2486A} is the primary component of a multi-star system comprising at least three stellar bodies. This K5V star is orbited by an M4V and an L5 star, as reported by \cite{gomes13}. In our investigation, BD+21 2486A exhibited signs of chromospheric activity. However, similar to HD 98712A, this activity is likely due to the influence of nearby companions, rather than indicating stellar youth.
\newline

\noindent 
\textit{RKS1510-1622}: {\bf HD 134439} is uniquely characterized by its kinematic and spectroscopic properties. HD 134439 stands as the lone halo star in our sample. It is a high proper motion, metal-poor subdwarf that likely originated from a dwarf galaxy merger, exhibiting distinct kinematics that place it in the halo region of our Toomre diagram in Figure \ref{fig:toomre_diagram}. Spectroscopically, as presented in Appendix~\ref{append:append_spectrallibrary}  its unique attributes include the absence of prominent Fe I lines typical of K dwarfs and a much narrower Na I doublet, earning it a unique spot outside the ESM library's boundaries. Its distinguishing features underscore the rich diversity in the stellar population. 
\newline

\noindent \textit{RKS1819$-$0156}: {\bf HD 168442 or GJ 710}, currently categorized as a calm K dwarf in our study (see Table \ref{tab:stellar_properties}), has a fascinating future in store. Projections indicate that in about 1.3 million years, this star will pass near the Sun at a distance of only 0.0636 parsecs \citep{bailerjones22}. At such proximity, GJ 710 would shine as brightly as the most luminous planets in our sky. Additionally, its apparent motion across our sky would also be noticeable, peaking at about one arcminute per year \citep{delafuente18}.
\newline

\noindent
\textit{RKS1833-1626}: {\bf HD 171075} another newly uncovered SB2. The binary nature of HD 171075 is marked by distinct double-line features in the CHIRON spectra. A double-line Ca I feature at 6717 \AA, in the spectral window from 6705 to 6720 \AA, is a clear indication of its binary status. It is ensured that the light from both stars in the system is sent to the spectrograph, considering the 2.7\arcsec size of the fiber as projected on the sky and the significantly smaller star separation. Please refer to Appendix~\ref{append:append_spectrallibrary} for snapshot of the spectrum we observed.
\newline

\noindent \textit{RKS2041-2219}: {\bf HD 196998 or DG Cap} is a known active star and is also one of the seven RVV K dwarfs that were analyzed with our benchmark sample. As detailed in $\S$\ref{subsec:galactic_populations}, {\bf DG Cap} is uniquely positioned as one of two active stars from our sample within the Milky Way Galaxy's thick disk population. It is described in more detail in \cite{hubbardjames22}.
\newline

\section{Spectral Gallery of Special K Dwarfs}\label{append:append_spectrallibrary}

This Appendix provides a collection of 53 noteworthy K dwarfs identified during our analysis. These stars exhibit a range of interesting features, including chromospheric activity, youth indicators, metallicity differences, and binary signatures. They are categorized into five groups (A through E) based on their spectral properties. Each group contains ten stars, plotted with identical wavelength panels that highlight the H$\alpha$ \& Li I features for consistency and comparison.

Groups A--D predominantly include young and/or chromospherically active stars, while Group E includes peculiar systems such as newly discovered spectroscopic binaries, Hyades cluster members, a candidate white dwarf companion system, and a halo star. Catalog names, RKS identifiers, and basic classifications (e.g., active, young, SB2) are annotated on each figure panel and caption.

The spectral plots provided here offer a visual resource to complement the quantitative analyses presented in the main text. Readers are encouraged to use these visualizations to explore subtle differences across the range of effective temperatures, metallicities, and stellar activity status present among the nearby K dwarf population.

\begin{figure}[hp] 
  \centering
  \includegraphics[width=0.7\textwidth]{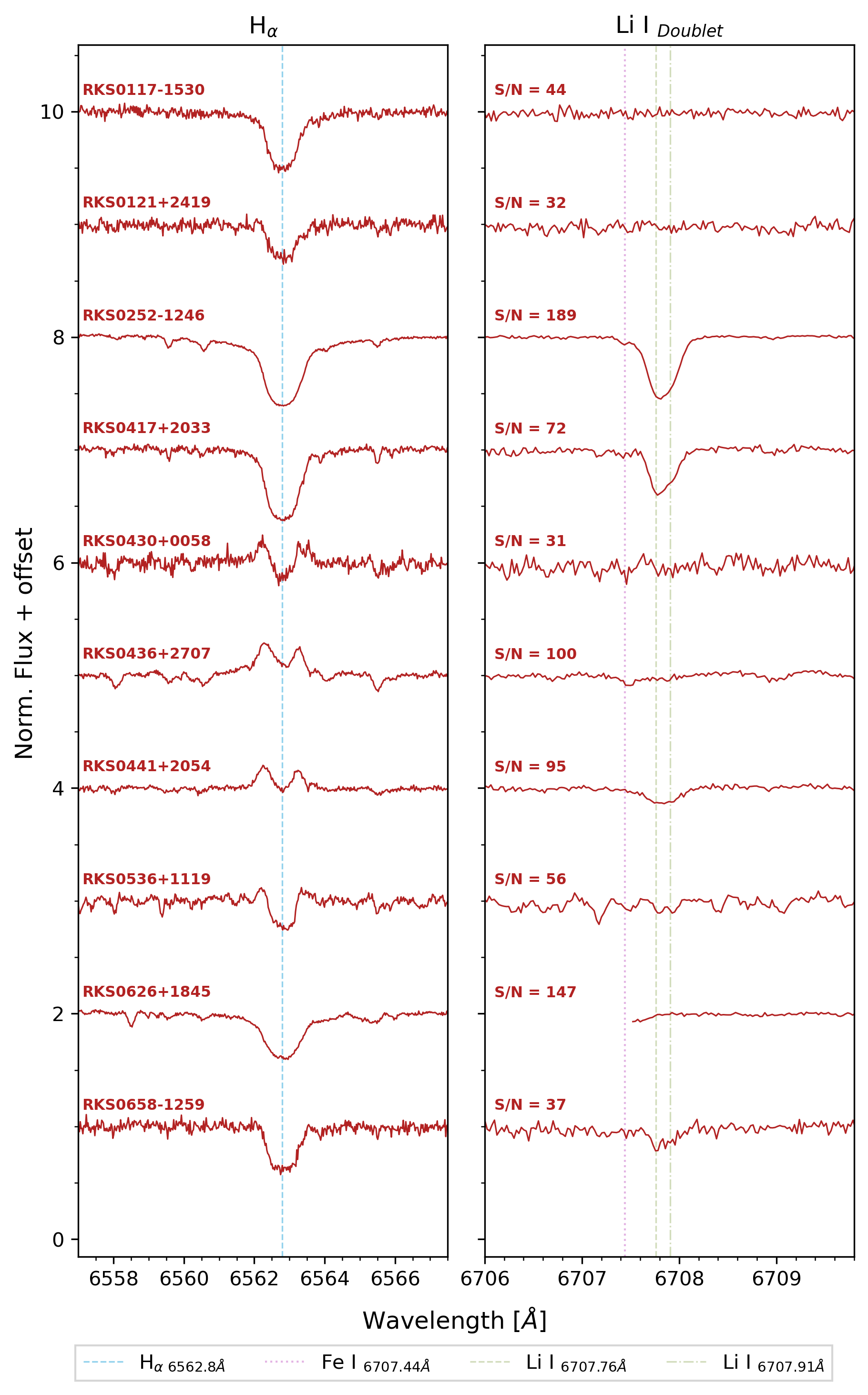}
  \caption{Group A: Spectra of young and active K dwarfs. Stars shown: RKS0117-1530 (A), RKS0121+2419 (A), RKS0252-1246 (Y+A), RKS0417+2033 (Y), RKS0430+0058 (A), RKS0436+2707 (A), RKS0441+2054 (Y+A), RKS0536+1119 (A), RKS0626+1845 (A), RKS0658-1259 (Y+A).}
  \label{fig:appendix_53_specialones}
\end{figure}

\begin{figure}[htbp] 
  \centering
  \includegraphics[width=0.7\textwidth]{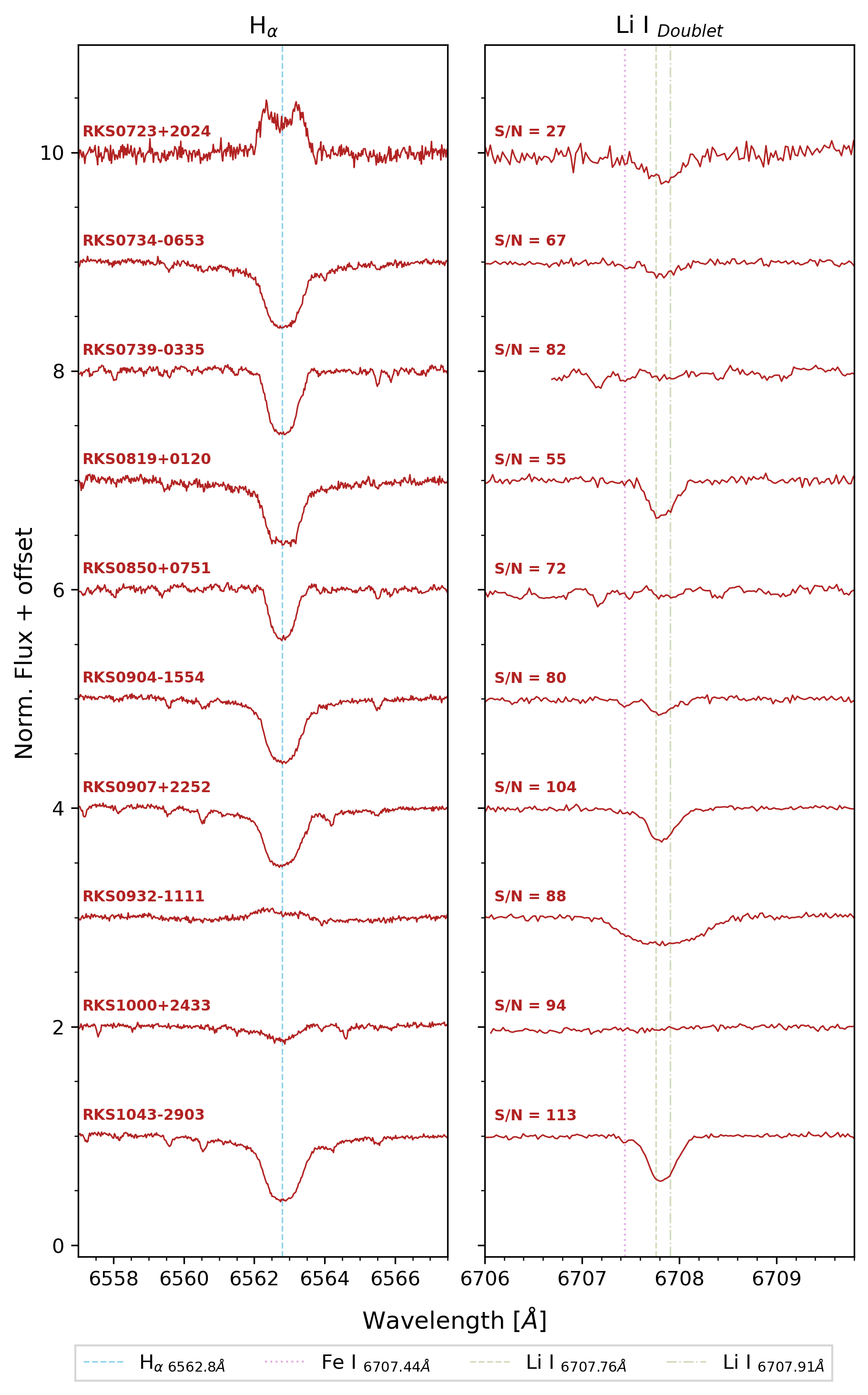}
  \caption{Group B: Spectra of young and active K dwarfs. Stars shown: RKS0723+2024 (Y+A), RKS0734-0653 (Y), RKS0739-0335 (A), RKS0819+0120 (Y+A), RKS0850+0751 (A), RKS0904-1554 (Y), RKS0907+2252 (Y+A), RKS0932-1111 (Y+A), RKS1000+2433 (A), RKS1043-2903 (Y+A).}
  \label{fig:groupB}
\end{figure}

\begin{figure}[htbp]
  \centering
  \includegraphics[width=0.7\textwidth]{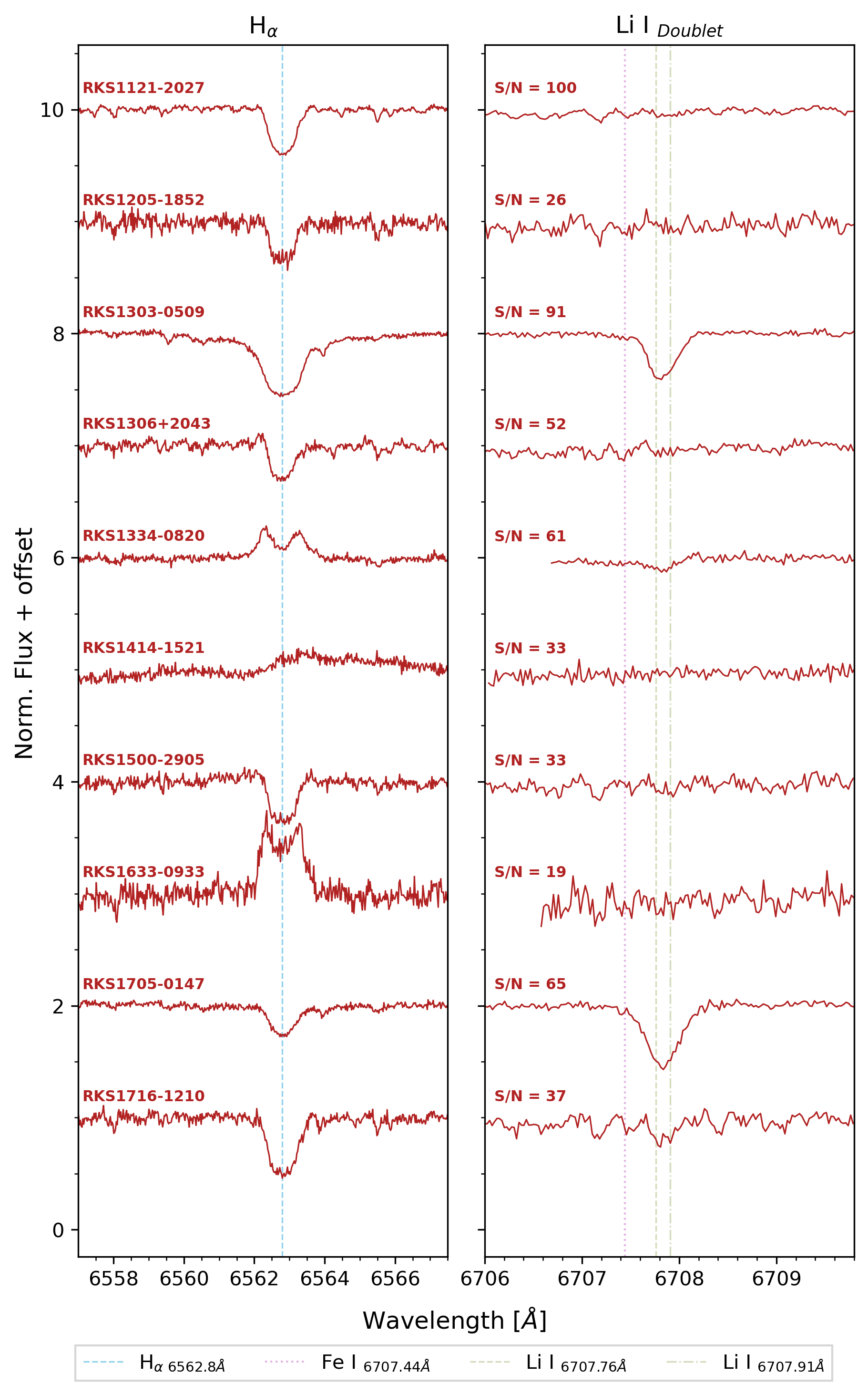}
  \caption{Group C: Spectra of young and active K dwarfs. Stars shown: RKS1121-2027 (A), RKS1205-1852 (A), RKS1303-0509 (Y), RKS1306+2043 (A), RKS1334-0820 (Y+A), RKS1414-1521 (A), RKS1500-2905 (A), RKS1633-0933 (A), RKS1705-0147 (Y+A), RKS1716-1210 (Y).}
  \label{fig:groupC}
\end{figure}

\begin{figure}[htbp]
  \centering
  \includegraphics[width=0.7\textwidth]{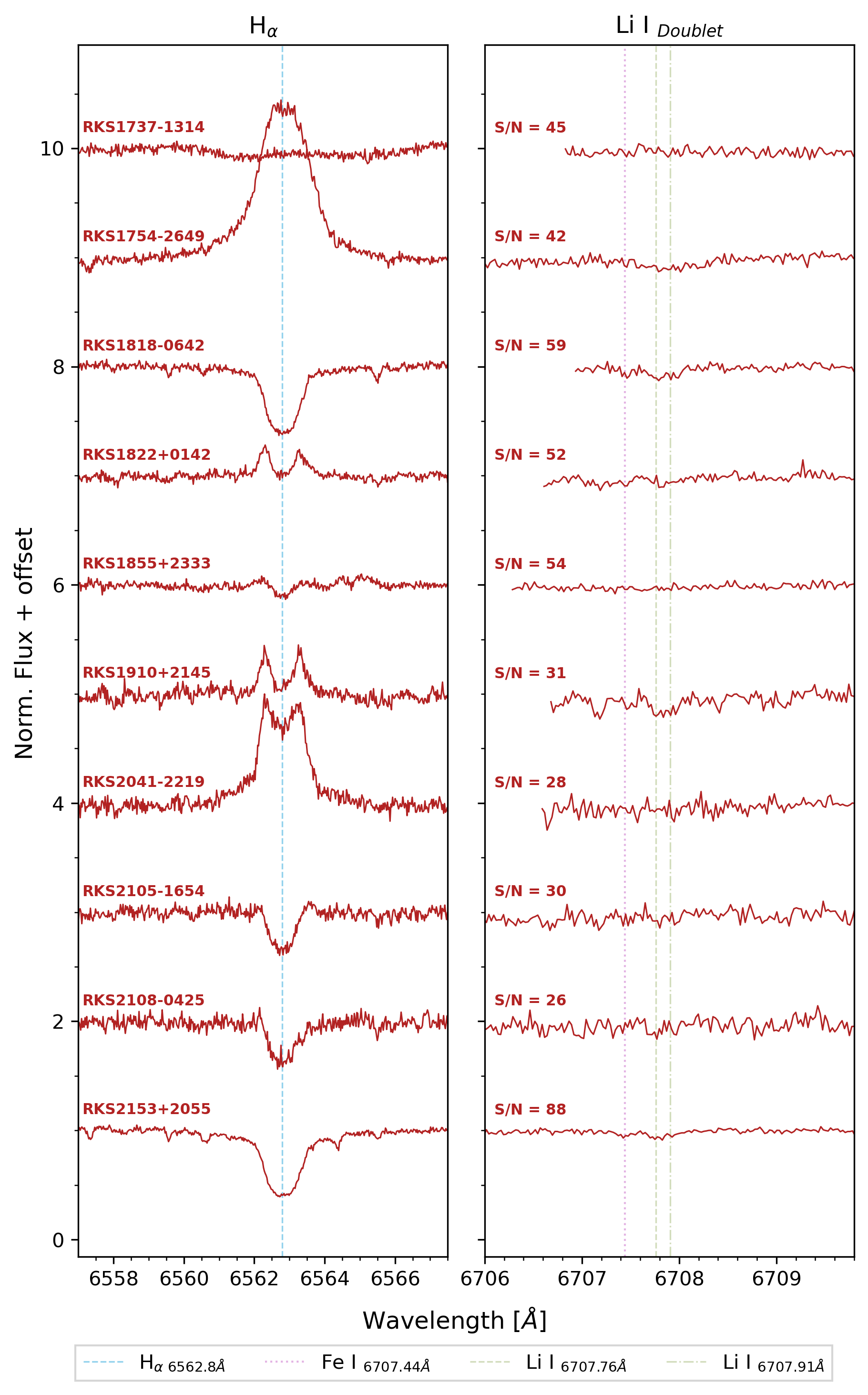}
  \caption{Group D: Spectra of young and active K dwarfs. Stars shown: RKS1737-1314 (A), RKS1754-2649 (Y+A), RKS1818-0642 (Y), RKS1822+0142 (A), RKS1855+2333 (A), RKS1910+2145 (Y+A), RKS2041-2219 (A), RKS2105-1654 (A), RKS2108-0425 (A), RKS2153+2055 (Y+A).}
  \label{fig:groupD}
\end{figure}

\begin{figure}[htbp]
  \centering
  \includegraphics[width=0.7\textwidth]{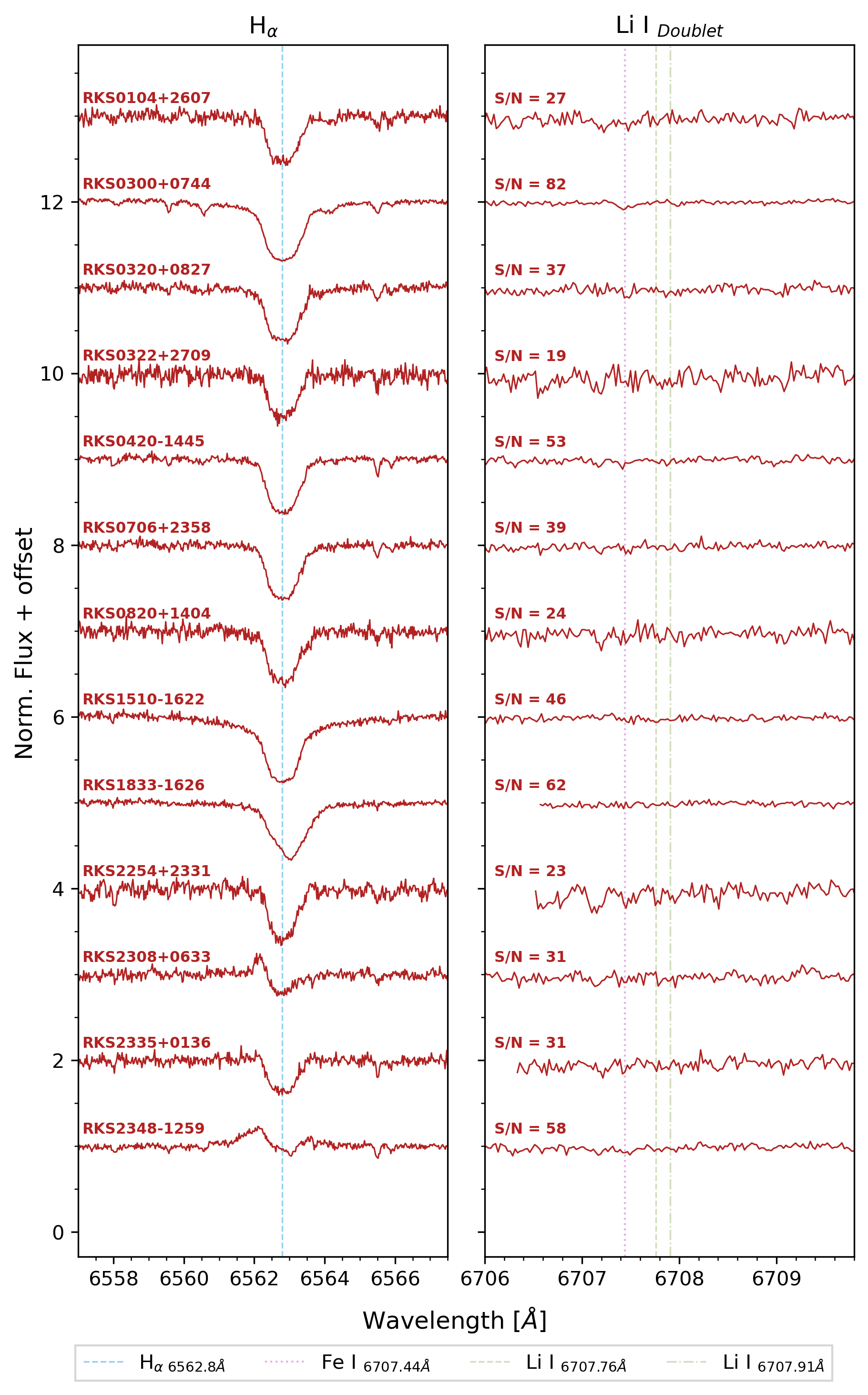}
  \caption{Group E: Spectra of young and active K dwarfs, plus moving group members and peculiar systems. Stars shown: RKS0104+2607 (MG Hyades), RKS0300+0744 (MG Hyades), RKS0320+0827 (MG Hyades), RKS0322+2709 (MG Hyades), RKS0420-1445 (MG Hyades), RKS0706+2358 (MG AB Dor), RKS0820+1404 (MG AB Dor), RKS1510-1622 (Halo), RKS1833-1626 (New SB2), RKS2254+2331 (MG Hyades), RKS2308+0633 (A), RKS2335+0136 (A), RKS2348-1259 (A).}
  \label{fig:groupE}
\end{figure}

\clearpage

\section{Astrometry and Photometry Data for the Survey Sample K Dwarfs}
\label{append:append_generaltable}
\begin{longrotatetable}
\begin{deluxetable*}{lcccrrccccc}
\tablecaption{Primary K dwarf Sample: Astrometry and Photometry Data \label{tab:astrometry}}
\tablewidth{0pt}
\tablehead{
\multicolumn{2}{c}{Star Identifiers} & \multicolumn{2}{c}{Coordinates} & \multicolumn{2}{c}{$\mu$} & & & \multicolumn{3}{c}{Photometry} \\
\cline{1-2} \cline{3-4} \cline{5-6} \cline{9-11} \\
RKS ID & Simbad Name & R.A. & Decl. & $\mu_{R.A}$ & $\mu_{Decl}$ & $\pi$ & Dist. & $B_{Gaia DR3}$ & $B_G-K_S$ & $M_{B_G}$ \\
& & (hh mm ss) & ($\pm$dd mm ss) & (mas yr$^{-1}$) & (mas yr$^{-1}$) & (mas) & (pc) & (mag) & (mag) & (mag)
}
\startdata
RKS0000$+$1659 & HD 224808 & 00 00 48.1 & $+$16 59 17 & $-$52.2 & $-$307.39 & 33.26 & 30.07 & 9.02 & 2.68 & 6.63 \\
RKS0001$-$1656 & BD$-$17 6862 & 00 01 25.8 & $-$16 56 54 & 293.9 & $-$253.44 & 31.76 & 31.49 & 11.00 & 3.78 & 8.51 \\
RKS0007$-$2349 & HD 283 & 00 07 32.5 & $-$23 49 07 & 320.87 & 91.24 & 30.49 & 32.79 & 8.91 & 2.23 & 6.33 \\
RKS0012$+$2142 & G 131$-$35 & 00 12 33.5 & $+$21 42 48 & 190.42 & $-$287.33 & 31.94 & 31.31 & 11.97 & 3.93 & 9.49 \\
RKS0012$+$2705 & BD$+$26 8 & 00 12 04.0 & $+$27 05 56 & 283.17 & $-$82.13 & 38.59 & 25.92 & 8.95 & 2.65 & 6.88 \\
RKS0016$-$1435 & BD$-$15 36 & 00 16 11.0 & $-$14 35 27 & 0.79 & $-$6.44 & 30.71 & 32.56 & 9.81 & 3.01 & 7.25 \\
RKS0017$+$2057 & G 131$-$51 & 00 17 59.1 & $+$20 57 24 & $-$252.05 & $-$366.92 & 36.10 & 27.70 & 11.34 & 3.93 & 9.13 \\
RKS0019$-$0303 & LP 644$-$95 & 00 19 12.3 & $-$03 03 13 & $-$26.88 & $-$217.81 & 32.06 & 31.19 & 11.16 & 3.70 & 8.69 \\
\enddata

\tablecomments{This table presents the complete sample of 589 primary K dwarf stars in the equatorial 33.3 pc volume-limited sample. Column descriptions: RKS ID is the RECONS K Star identifier; R.A. and Decl. are J2000.0 coordinates; $\mu_{R.A.}$ and $\mu_{Decl.}$ are proper motions from Gaia DR3 with uncertainties; $\pi$ is parallax with uncertainty; Dist. is distance calculated from parallax; $B_{Gaia}$ is the Gaia G-band magnitude; $B_G-K_S$ is the color index using 2MASS $K_S$ photometry; $M_{B_G}$ is the absolute G-band magnitude. This represents the final sample used for all analyses presented in this work.}

\end{deluxetable*}
\end{longrotatetable}

\clearpage

\section{Kinematic Properties of Survey Sample K Dwarfs}\label{append:append_kinematicstable}
\begin{longtable}{lrrrrrrcl}
\caption{Primary K dwarf Sample: Kinematic Properties} \label{tab:table_kinematics} \\
\hline\hline
RKS ID & $\gamma_{\rm Gaia~DR3}$ & $\gamma_{\rm REC}$ & $\sigma_{\gamma_{\rm REC}}$ & $U$ & $V$ & $W$ & Group & Disk \\
& (km s$^{-1}$) & (km s$^{-1}$) & (km s$^{-1}$) & (km s$^{-1}$) & (km s$^{-1}$) & (km s$^{-1}$) & & \\
\hline
\endfirsthead

\multicolumn{9}{c}%
{{\bfseries Table \thetable\ continued from previous page}} \\
\hline
RKS ID & $\gamma_{\rm Gaia~DR3}$ & $\gamma_{\rm REC}$ & $\sigma_{\gamma_{\rm REC}}$ & $U$ & $V$ & $W$ & Group & Disk \\
& (km s$^{-1}$) & (km s$^{-1}$) & (km s$^{-1}$) & (km s$^{-1}$) & (km s$^{-1}$) & (km s$^{-1}$) & & \\
\hline
\endhead

\hline \multicolumn{9}{r}{{Continued on next page}} \\
\endfoot

\hline
\endlastfoot

RKS0000$+$1659 & $-$17.92   & $-$17.65    & 0.04        & 29.50   & $-$33.84  & $-$16.46 & Field  &  Thin disk   \\
RKS0001$-$1656 & 0.47     & 0.58      & 0.04        & $-$20.08  & $-$51.95  & $-$15.93 & Field  &  Thin disk   \\
RKS0007$-$2349 & $-$43.17   & $-$43.08    & 0.04        & $-$55.34  & $-$16.44  & 34.82  & Field  &  Thin disk   \\
RKS0012$+$2142 & $-$5.02    & $-$5.01     & 0.05        & $-$5.64   & $-$39.08  & $-$32.91 & Field  &  Thin disk   \\
RKS0012$+$2705 & 12.50    & 12.57     & 0.04        & $-$30.26  & $-$11.37  & $-$20.62 & Field  &  Thin disk   \\
RKS0016$-$1435 & $-$13.51   & $-$13.62    & 0.04        & 0.26    & $-$4.41   & 12.92  & Field  &  Thin disk   \\
RKS0017$+$2057 & $-$49.24   & $-$49.51    & 0.05        & 62.59   & $-$44.15  & 1.22   & Field  &  Thick disk   \\
RKS0019$-$0303 & $-$20.92   & $-$20.97    & 0.05        & 21.36   & $-$31.58  & 6.30   & Field  &  Thin disk   \\
\end{longtable}






\clearpage

\section{Table of Stellar Properties for Survey Sample K Dwarfs}
\label{append:append_stellarpropertiestable}

\vspace{2ex}
{\raggedright \textbf{Table Key:} \par}

{\raggedright EW[H$\alpha$]: Equivalent width of the H$\alpha$ line at 6563 \text{\normalfont\AA} \par}
{\raggedright EW[Li I]: Equivalent width of the Li I line at 6707.8 \text{\normalfont\AA} \par}	
{\raggedright NM: Not Measured \par}

\vspace{2ex}

{\raggedright \textbf{In the Status column:} \par}
{\raggedright M represents Mature K dwarfs identified by this work. \par}
{\raggedright Y represents Youth, \par}
{\raggedright A represents Active, \par}
{\raggedright Y$+$A represents both Youth and Active, \par}

\footnotesize
\begin{longtable}{lcccccccccccc}
\caption{Primary K dwarf Sample: Stellar Properties} \label{tab:stellar_properties} \\
\hline
RKS ID & $T_{\rm eff}$ & $\sigma_{T_{\rm eff}}$ & [Fe/H] & $\sigma_{\rm [Fe/H]}$ & $\log g$ & $\sigma_{\log g}$ & $v \sin i$ & EW(H$\alpha$) & S/N(H$\alpha$) & EW(Li I) & S/N(Li I) & Status \\
& (K) & (K) & (dex) & (dex) & (dex) & (dex) & (km s$^{-1}$) & (\AA) & & (\AA) & & \\
\hline
\endfirsthead

\multicolumn{13}{c}%
{{\bfseries Table \thetable\ continued from previous page}} \\
\hline
RKS ID & $T_{\rm eff}$ & $\sigma_{T_{\rm eff}}$ & [Fe/H] & $\sigma_{\rm [Fe/H]}$ & $\log g$ & $\sigma_{\log g}$ & $v \sin i$ & EW(H$\alpha$) & S/N(H$\alpha$) & EW(Li I) & S/N(Li I) & Status \\
& (K) & (K) & (dex) & (dex) & (dex) & (dex) & (km s$^{-1}$) & (\AA) & & (\AA) & & \\
\hline
\endhead

\hline \multicolumn{13}{r}{{Continued on next page}} \\
\endfoot

\hline
\endlastfoot

RKS0000$+$1659 & 4896 & 101 & $-$0.12 & 0.10 & 4.52 & 0.68 & $<$5 & 0.98 & 88 & 0.00 & 88 & M\\
RKS0001$-$1656 & 4017 & 393 & 0.05 & 0.23 & 4.67 & 0.70 & $<$5 & 0.51 & 32 & 0.00 & 30 & M\\
RKS0007$-$2349 & 5212 & 109 & $-$0.50 & 0.10 & 4.54 & 0.68 & $<$5 & 1.14 & 71 & 0.00 & 71 & M\\
RKS0012$+$2142 & 3764 & 111 & $-$0.18 & 0.09 & 4.75 & 0.71 & $<$5 & 0.43 & 23 & 0.00 & 21 & M\\
RKS0012$+$2705 & 4805 & 100 & $-$0.02 & 0.09 & 4.50 & 0.68 & $<$5 & 0.88 & 64 & 0.00 & 62 & M\\
RKS0016$-$1435 & 4580 & 100 & 0.06 & 0.09 & 4.57 & 0.69 & $<$5 & 0.79 & 55 & 0.00 & 54 & M\\
RKS0017$+$2057 & 3806 & 106 & $-$0.01 & 0.14 & 4.70 & 0.71 & $<$5 & 0.45 & 36 & 0.00 & 34 & M\\
RKS0019$-$0303 & 3949 & 100 & $-$0.23 & 0.11 & 4.71 & 0.71 & $<$5 & 0.51 & 27 & 0.00 & 25 & M\\
\end{longtable}

\clearpage

\section{Table of CHIRON Observations}
\label{append:tablechironobs}

\footnotesize
\begin{longtable}{l r r r c c c}
\caption{List of Spectra Used in the Final Analysis} \label{tab:table_obslog} \\
\toprule
RKS ID                          & \multicolumn{1}{c}{R.A.}     & \multicolumn{1}{c}{Decl.}    & \multicolumn{1}{c}{V}    &  \multicolumn{1}{c}{Date \& Time of Obs}    & \multicolumn{1}{c}{Exp. Time}  & \multicolumn{1}{c}{SNR} \\
\multicolumn{1}{c}{...}         & \multicolumn{1}{c}{(hh mm ss)} & \multicolumn{1}{c}{(dd mm ss)} & \multicolumn{1}{c}{(mag)}  & \multicolumn{1}{c}{(UTC)}             & \multicolumn{1}{c}{(s)}          & \multicolumn{1}{c}{...} \\
\midrule
\endfirsthead

\multicolumn{7}{c}%
{\tablename\ \thetable\ $-$$-$ \textit{Continued from previous page}} \\
\toprule
RKS ID                  & \multicolumn{1}{c}{R.A.}     & \multicolumn{1}{c}{Decl.}    & \multicolumn{1}{c}{V}             &  \multicolumn{1}{c}{Date \& Time of Obs}    & \multicolumn{1}{c}{Exp. Time} & \multicolumn{1}{c}{SNR} \\
\multicolumn{1}{c}{...}         & \multicolumn{1}{c}{(hh mm ss)} & \multicolumn{1}{c}{(dd mm ss)} & \multicolumn{1}{c}{(mag)}  & \multicolumn{1}{c}{(UTC)}             & \multicolumn{1}{c}{(s)}          & \multicolumn{1}{c}{...} \\
\midrule
\midrule
\endhead

\midrule
\multicolumn{7}{r}{\textit{Continued on next page}} \\
\endfoot

\bottomrule
\endlastfoot

RKS0000$+$1659 & 00 00 48.1 & $+$16 59 17 & 8.8  & 2021$-$12$-$06 0:34:29  & 900  & 88  \\
RKS0001$-$1656 & 00 01 25.8 & $-$16 56 54 & 10.8 & 2018$-$12$-$01 0:59:52  & 900  & 31  \\
RKS0007$-$2349 & 00 07 32.5 & $-$23 49 07 & 8.7  & 2020$-$12$-$14 1:22:54  & 900  & 71  \\
RKS0012$+$2142 & 00 12 33.5 & $+$21 42 48 & 11.8 & 2021$-$08$-$13 7:17:31  & 1200 & 29  \\
RKS0012$+$2705 & 00 12 04.0 & $+$27 05 56 & 8.7  & 2018$-$11$-$08 2:21:41  & 900  & 63  \\
RKS0016$-$1435 & 00 16 11.0 & $-$14 35 27 & 10.0 & 2021$-$08$-$13 6:58:21  & 900  & 55  \\
RKS0017$+$2057 & 00 17 59.1 & $+$20 57 24 & 11.0 & 2021$-$07$-$27 7:42:39  & 1800 & 36  \\
RKS0019$-$0303 & 00 19 12.3 & $-$03 03 13 & 10.9 & 2018$-$12$-$30 1:03:13  & 900  & 31  \\
\end{longtable}

\end{document}